\documentclass[12pt,a4paper]{article}

\usepackage[height=22cm, width=16cm, centering]{geometry}

\usepackage{amssymb}
\usepackage{amsmath,latexsym}
\usepackage{graphicx}
\usepackage{color}

\def\thefootnote{\fnsymbol{footnote}}

\newcommand{\be}{\begin{equation}}
\newcommand{\ee}{\end{equation}}
\newcommand{\bea}{\begin{eqnarray}}
\newcommand{\eea}{\end{eqnarray}}
\newcommand{\bdm}{\begin{displaymath}}
\newcommand{\edm}{\end{displaymath}}

\newcommand{\FF}{{\cal F}_{{\pi^0}^*\gamma^*\gamma^*}}
\newcommand{\FFa}{{\cal F}_{\pi^0\gamma^*\gamma^*}}
\newcommand{\FFeta}{{\cal F}_{\eta\gamma^*\gamma^*}}
\newcommand{\FFetaprime}{{\cal F}_{\eta^\prime\gamma^*\gamma^*}}
\newcommand{\FFP}{{\cal F}_{{\rm P}\gamma^*\gamma^*}}

\newcommand{\HLbLpi}{\mathrm{HLbL};\pi^0}
\newcommand{\HLbLpione}{\mathrm{HLbL};\pi^0(1)}
\newcommand{\HLbLpitwo}{\mathrm{HLbL};\pi^0(2)}
\newcommand{\HLbLP}{\mathrm{HLbL; P}}
\newcommand{\HLbLPone}{\mathrm{HLbL; P(1)}}
\newcommand{\HLbLPtwo}{\mathrm{HLbL; P(2)}}

\newcommand{\VVP}{\langle V \! V \! P\rangle}
\newcommand{\order}{{\cal O}}

\DeclareMathOperator{\arccot}{arccot}

\begin{document}

\begin{titlepage}

\begin{flushright}
{\small 
 February 10, 2016 \\
 MITP/16-018
}
\end{flushright}

\vspace*{0.5cm}

\begin{center}
{\Large {\bf 
On the precision of a data-driven estimate of \\  
hadronic light-by-light scattering in the muon $g-2$: \\[0.25cm]
pseudoscalar-pole contribution}} \\[1cm]
Andreas Nyf\/feler\footnote{nyf\/feler@kph.uni-mainz.de} \\[0.5cm]
Institut f\"ur Kernphysik and PRISMA Cluster of Excellence, \\ 
Johannes Gutenberg-Universit\"at Mainz,   
D-55128 Mainz, Germany 
\end{center}

\vspace*{0.5cm} 
\begin{abstract} 
  The evaluation of the numerically dominant pseudoscalar-pole
  contribution to hadronic light-by-light scattering in the muon $g-2$
  involves the pseudoscalar-photon transition form factor ${\cal
    F}_{{\rm P}\gamma^*\gamma^*}(-Q_1^2, -Q_2^2)$ with ${\rm P} =
  \pi^0, \eta, \eta^\prime$ and, in general, two off-shell photons
  with spacelike momenta $Q_{1,2}^2$. We show, in a largely
  model-independent way, that for $\pi^0~(\eta, \eta^\prime)$ the
  region of photon momenta below about $1~(1.5)~\mbox{GeV}$ gives the
  main contribution to hadronic light-by-light scattering. We then
  discuss how the precision of current and future measurements of the
  single- and double-virtual transition form factor in different
  momentum regions impacts the precision of a data-driven estimate of
  this contribution to hadronic light-by-light scattering. Based on
  Monte Carlo simulations for a planned first measurement of the
  double-virtual form factor at BESIII, we find that for the $\pi^0,
  \eta, \eta^\prime$-pole contributions a precision of $14\%, 23\%,
  15\%$ seems feasible. Further improvements can be expected from
  other experimental data and also from the use of dispersion
  relations for the different form factors themselves.
\end{abstract}


\end{titlepage}


\renewcommand{\thefootnote}{\arabic{footnote}}
\setcounter{footnote}{0}

\section{Introduction}

The anomalous magnetic moment of the muon $a_\mu = (g-2) / 2$ has
served for many years as important test of the Standard Model (SM) of
particle physics, see Refs.~\cite{Jegerlehner_Book, JN_09, g-2_review,
  PDG_2015} which review theory and experiment and contain many
references to earlier work. The contributions from the different
sectors in the SM and the current experimental value, largely
dominated by the measurement at Brookhaven~\cite{g-2_exp}, corrected
for a small shift in the ratio of the magnetic moments of the muon and
the proton~\cite{shift_ratio}, have been collected in
Table~\ref{Tab:g-2_status}. For the muon $g-2$, all sectors of the SM
contribute significantly at the current level of precision.  The QED
contribution dominates numerically, but it is very precisely known up
to 5-loop order~\cite{Aoyama_et_al_12}. Also the electroweak
contribution is under control at the two-loop level, including a small
hadronic uncertainty and estimates of leading three-loop
contributions~\cite{Gnendiger_et_al_13}. The main source of
uncertainty originates from the hadronic contributions from vacuum
polarization (HVP) and light-by-light scattering (HLbL) at various
orders in the electromagnetc coupling $\alpha$. Comparing SM theory
and experiment, a discrepancy of $3-5$ standard deviations is observed
for several years now.\footnote{The absolute and relative size of the
  deviation depends on the treatment of the hadronic contributions and
  how aggressively the errors are estimated. See
  Ref.~\cite{HVP_recent_estimates} for other recent evaluations of the
  HVP contribution.} This deviation could be a sign of New Physics
beyond the SM~\cite{Jegerlehner_Book, JN_09, g-2_review, PDG_2015}, but the
large hadronic uncertainties make it difficult to draw firm
conclusions. These uncertainties need to be reduced and better
controlled~\cite{g-2_hadronic}, also in view of planned future muon
$g-2$ experiments at Fermilab and J-PARC which will try to reduce the
experimental error by a factor of four to about $\delta a_\mu^{\rm
  exp} = 16 \times 10^{-11}$~\cite{future_g-2_exp}.

\begin{table}[h]  

  \caption{Contributions to the muon $g-2$ from the different sectors in
    the SM and comparison of theory and experiment.}   

\label{Tab:g-2_status}

\begin{center}
\renewcommand{\arraystretch}{1.2}
\begin{tabular}{|l|r@{}l@{$\,\pm\,$}r@{}l|c|}
\hline
\multicolumn{1}{|c|}{Contribution}  & \multicolumn{4}{c}{$a_{\mu}
  \times 10^{11}$} & \multicolumn{1}{|c|}{Reference} \\  
\hline
QED (leptons + photons) & 116~584~718 & .853 &  0 & .036 &
\cite{Aoyama_et_al_12} \\  
Electroweak   &         153 & .6   &  1 & .0   & \cite{Gnendiger_et_al_13} \\
HVP: LO       &        6907 & .5   & 47 & .2   &
\cite{Jegerlehner_Szafron_11} \\ 
\hspace*{1.15cm}NLO  & -100 & .3   &  2 & .2   &
\cite{Jegerlehner_Szafron_11} \\  
\hspace*{1.15cm}NNLO &   12 & .4   &  0 & .1   & \cite{Kurz_et_al_14} \\ 
HLbL          &         116 &      & 40 &      & \cite{N_09, JN_09} \\ 
\hspace*{1.15cm}NLO   &   3 &      &  2 &      & \cite{HLbL_at_NLO_14} \\ 
Theory (SM)   & 116~591~811 &      & 62 &      & $-$ \\
\hline
Experiment    & 116~592~089 &      & 63 &      & \cite{g-2_exp,
  shift_ratio} \\  
Experiment - Theory ($3.1~\sigma$) 
              &  278 &      & 88 &      &  $-$ \\
\hline
\end{tabular}
\end{center} 

\end{table}

While the HVP contribution can be improved systematically with
measurements of the cross section $\sigma(e^+ e^- \to
\mbox{hadrons})$, the often used estimates for HLbL
\bea
a_\mu^{\mathrm{HLbL}} & = & (105 \pm 26) \times 10^{-11}, \qquad
\mbox{\cite{PdeRV_09}} \label{HLbL_PdeRV} \\ 
a_\mu^{\mathrm{HLbL}} & = & (116 \pm 40) \times 10^{-11}, \qquad
\mbox{\cite{N_09,JN_09}}  \label{HLbL_JN} 
\eea
are both largely based on the same model calculations~\cite{HKS, HK,
  BPP, KN_02, MV_04}, which suffer from partly uncontrollable
uncertainties. Therefore the error estimates in
Eqs.~(\ref{HLbL_PdeRV}) and (\ref{HLbL_JN}) are essentially just
guesses. Note that the central values probably need to be shifted
downwards to $a_\mu^{\mathrm{HLbL}} = (98 \pm 26) \times 10^{-11}$ for
\cite{PdeRV_09} and $a_\mu^{\mathrm{HLbL}} = (102 \pm 40) \times
10^{-11}$ for \cite{N_09, JN_09}, since recent
reevaluations~\cite{Pauk_Vanderhaeghen_single_mesons,
  Jegerlehner_MITP_14, Jegerlehner_FCCP_15} of the axial-vector
contribution yield a smaller value $a_\mu^{\mathrm{HLbL; axial}} = (8
\pm 3) \times 10^{-11}$ compared to the result $a_\mu^{\mathrm{HLbL;
    axial}} = (22 \pm 5) \times 10^{-11}$ obtained in
Ref.~\cite{MV_04} and used in Refs.~\cite{N_09,
  JN_09}. Ref.~\cite{PdeRV_09} had already used a somewhat smaller
value $a_\mu^{\mathrm{HLbL; axial}} = (15 \pm 10) \times
10^{-11}$. Furthermore, after the publication of Refs.~\cite{PdeRV_09,
  N_09,JN_09}, there were claims in Refs.~\cite{DSE,
  nonlocal_chiQM_quark_loop} using different models that the dressed
quark-loop contribution might be around $110 \times 10^{-11}$, i.e.\
much bigger than for instance the value $(21 \pm 3) \times 10^{-11}$
estimated in Ref.~\cite{BPP}. Moreover, in Ref.~\cite{pion-loop} it
was argued that also the pion-loop contribution could be potentially
bigger in absolute size, $-(11 - 71) \times 10^{-11}$, compared to
$-(19 \pm 13) \times 10^{-11}$ in Ref.~\cite{BPP}. See
Ref.~\cite{Bijnens_FCCP_15} for an analysis of these claims and a
brief review on other recent developments in HLbL.

There are attempts ongoing to calculate the HLbL contribution to the
muon $g-2$ from first principles in Lattice QCD. A first, still
incomplete, result was obtained recently in
Ref.~\cite{Lattice_HLbL_Blum_et_al} which contains references to
earlier work in the last years. Another approach was proposed in
Ref.~\cite{Lattice_HLbL_Mainz}. It remains to be seen, how fast
reliable estimates for HLbL can be obtained within Lattice QCD, where
all systematic uncertainties of the extrapolations to physical quark
masses (physical pion masses), to the continuum and to infinite volume
are fully under control and also all quark-disconnected contributions
are included.

In this situation, a dispersive approach to HLbL was proposed recently
in Refs.~\cite{HLbL_DR_Bern_Bonn, HLbL_DR_Mainz}, which tries, in the
spirit of the HVP calculation, to relate the presumably numerically
dominant contributions from the pseudoscalar-{\it poles} and the
pion-loop with {\it on-shell} intermediate pseudoscalar states to, in
principle, measurable form factors and cross-sections with off-shell
photons:
\bea 
\gamma^* \gamma^* & \to & \pi^0, \eta, \eta^\prime, \\
\gamma^* \gamma^* & \to & \pi^+ \pi^-, \pi^0 \pi^0. 
\eea 
The two dispersive approaches differ
somewhat. Ref.~\cite{HLbL_DR_Bern_Bonn} considers first the four-point 
function $\langle VVVV \rangle$ with three off-shell and one on-shell
photons, then identifies the intermediate on-shell hadronic states and
then projects on the muon $g-2$. On the other hand,
Ref.~\cite{HLbL_DR_Mainz} writes down directly a dispersion relation
for the Pauli form factor $F_2(k^2)$ and evaluates the imaginary part
of $F_2(k^2)$ from the various multi-particle cuts with hadrons and
photons in the Feynman diagrams, and then calculates $a_\mu = F_2(0)$.

The hope is that this data-driven estimate for HLbL will allow a
$10\%$ precision for these contributions, with a reliable and
controllable error related to the experimental measurement precision,
and that the remaining, hopefully smaller contributions, e.g.\ from
axial-vectors ($3\pi$-intermediate state), other heavier states and a
dressed quark-loop, properly matched to perturbative QCD and avoiding
double-counting, can be obtained within models with about $30\%$
uncertainty to reach an overall, reliable precision goal of about
$20\%~(\delta a_\mu^{\mathrm{HLbL}} \approx 20 \times 10^{-11})$.

\begin{figure}[h!]

\includegraphics[width=\textwidth]{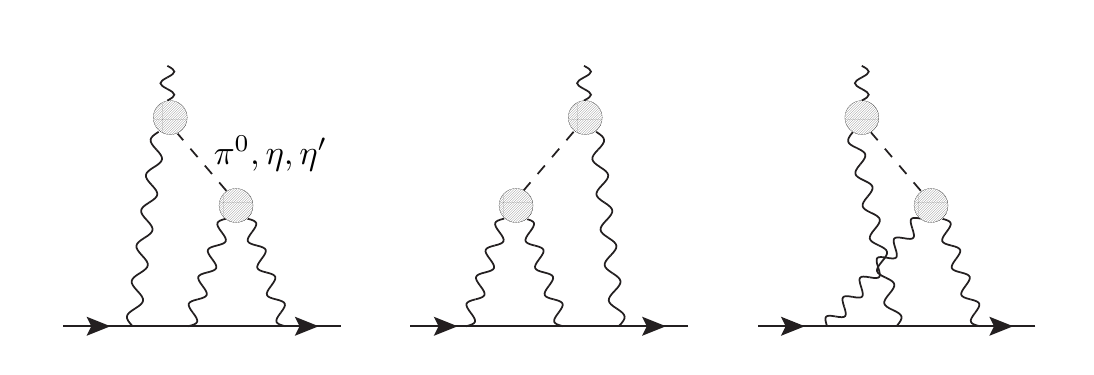}    

\caption{The pseudoscalar-pole contribution to hadronic light-by-light
  scattering. The shaded blobs represent the transition form factor
  ${\cal F}_{{\rm P} \gamma^* \gamma^*}(q_1^2,q_2^2)$ where ${\rm P} =
  \pi^0, \eta, \eta^\prime$.}  

\label{Fig:HLbL_PS-poles} 

\end{figure}

In this paper we will concentrate on the dispersive approach to the
pseudoscalar-pole contribution to HLbL which is numerically dominant
according to most model calculations. It arises from the one-particle
intermediate states of the light pseudoscalars $\pi^0, \eta,
\eta^\prime$ shown in the Feynman diagrams in
Fig.~\ref{Fig:HLbL_PS-poles}.
The blobs in the Feynman diagrams represent the double-virtual
transition form factor ${\cal F}_{{\rm P} \gamma^*
  \gamma^*}(q_1^2,q_2^2)$ where ${\rm P} = \pi^0, \eta, \eta^\prime$.
See Ref.~\cite{TFF_mini_review} for a recent brief overview on
transition form factors (TFF), many more details can be found in the
older review~\cite{Landsberg}.\footnote{More generally, one can define
  a pseudoscalar-{\it exchange} contribution to HLbL which involves a
  form factor with off-shell pseudoscalars ${\cal F}_{{\rm P}^*
    \gamma^* \gamma^*}((q_1 + q_2)^2, q_1^2,q_2^2)$~\cite{HKS, BPP,
    Jegerlehner_essentials, N_09, JN_09}, but then the contribution to
  HLbL is model-dependent. In particular, it will depend on the
  interpolating field used for the pseudoscalars.}

In order to simplify the notation, we will now discuss mainly the
neutral pion-pole contribution. The generalization to the pole
contributions of $\eta$ and $\eta^\prime$ is straightforward.  The
pion-photon transition form factor $\FFa(q_1^2,q_2^2)$ is defined by
the following vertex function in QCD:
\be  \label{TFF} 
i \int d^4 x \, e^{i q_1 \cdot x}  \langle 0 | T \{ j_\mu(x) j_\nu(0) \}|
\pi^0(q_1 + q_2) \rangle 
= \varepsilon_{\mu\nu\alpha\beta} \, q_1^\alpha \, q_2^\beta \,  
 \FFa(q_1^2, q_2^2) \, . 
\ee 
Here $j_\mu(x) = ({\overline \psi} \hat Q \gamma_\mu \psi)(x)$ is the
light quark part of the electromagnetic current (${\overline \psi}
\equiv ({\overline u}, {\overline d}, {\overline s}$) and $\hat Q =
\mbox{diag}(2,-1,-1)/3$ is the charge matrix). The form factor
describes the interaction of an on-shell neutral pion with two
off-shell photons with four-momenta $q_1$ and $q_2$. It is Bose
symmetric, $\FFa(q_1^2,q_2^2) = \FFa(q_2^2,q_1^2)$, because the two
photons are indistinguishable.  The form factor for real photons is
related to the decay width into two photons: $\FFa^2(0,0) = 4
\Gamma(\pi^0 \to \gamma\gamma) / (\pi \alpha^2 m_\pi^3)$. Often the
normalization with the chiral anomaly is used $\FFa(0,0) = - N_c
/(12\pi^2 F_\pi)$.

If one evaluates only the pion-pole contribution of the Feynman
diagrams and projects on the muon $g-2$, one obtains the
result~\cite{KN_02}
\bea 
\label{amupi0_start}
a_\mu^{\HLbLpi} & = & \left( \frac{\alpha}{\pi} \right)^3 \left[
  a_\mu^{\HLbLpione} + a_\mu^{\HLbLpitwo} \right],  \\ 
a_\mu^{\HLbLpione} & = & 
\int \frac{d^4 q_1}{(2\pi)^4} \frac{d^4 q_2}{(2\pi)^4}
\,\frac{1}{q_1^2 q_2^2 (q_1 + q_2)^2[(p+ q_1)^2 - m_\mu^2][(p - q_2)^2 -
    m_\mu^2]} \nonumber \\
&& 
\times 
\frac{ \FFa(q_1^2, (q_1 + q_2)^2) \ \ \FFa(q_2^2, 0)}{q_2^2 -
  m_{\pi}^2} \ \tilde T_1(q_1,q_2;p) \, ,  
\label{amupi0_start_1} \\
a_\mu^{\HLbLpitwo} & = & 
\int \frac{d^4 q_1}{(2\pi)^4} \frac{d^4 q_2}{(2\pi)^4}
\,\frac{1}{q_1^2 q_2^2 (q_1 + q_2)^2[(p+ q_1)^2 - m_\mu^2][(p - q_2)^2 -
    m_\mu^2]} \nonumber \\
&& 
\times 
\frac{ \FFa(q_1^2, q_2^2) \ \ \FFa((q_1 + q_2)^2, 0)}{(q_1 +
  q_2)^2 - m_{\pi}^2} \ \tilde T_2(q_1,q_2;p) \, , 
\label{amupi0_start_2}
\eea 
where $p^2 = m_\mu^2$ (on-shell muon) and the external photon has now
zero four-momentum (soft photon). The kinematic functions $\tilde
T_{1,2}(q_1,q_2;p)$ are reproduced in
Appendix~\ref{App:weight_functions}.  The first and the second graphs
in Fig.~\ref{Fig:HLbL_PS-poles} give rise to identical contributions,
leading to the term with $\tilde T_1$. The third graph yields the
contribution involving $\tilde T_2$.

There have been objections~\cite{Vainshtein_FCCP_15} raised recently
about the implementation of the dispersive approach for the pion-pole
contribution in Refs.~\cite{HLbL_DR_Bern_Bonn,
  HLbL_DR_Mainz}. According to the arguments in Refs.~\cite{MV_04,
  Vainshtein_FCCP_15} there should be no form factor at the external
vertex with the soft-photon. This amounts to setting the
single-virtual form factors $\FFa(q^2, 0)$ in
Eqs.~(\ref{amupi0_start_1}) and (\ref{amupi0_start_2}) to a
constant. Maybe the disagreement arises whether one interprets the
diagrams in Fig.~\ref{Fig:HLbL_PS-poles} as genuine Feynman diagrams
contributing to the muon $g-2$ or as unitarity diagrams often used in
the context of dispersive approaches. In Ref.~\cite{HLbL_DR_Mainz} it
was shown, however, that when one writes down a dispersion relation
for the Pauli form factor $F_2(k^2)$ and evaluates the imaginary part
of $F_2(k^2)$ from the various two-particle and three-particle cuts in
the Feynman diagrams, and then calculates $a_\mu = F_2(0)$, one
obtains for a simple Vector Meson Dominance (VMD) model for the form
factor exactly the expressions in Eqs.~(\ref{amupi0_start_1}) and
(\ref{amupi0_start_2}) {\it with} a form factor at the external
vertex. We will therefore use the prescription from Refs.~\cite{KN_02,
  HLbL_DR_Bern_Bonn, HLbL_DR_Mainz} to study the pseudoscalar-pole
contribution $a_\mu^{\HLbLP}$ to HLbL.

Most model evaluations of $a_\mu^{\HLbLpi}$ (pion-pole defined in
different ways or pion-exchange with off-shell-pion form factors) and
$a_\mu^{\HLbLP}$, with ${\rm P} = \pi^0, \eta, \eta^\prime$, agree at
the level of $15\%$, but the full range of estimates (central values)
is much larger~\cite{HLbL_P_literature}: 
\bea 
a_{\mu;{\rm models}}^{\HLbLpi} & = & (50 - 80) \times 10^{-11} 
\ = \ (65 \pm 15) \times 10^{-11} \quad (\pm 23\%),  \label{range_HLbLpi0}
\\ 
a_{\mu;{\rm models}}^{\HLbLP} & = & (59 - 114) \times 10^{-11} 
\ = \ (87 \pm 27) \times 10^{-11} \quad (\pm 31\%). \label{range_HLbLP}
\eea

This situation has to be improved without relying too much on various
models, in particular before the new muon $g-2$ experiment at Fermilab
yields results with a four-fold improvement over the Brookhaven
experiment in a few years~\cite{future_g-2_exp}.  In this paper we
therefore study, as model-independently as possible, which are the
most important momentum regions for the pseudoscalar-pole contribution
$a_\mu^{\HLbLP}$. We also analyze what is the impact of the precision
of current and future measurements of the single-virtual $\FFP(q^2,0)$
and the double-virtual pseudoscalar transition form factor
$\FFP(q_1^2, q_2^2)$ in different momentum regions on the uncertainty
of a data-driven estimate of this contribution to HLbL.  We hope that
this information will be a valuable guide to help the experimental
community to plan, design and analyze the measurements of decay rates,
form factors and cross-sections of the light pseudoscalars $\pi^0,
\eta, \eta^\prime$ and their interactions with off-shell photons. In a
way, our approach is a generalization of the pie-charts often shown
for the HVP contribution and its error as a function of the
center-of-mass energy $\sqrt{s}$ in $\sigma(e^+ e^- \to
\mbox{hadrons})$~\cite{JN_09}. However, since HLbL involves different
amplitudes~\cite{BPP, HLbL_DR_Bern_Bonn, General_4pt_functions} which
depend on several invariant momenta, the situation is of course more
complicated than for the HVP contribution.

This paper is organized as follow. Section~\ref{Sec:3d_integral}
recalls the three-dimensional integral representation for the
pseudoscalar-pole contribution $a_\mu^\HLbLP$ derived in
Ref.~\cite{JN_09} which separates model-independent weight functions
$w_{1,2}(Q_1, Q_2, \cos\theta)$ from the dependence on the form factor
$\FFP(-Q_1^2, -Q_2^2)$ for spacelike (Euclidean) momenta with
magnitude $Q_{1,2}$ and angle $\theta$ between the momentum
vectors. In Section~\ref{Sec:weight_functions} the weight functions
for $\pi^0$, $\eta$ and $\eta^\prime$ are analyzed in detail. Several
three-dimensional plots (as function of the two momenta $Q_{1,2}$) and
one-dimensional plots (as function of the angle $\theta$) are shown
and the maxima and minima of the weight functions are determined. The
relevant momentum regions for $a_\mu^\HLbLP$ in different bins in the
$(Q_1, Q_2)$-plane are identified in
Section~\ref{Sec:momentum_regions} within two simple models for the
TFF, since the loop integral in Eq.~(\ref{amupi0_start_1}) diverges
without a form factor which dampens the large momentum
region. Section~\ref{Sec:TFF_experiment} summarizes the experimental
status on the precision of measurements of the TFF for $\pi^0, \eta$
and $\eta^\prime$. This is based on data for the two-photon decay
width $\Gamma({\rm P} \to \gamma\gamma)$, the slope of the form factor
at zero momentum and data for the single-virtual form factor
$\FFP(-Q^2,0)$ in the spacelike and timelike momentum region. For the
double-virtual form factor $\FFP(-Q_1^2, -Q_2^2)$ there is currently
no experimental data available. We use the results of a Monte Carlo
(MC) simulation~\cite{BESIII_private} for planned measurements of this
form factor at the BESIII detector to estimate the potential
precision which could be reached in the next few
years. Section~\ref{Sec:impact} then discusses the impact of the
experimental uncertainties for the TFF on $a_\mu^\HLbLP$ and points
out in which specific momentum regions (momentum bins) a high
experimental precision of TFF is needed for a precise data-driven
estimate of the pseudoscalar-pole contribution to HLbL. Finally,
Section~\ref{Sec:conclusions} presents a summary of our findings and
the conclusions. In Appendix~\ref{App:weight_functions} we reproduce
the formulae for the kinematic functions $\tilde T_{1,2}$ in the
loop-integrals in Eqs.~(\ref{amupi0_start_1}) and
(\ref{amupi0_start_2}) from Ref.~\cite{KN_02} and the weight functions
$w_{1,2}$ from Ref.~\cite{JN_09}. We give the Taylor expansions for
the weight functions in various limits (small and large momenta,
collinear momenta). A brief summary of the two form factor models that
we use in our numerical analysis can be found in
Appendix~\ref{App:TFF_models}.

\section{Three-dimensional integral representation for
  the pseudoscalar-pole contribution $a_\mu^\HLbLP$}  
\label{Sec:3d_integral}

We concentrate again mainly on the pion-pole contribution in this
Section.  After a Wick rotation to Euclidean momenta $Q_i, i=1,2,$ and
averaging over the direction of the muon momentum $p$ using the method
of Gegenbauer polynomials (hyperspherical
approach)~\cite{hyperspherical}, one can perform for arbitrary form
factors in the two-loop integrals~(\ref{amupi0_start_1}) and
(\ref{amupi0_start_2}) all angular integrations, except one over the
angle $\theta$ between the four-momenta $Q_1$ and $Q_2$ which also
appears through $Q_1 \cdot Q_2$ in the form factors. In this way one
obtains the following three-dimensional integral representation for
the pion-pole contribution with on-shell-pion transition form
factors~\cite{JN_09}
\bea 
a_\mu^{\HLbLpione} & = & \int_0^\infty \!dQ_1 \int_0^\infty \!dQ_2
\int_{-1}^{1} \!d\tau \, \,  w_1(Q_1,Q_2,\tau) \nonumber \\
& & \times \, \FFa(-Q_1^2, -(Q_1 + Q_2)^2) \, \, 
\FFa(-Q_2^2,0), \label{amupi0_1} \\  
a_\mu^{\HLbLpitwo} & = & \int_0^\infty \!dQ_1 \int_0^\infty \!dQ_2
\int_{-1}^{1} \!d\tau \, \, w_2(Q_1,Q_2,\tau) \nonumber \\
& & \times \, \FFa(-Q_1^2, -Q_2^2) \, \, \FFa(-(Q_1+Q_2)^2,0).
\label{amupi0_2}
\eea 
The integrations in Eqs.~(\ref{amupi0_1}) and (\ref{amupi0_2}) run
over the lengths of the two Euclidean four-momenta $Q_1$ and $Q_2$ and
the angle $\theta$ between them $Q_1 \cdot Q_2 = Q_1 Q_2 \cos\theta$.
We have written $Q_i \equiv |(Q_i)_{\mu}|, i=1,2$, for the length of
the four-vectors. Following Ref.~\cite{HLbL_DR_Bern_Bonn} we changed
the notation used in Ref.~\cite{JN_09} and write $\tau = \cos\theta$
in order to avoid confusion with the Mandelstam variable $t$ in the
context of the dispersive approach.

The weight functions which appear in the integrals ~(\ref{amupi0_1})
and (\ref{amupi0_2}) are given by 
\bea 
w_1(Q_1,Q_2,\tau)  & = & \left(- \frac{2\pi}{3} \right)
\sqrt{1-\tau^2} \, \frac{Q_1^3 Q_2^3}{Q_2^2 + m_\pi^2} \,
I_1(Q_1,Q_2,\tau), 
\label{w1} \\ 
w_2(Q_1,Q_2,\tau)  & = & \left(- \frac{2\pi}{3} \right)
\sqrt{1-\tau^2} \, \frac{Q_1^3 Q_2^3}{(Q_1+Q_2)^2 + m_\pi^2} \,
I_2(Q_1,Q_2,\tau),  
\label{w2}
\eea 
where the functions $I_{1,2}(Q_1,Q_2,\tau)$ have been calculated in
Ref.~\cite{JN_09} and are given in
Appendix~\ref{App:weight_functions}. From our definition of the form
factor in Eq.~(\ref{TFF}) it follows that the weight functions
$w_{1,2}(Q_1,Q_2,\tau)$ are dimensionless. Furthermore $w_{2}(Q_1,Q_2,\tau)$
is symmetric under $Q_1 \leftrightarrow Q_2$~\cite{JN_09}. Finally,
$w_{1,2}(Q_1,Q_2,\tau) \to 0$ for $Q_{1,2} \to 0$ and for $\tau \to \pm
1$. The precise behavior of $w_{1,2}(Q_1,Q_2,\tau)$ for $Q_{1,2} \to 0$
and $\tau \to \pm 1$, as well as for $Q_{1,2} \to \infty$, can be found
in Appendix~\ref{App:weight_functions}.

The three-dimensional integral representation in Eqs.~(\ref{amupi0_1})
and (\ref{amupi0_2}) separates the generic kinematics in the pion-pole
contribution to HLbL, described by the model-independent weight
functions $w_{1,2}(Q_1,Q_2,\tau)$,\footnote{Note that the weight
  functions $w_{1,2}(Q_1,Q_2,\tau)$ also describe the relevant
  momentum regions in the case where one defines the pion-pole
  contribution according to Refs.~\cite{MV_04, Vainshtein_FCCP_15} and
  sets the single-virtual form factor at the external vertex to a
  constant. Moreover, these weight functions are also relevant if one
  evaluates the pion-exchange contribution with model-dependent
  off-shell pion form factors $\FF(-(Q_1 + Q_2)^2,
  -Q_1^2,-Q_2^2)$~\cite{N_09, JN_09}.} from the dependence on the
single- and double-virtual form factors $\FFa(-Q^2,0)$ and
$\FFa(-Q_1^2,-Q_2^2)$ in the spacelike (Euclidean) region, which can
in principle be measured, obtained from a dispersion
relation~\cite{DR_pion_TFF} (for $\eta, \eta^\prime$ in
Refs.~\cite{DR_eta_etaprime_TFF, DR_eta_TFF_a2,
  DR_eta_TFF_double_virtual}), or, as has been done so far, modelled.
We will discuss the experimental situation concerning the single- and
double-virtual TFF of $\pi^0, \eta, \eta^\prime$ below in
Section~\ref{Sec:TFF_experiment}.

In Ref.~\cite{KN_02} a two-dimensional integral representation for
$a_\mu^{\HLbLpi}$ was derived which also allowed a separation between
certain weight functions and the form factors, however, the derivation
was only possible for a class of VMD-like form factors based on
large-$N_c$ QCD. The constant Wess-Zumino-Witten (WZW)~\cite{WZW} form
factor, as defined in Eq.~(\ref{normalization_anomaly}), also falls
into this class and since in this case there is no dependence on the
angle~$\theta$ from the form factors in Eqs.~(\ref{amupi0_1}) and
(\ref{amupi0_2}), one can obtain two of the weight functions in
Ref.~\cite{KN_02} from the weight functions derived in
Ref.~\cite{JN_09} by integrating over the angles
\bea 
w_{f_1}(Q_1, Q_2) & = & \int_{-1}^{1} d\tau \, w_1(Q_1,
Q_2,\tau), \label{wf1} \\  
\left. w_{g_2}(m_\pi, Q_1, Q_2) \right|_{{\rm symm}}& = & \int_{-1}^{1}
d\tau \, w_2(Q_1, Q_2, \tau),  \label{wg2} 
\eea  
where $\left. w_{g_2}(m_\pi, Q_1,Q_2) \right|_{{\rm symm}} =
\left[w_{g_2}(m_\pi, Q_1, Q_2) + w_{g_2}(m_\pi, Q_2, Q_1) \right] /
2$. Only this symmetric part of the function $w_{g_2}(m_\pi, Q_1,
Q_2)$ given in Ref.~\cite{KN_02} contributes in the corresponding
two-dimensional integral representation for $a_\mu^{\HLbLpitwo}$ since
it is multiplied by $\FFa(-Q_1^2, -Q_2^2)$. This symmetrization
removes the negative regions visible in the plots for the function
$w_{g_2}(m_\pi, Q_1, Q_2)$ shown in Ref.~\cite{KN_02}.  The plots of
the symmetrized function will then look very similar to those for
$w_2(Q_1, Q_2, \tau)$ shown in Fig.~\ref{Fig:w_i_pion} below. We have
checked the relations (\ref{wf1}) and (\ref{wg2}) numerically.

As one can see from the expressions for the weight functions $w_{1,2}$
in Eqs.~(\ref{w1}) and (\ref{w2}), the only explicit dependence on the
pseudoscalar appears via the mass in the pseudoscalar
propagators. This will be important with respect to the most relevant
momentum regions for the HLbL contribution to the muon $g-2$. Of
course, also the form factors will depend on the type of the
pseudoscalar. First of all the normalization of the form factors is
related to the decay width $\Gamma({\rm P} \to \gamma\gamma)$, and,
secondly, there is a difference in the momentum dependence, since
e.g.\ in a VMD model, the relevant vector meson masses will be
different for $\pi^0, \eta$ and $\eta^\prime$ (e.g.\ $\rho$-meson
versus $\phi$-meson).

\section{Model-independent weight functions $w_{1,2}(Q_1,Q_2,\tau)$}
\label{Sec:weight_functions}

\subsection{Weight functions for $\pi^0$}

\begin{figure}[t!]

\centerline{\includegraphics[width=0.5\textwidth]{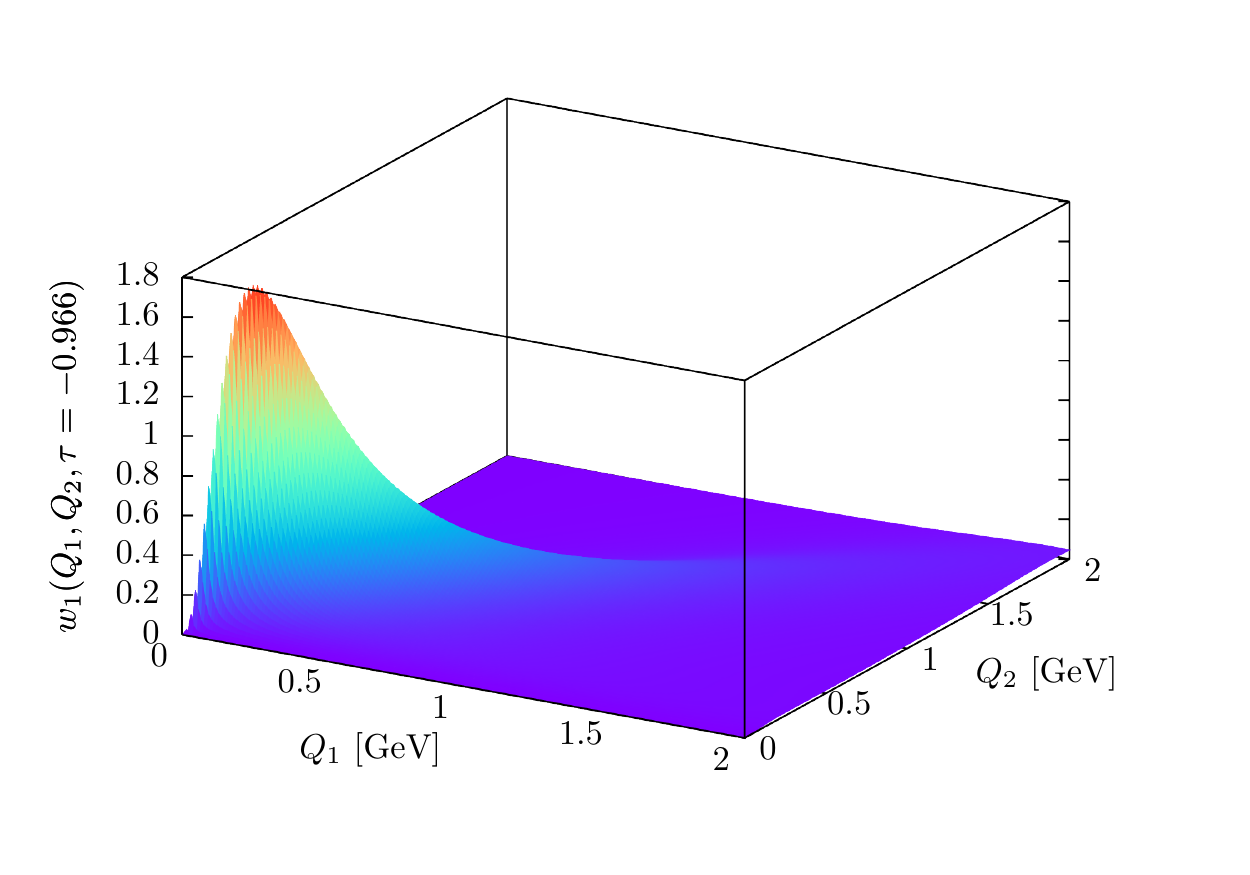}
\includegraphics[width=0.5\textwidth]{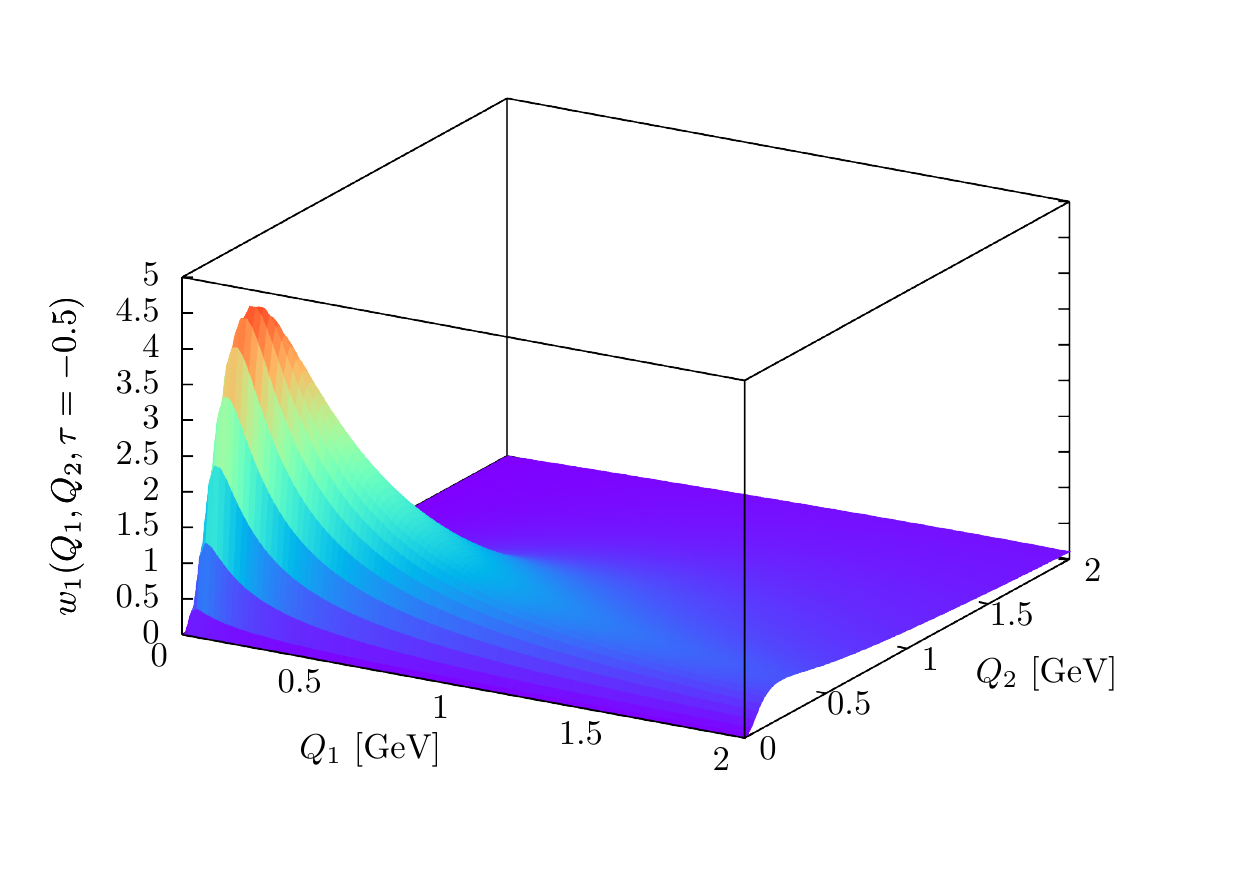}}

\centerline{\includegraphics[width=0.5\textwidth]{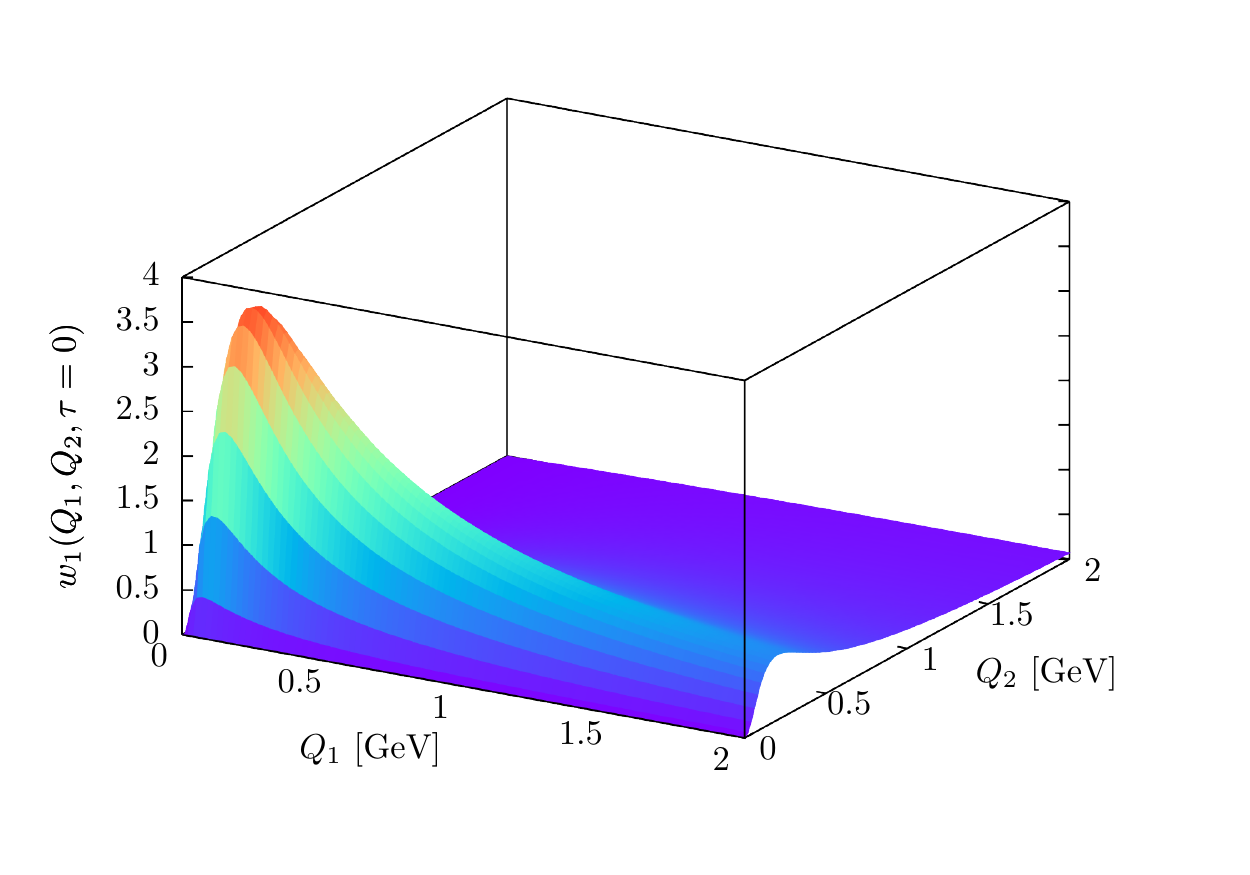}
\includegraphics[width=0.5\textwidth]{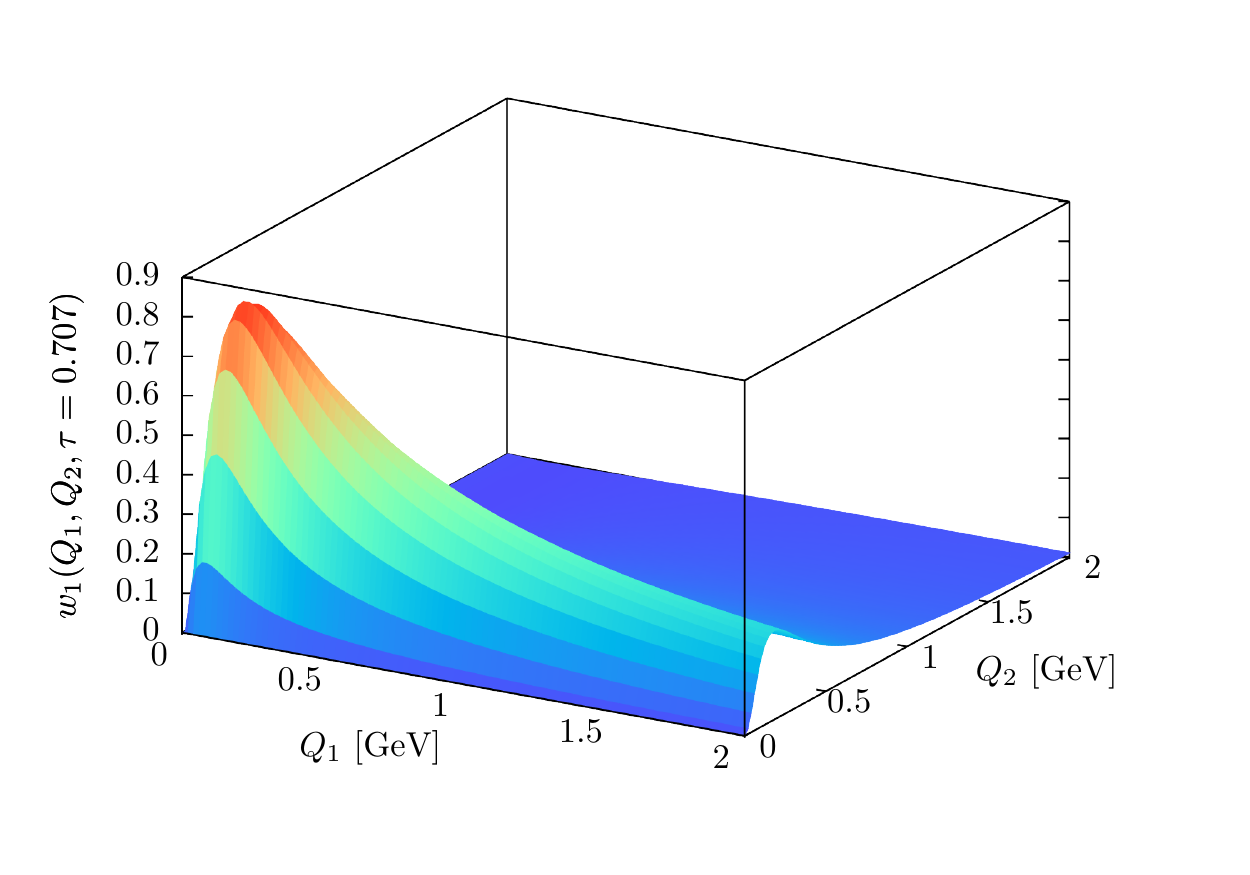}}

\centerline{\includegraphics[width=0.5\textwidth]{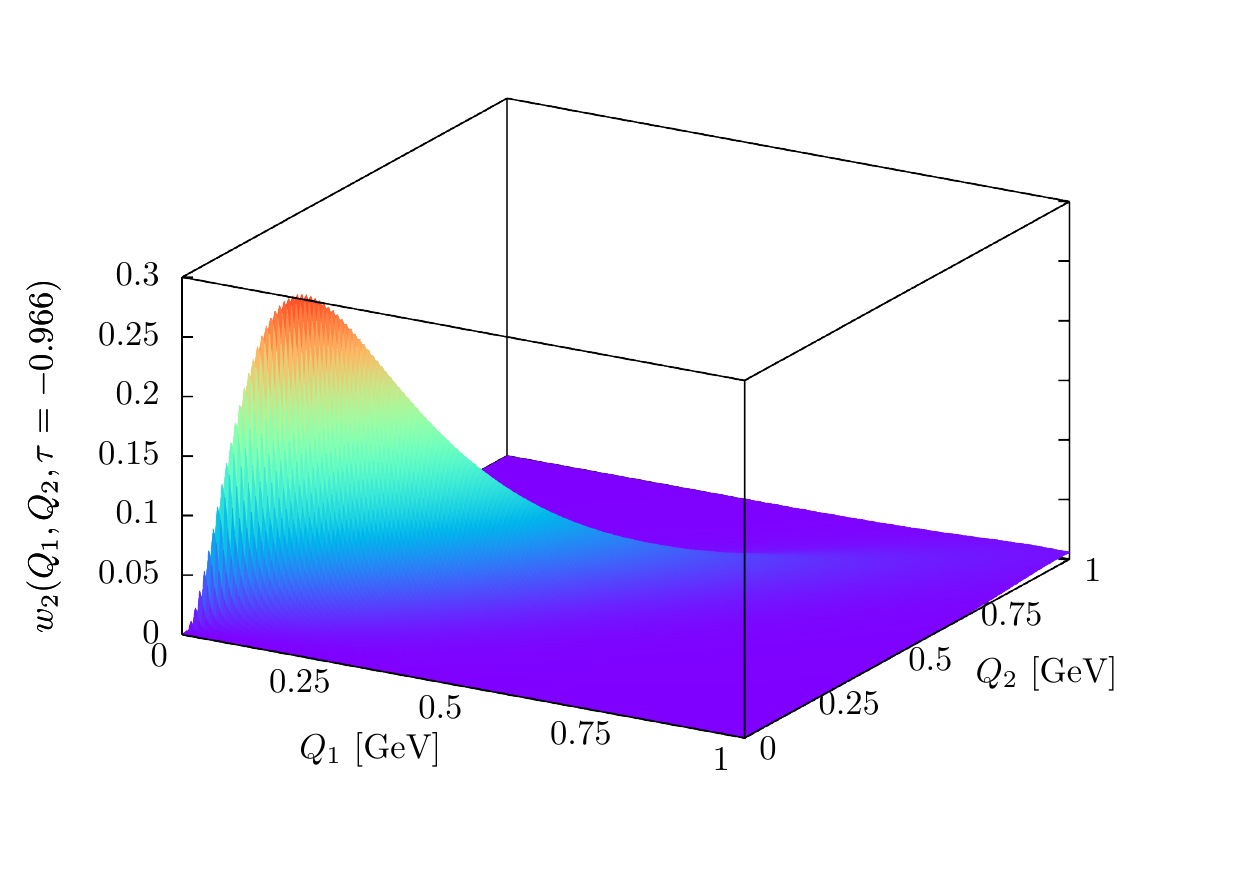}
\includegraphics[width=0.5\textwidth]{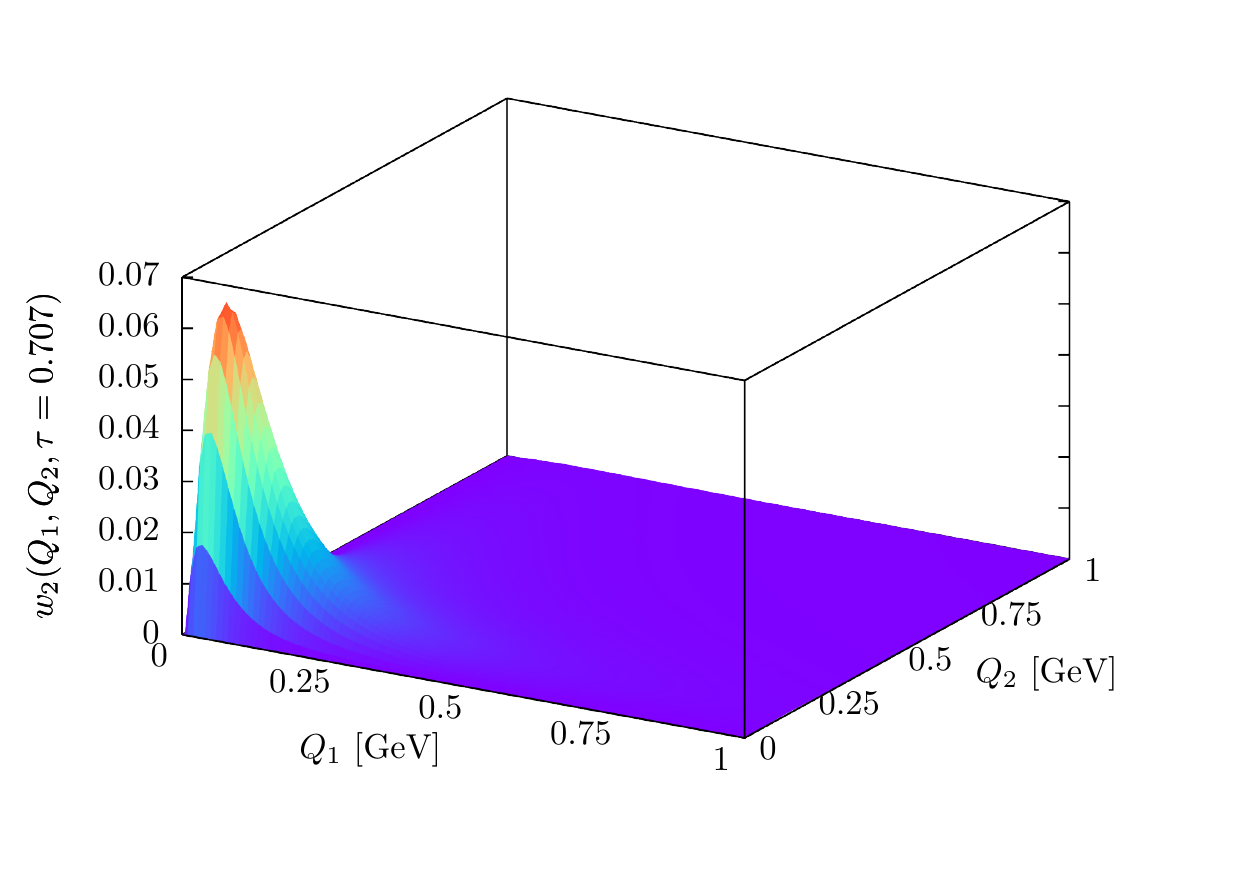}}

\caption{Weight functions $w_1(Q_1,Q_2,\tau)$ and $w_2(Q_1,Q_2,\tau)$
  for the pion as function of the Euclidean momenta $Q_1$ and $Q_2$
  for a selection of values of $\tau = \cos\theta$. Top left: $w_1$
  for $\theta = 165^\circ~(\tau = -0.966)$, top right: $w_1$ for
  $\theta = 120^\circ~(\tau = -0.5)$, middle left: $w_1$ for $\theta =
  90^\circ~(\tau = 0)$, middle right: $w_1$ for $\theta =
  45^\circ~(\tau = 0.707)$. Bottom left: $w_2$ for $\theta =
  165^\circ~(\tau = -0.966)$, bottom right: $w_2$ for $\theta =
  45^\circ~(\tau = 0.707)$. Note the different range in $Q_{1,2}$ for
  $w_2$. The plots of $w_2$ for $\theta = 120^\circ$ and $90^\circ$
  look similar to the one shown for $\theta = 45^\circ$, but the peaks
  are slightly broader.}

\label{Fig:w_i_pion}

\end{figure}

In Fig.~\ref{Fig:w_i_pion} we have plotted the weight functions
$w_1(Q_1,Q_2,\tau)$ and $w_2(Q_1,Q_2,\tau)$ for the pion as function
of $Q_1$ and $Q_2$ for a selection of values of $\theta$. Note that
although the weight functions rise very quickly to the maxima in the
plots in Figs.~\ref{Fig:w_i_pion}, the slopes along the two axis and
along the diagonal $Q_1 = Q_2$ actually vanish for both functions, see
Eq.~(\ref{slopes}).  We stress again that these weight functions are
completely independent of any models for the form factors. In this
respect these three-dimensional plots differ from similar plots in
Ref.~\cite{other_3D_plots} which show, in the context of specific
models, for various contributions to HLbL the full integrand in the
$(Q_1,Q_2)$-plane after the angular integrations, including the form
factors.

\begin{table}[h!] 

  \caption{Values of the maxima of the weight functions
    $w_1(Q_1,Q_2,\tau)$ and $w_2(Q_1,Q_2,\tau)$ for the pion and
    locations of the maxima in the $(Q_1,Q_2)$-plane for a selection
    of angles $\theta$ with decreasing $\theta$ (increasing $\tau =
    \cos\theta$).}      

\label{Tab:Maxima_weightfunctions}

\begin{center} 
\renewcommand{\arraystretch}{1.25}
\begin{tabular}{|r@{$^\circ$}l|r@{.}l|c|c|r@{.}l|c|}
\hline 
\multicolumn{2}{|c|}{$\theta~(\tau = \cos\theta)$} &
\multicolumn{2}{|c|}{Max.\ $w_1$} & $Q_1$ [GeV] & $Q_2$ [GeV] &
\multicolumn{2}{|c|}{Max.\ $w_2$} & $Q_1 = Q_2$ [GeV] \\ 
\hline 
$175$ & ~(-0.996) & ~0 & 592 & 0.163 & 0.163 & ~~0 & 100 & 0.142 \\
$165$ & ~(-0.966) &  1 & 734 & 0.164 & 0.162 &   0 & 277 & 0.132 \\
$150$ & ~(-0.866) &  3 & 197 & 0.166 & 0.158 &   0 & 441 & 0.114 \\ 
$135$ & ~(-0.707) &  4 & 176 & 0.171 & 0.153 &   0 & 494 & 0.099 \\
$120$ & ~(-0.5)   &  4 & 559 & 0.176 & 0.146 &   0 & 471 & 0.087 \\
$105$ & ~(-0.259) &  4 & 349 & 0.182 & 0.139 &   0 & 403 & 0.078 \\
$90$  & ~(0.0)    &  3 & 664 & 0.187 & 0.130 &   0 & 312 & 0.070 \\
$75$  & ~(0.259)  &  2 & 702 & 0.189 & 0.122 &   0 & 218 & 0.063 \\
$60$  & ~(0.5)    &  1 & 691 & 0.187 & 0.114 &   0 & 132 & 0.057 \\
$45$  & ~(0.707)  &  0 & 840 & 0.180 & 0.106 &   0 & 064 & 0.050 \\
$30$  & ~(0.866)  &  0 & 283 & 0.168 & 0.099 &   0 & 021 & 0.043 \\ 
$15$  & ~(0.966)  &  0 & 0385 & 0.154 & 0.092 &  0 & 0027 & 0.037 \\ 
$5 $  & ~(0.996)  &  0 & 0015 & 0.147 & 0.089 &  0 & 000092 & 0.037 \\ 
\hline 
\end{tabular}
\end{center} 

\end{table}

We can immediately see that for both weight functions the low-momentum
region, $Q_{1,2} \leq 0.5~\mbox{GeV}$, is the most important in the
corresponding integrals~(\ref{amupi0_1}) and (\ref{amupi0_2}) for
$a_\mu^{\HLbLpi}$.  For $w_1(Q_1,Q_2,\tau)$ there is a peak around
$Q_1 \sim 0.15 - 0.19~\mbox{GeV}$, $Q_2 \sim 0.09 -
0.16~\mbox{GeV}$. In Table~\ref{Tab:Maxima_weightfunctions} we have
collected the values of the maxima of $w_1(Q_1,Q_2,\tau)$ and the
locations of the maxima in the $(Q_1,Q_2)$-plane for a selection of
values of $\theta$. With decreasing $\theta$ (increasing $\tau$), the
value of the maximum grows until $\theta=120^\circ~(\tau = -0.5)$ and
then decreases again. Note that $w_1(Q_1,Q_2,\tau)$ can become
negative for $\theta \leq 75^\circ$ and for some values of $Q_{1,2}$,
but the minima are small in absolute size compared to the maxima.
Table~\ref{Tab:global_Max_Min} shows the global maximum and minimum of
the function $w_1(Q_1, Q_2, \tau)$ in the physically allowed region
$Q_{1,2} \geq 0$ and $-1 \leq \tau \leq 1$.

\begin{table}[h!] 

  \caption{Values and locations in $(Q_1, Q_2, \theta)$ of the global
    maxima and minima of the weight functions $w_1(Q_1,Q_2,\tau)$ and
    the global maxima of $w_2(Q_1,Q_2,\tau)$ for all three
    pseudoscalars $\pi^0, \eta, \eta^\prime$.}     

\label{Tab:global_Max_Min}

\begin{center} 
\renewcommand{\arraystretch}{1.25}
\begin{tabular}{|l|r@{.}l|c|c|r@{.}l|}
\hline 
& \multicolumn{2}{|c|}{Value} & $Q_1$ [GeV] & $Q_2$ [GeV] &
\multicolumn{2}{|c|}{$\theta~(\tau = \cos\theta)$} \\ 
\hline 
Max.\ $\left. w_1 \right|_{\pi^0}$ & 4 & 563   & 0.177 & 0.145 
& 118 & 1$^\circ~(-0.471)$ \\ 
Min.\ $\left. w_1 \right|_{\pi^0}$ & -0 & 0044 & 0.118 & 1.207 
& 45 & 7$^\circ~(0.698)$ \\ 
Max.\ $\left. w_2 \right|_{\pi^0}$ & 0 & 495   & 0.097 & 0.097 
& 133 & 1$^\circ~(-0.684)$ \\ 
\hline \hline 
Max.\ $\left. w_1 \right|_{\eta}$ & 0 & 813   & 0.322 & 0.310 
& 123 & 8$^\circ~(-0.556)$ \\ 
Min.\ $\left. w_1 \right|_{\eta}$ & -0 & 0037 & 0.129 & 1.368 
& 47 & 1$^\circ~(0.680)$ \\ 
Max.\ $\left. w_2 \right|_{\eta}$ & 0 & 044   & 0.124 & 0.124 
& 122 & 0$^\circ~(-0.531)$ \\ 
\hline \hline 
Max.\ $\left. w_1 \right|_{\eta^\prime}$ & 0 & 332   & 0.415 & 0.416 
& 124 & 8$^\circ~(-0.571)$ \\ 
Min.\ $\left. w_1 \right|_{\eta^\prime}$ & -0 & 0030 & 0.144 & 1.595 
& 48 & 7$^\circ~(0.661)$ \\ 
Max.\ $\left. w_2 \right|_{\eta^\prime}$ & 0 & 015   & 0.128 & 0.128 
& 120 & 9$^\circ~(-0.513)$ \\ 
\hline 
\end{tabular}
\end{center} 

\end{table}

For $\theta \leq 150^\circ~(\tau \geq -0.85)$, a ridge develops along the
$Q_1$ direction for $Q_2 \sim 0.18 - 0.26~\mbox{GeV}$ (maximum along
the line $Q_1 = 2~\mbox{GeV}$). This ridge leads for a constant WZW
form factor to an ultraviolet divergence $(\alpha/\pi)^3 {\cal C}
\ln^2(\Lambda/m_\mu)$~\cite{Melnikov_01} for some momentum cutoff
$\Lambda$ with ${\cal C} = 3 (N_c/(12\pi))^2 (m_\mu/F_\pi)^2 =
0.0248$~\cite{KN_02,Knecht_et_al_PRL_02}.  Of course, realistic form
factors fall off for large momenta $Q_{1,2}$, see
Fig.~\ref{Fig:LMD+V_vs_VMD} in Appendix~\ref{App:TFF_models}, and the
integral $a_\mu^{\HLbLpione}$ will be convergent.

The weight function $w_2(Q_1,Q_2,\tau)$ is about an order of magnitude
smaller than $w_1(Q_1,Q_2,\tau)$ as can be seen in
Fig.~\ref{Fig:w_i_pion}. There is no ridge, since the function is
symmetric under $Q_1 \leftrightarrow Q_2$. Here the peak is around
$Q_1 = Q_2 \sim 0.14~\mbox{GeV}$ for $\tau$ near $-1$. The value of
the maximum grows for decreasing $\theta$ (increasing $\tau$), until
$\theta = 135^\circ~(\tau = -0.707)$ and then decreases again. The
location of the peak thereby moves to much lower values, down to $Q_1
= Q_2 \sim 0.04~\mbox{GeV}$ when $\tau$ is near $+1$, see
Table~\ref{Tab:Maxima_weightfunctions} for more details about the peak
locations. The function $w_2(Q_1,Q_2,\tau)$ is always positive and its
global maximum is shown in Table~\ref{Tab:global_Max_Min}.

\begin{figure}[t!]

\centerline{\includegraphics[width=0.825\textwidth]{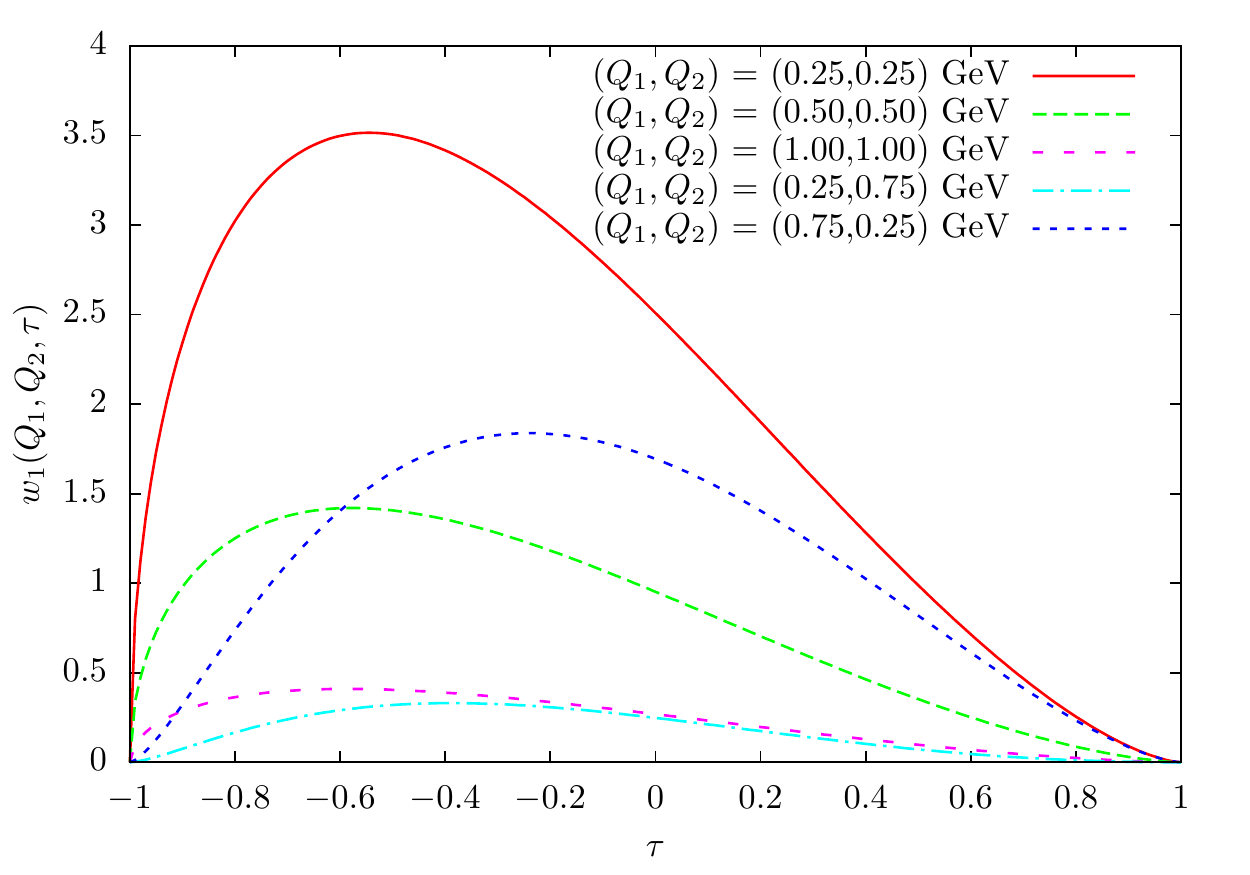}}

\centerline{\hspace*{-0.51cm}\includegraphics[width=0.86\textwidth]{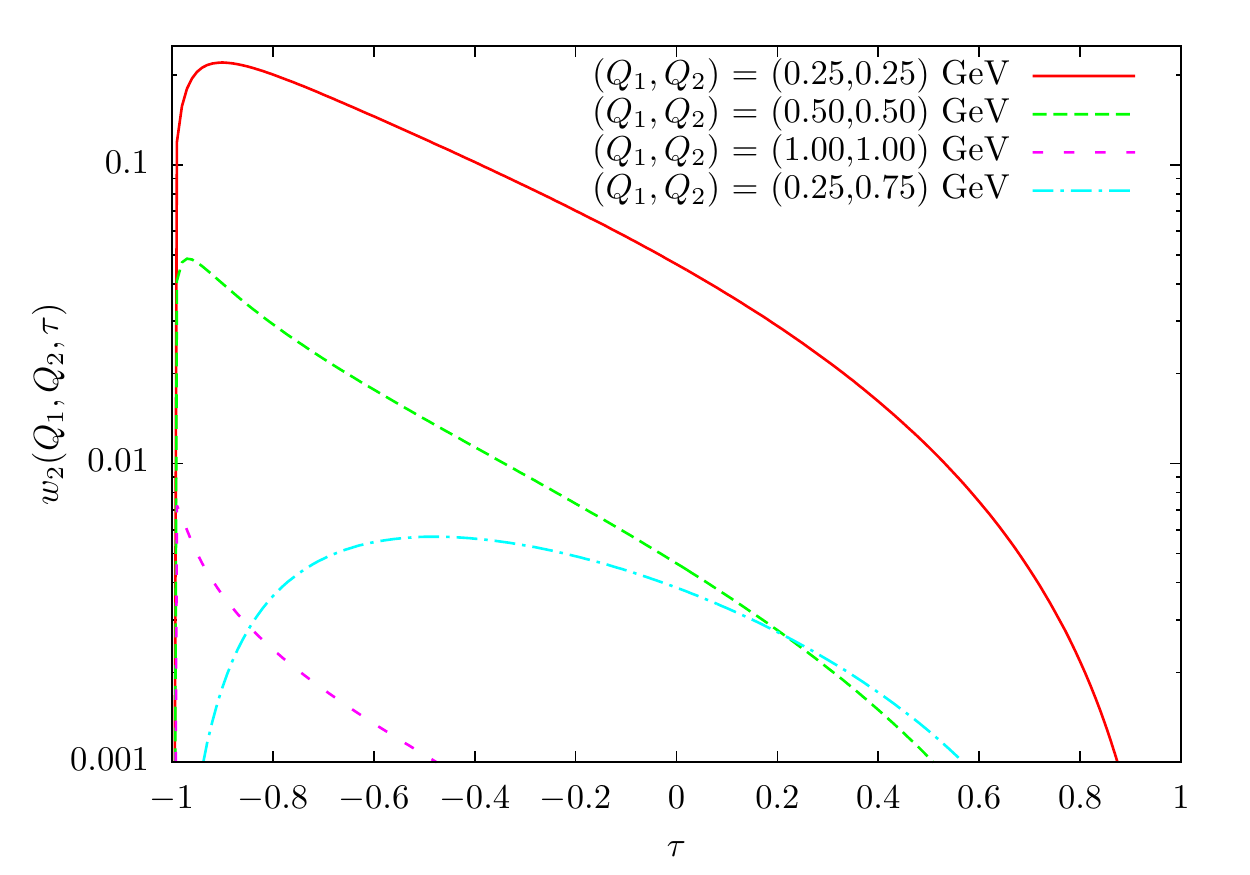}}

\caption{Weight functions $w_1(Q_1,Q_2,\tau)$ (top) and
  $w_2(Q_1,Q_2,\tau)$ (bottom) for the pion as function of $\tau =
  \cos\theta$ for a selection of $Q_1$ and $Q_2$ values. Note the
  logarithmic scale for $w_2$ and that the function is symmetric under
  the exchange $Q_1 \leftrightarrow Q_2$.} 

\label{Fig:weightfunctions_1Dplots}

\end{figure}

Figure~\ref{Fig:weightfunctions_1Dplots} shows the weight functions
$w_{1,2}(Q_1,Q_2,\tau)$ as a function of $\tau$ for some selected
values of $Q_1$ and $Q_2$. One can see that there is a strong
enhancement for $Q_1 = Q_2$ for negative $\tau$ when the original
four-vectors $(Q_1)_\mu$ and $(Q_2)_\mu$ become more and more
antiparallel. In particular, the functions have a rather large slope
for $\tau \to -1$ and the maximal value is always located at $\tau <
0$.\footnote{Note the factor $1/(Q_1+Q_2)^2$ from one of the photon
  propagators in the weight functions $w_{1,2}(Q_1,Q_2,\tau)$, see
  Eqs.~(\ref{I1}) and (\ref{I2}). According to
  Eqs.~(\ref{w1_tau_to_-1_diagonal}) and (\ref{w2_tau_to_-1_diagonal})
  the weight functions $w_{1,2}(Q,Q,\tau)$ along the diagonal $Q_1 =
  Q_2 = Q$ have infinite slope at $\tau = -1$.}  However, both weight
functions always vanish for $\tau = -1$, as mentioned before and
discussed in Appendix~\ref{App:weight_functions}. Furthermore the
weight functions get smaller for growing values of $Q_{1,2} >
0.5~\mbox{GeV}$, as is already visible in the three-dimensional plots
in Fig.~\ref{Fig:w_i_pion}.

\subsection{Weight functions for $\eta$ and $\eta^\prime$} 

\begin{figure}[t!]

\centerline{\includegraphics[width=0.5\textwidth]{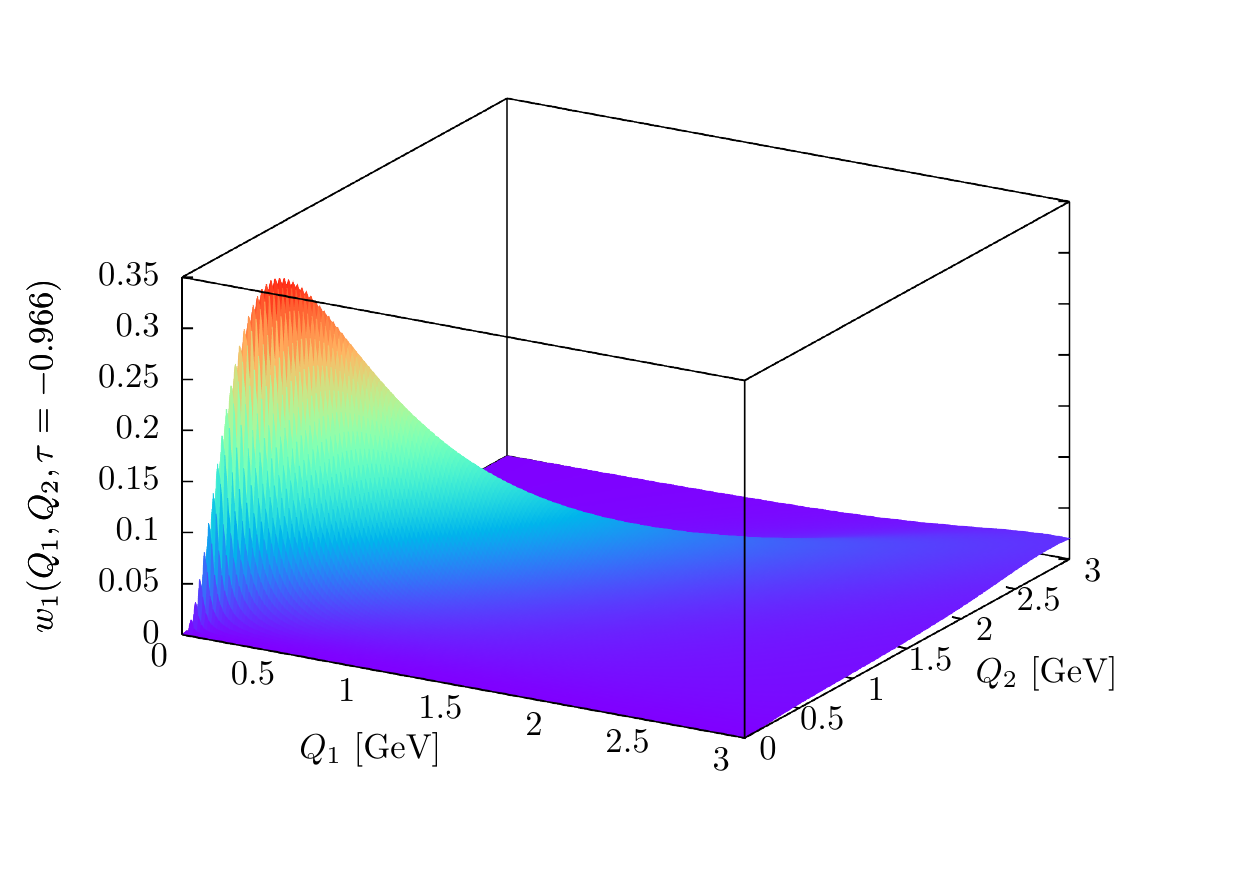}
\includegraphics[width=0.5\textwidth]{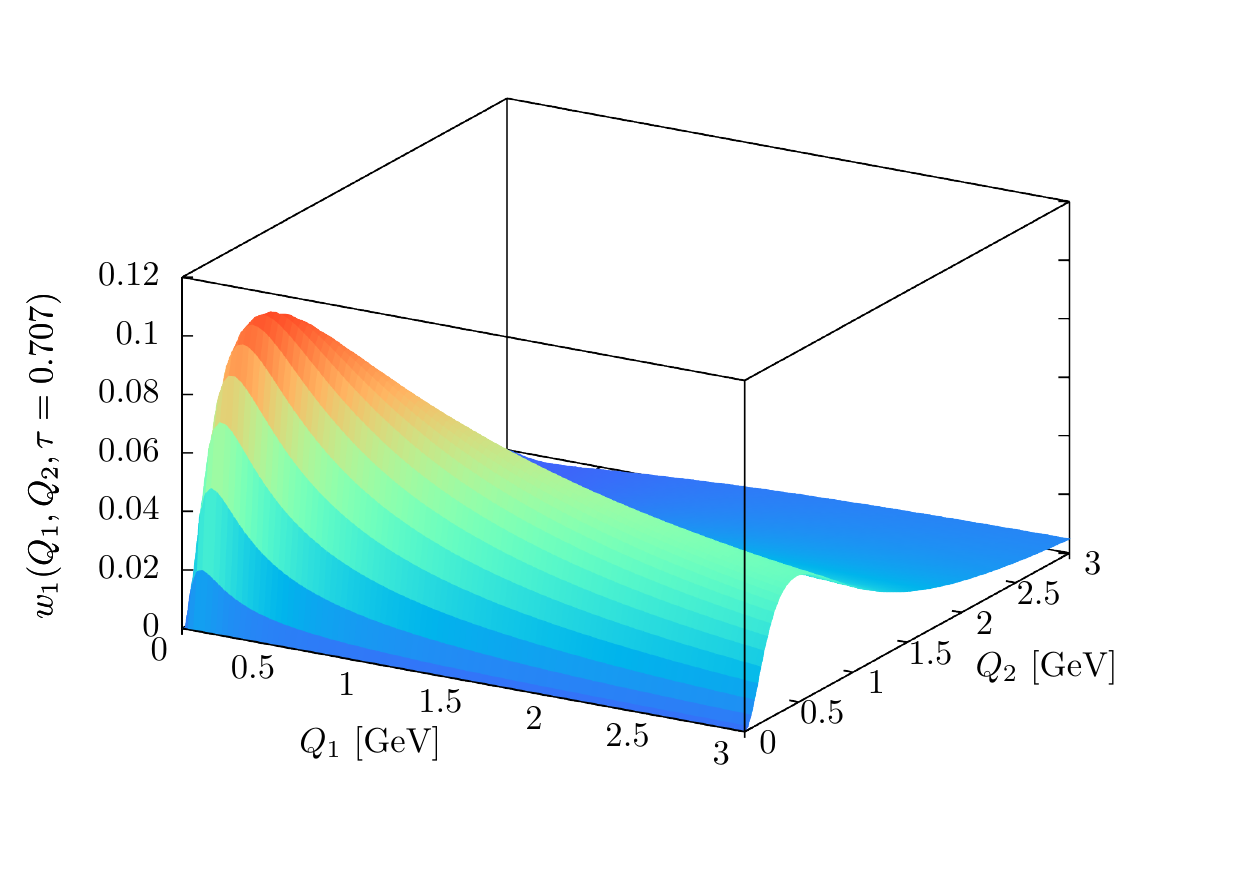}}

\centerline{\includegraphics[width=0.5\textwidth]{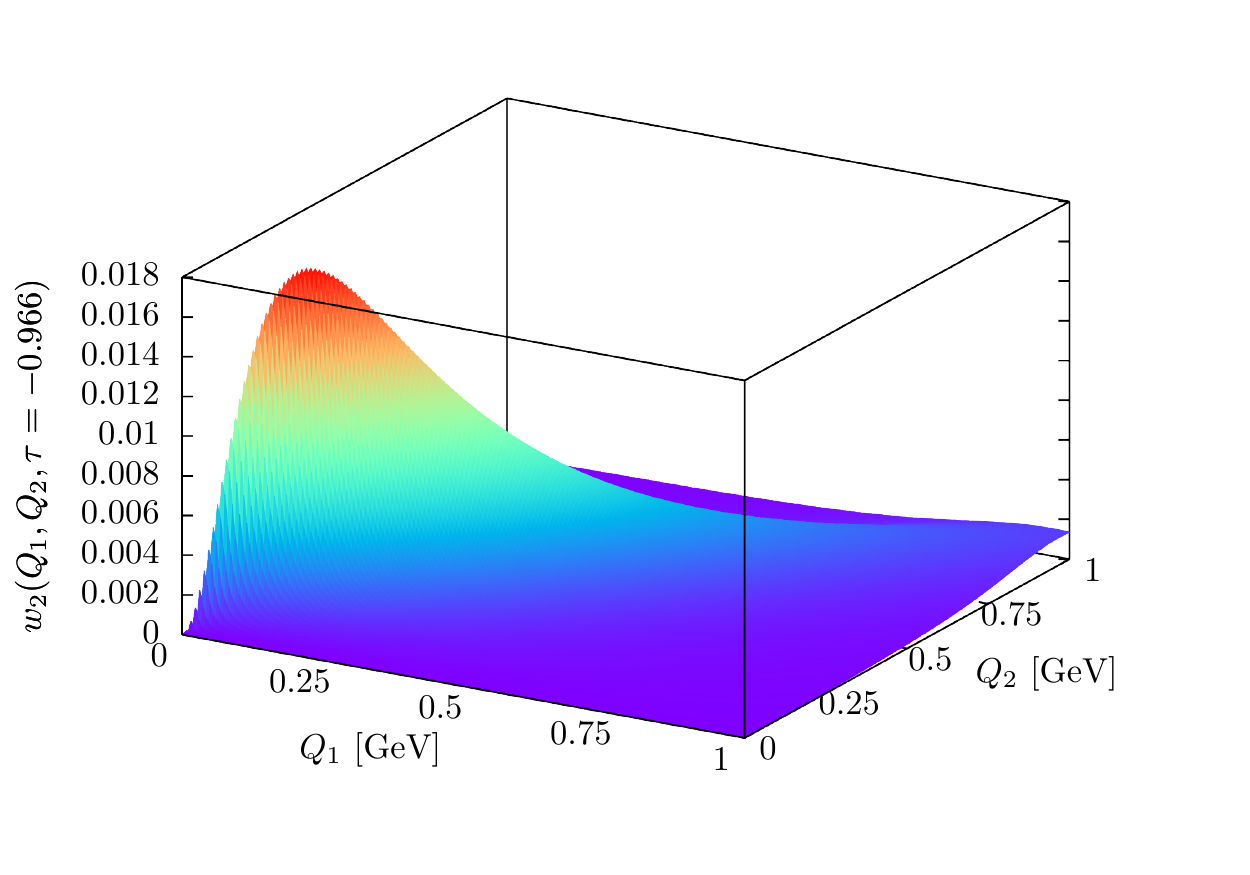}
\includegraphics[width=0.5\textwidth]{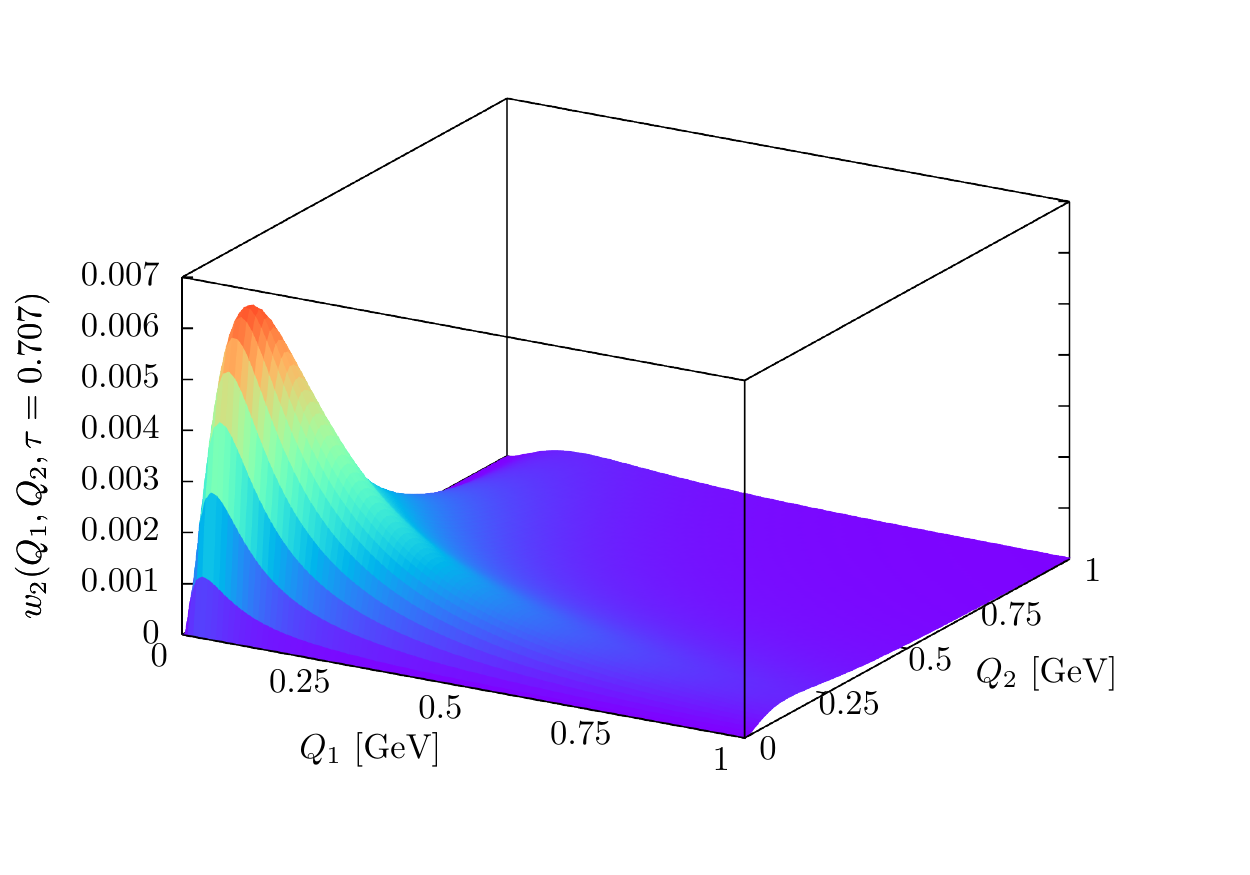}}

\centerline{\includegraphics[width=0.5\textwidth]{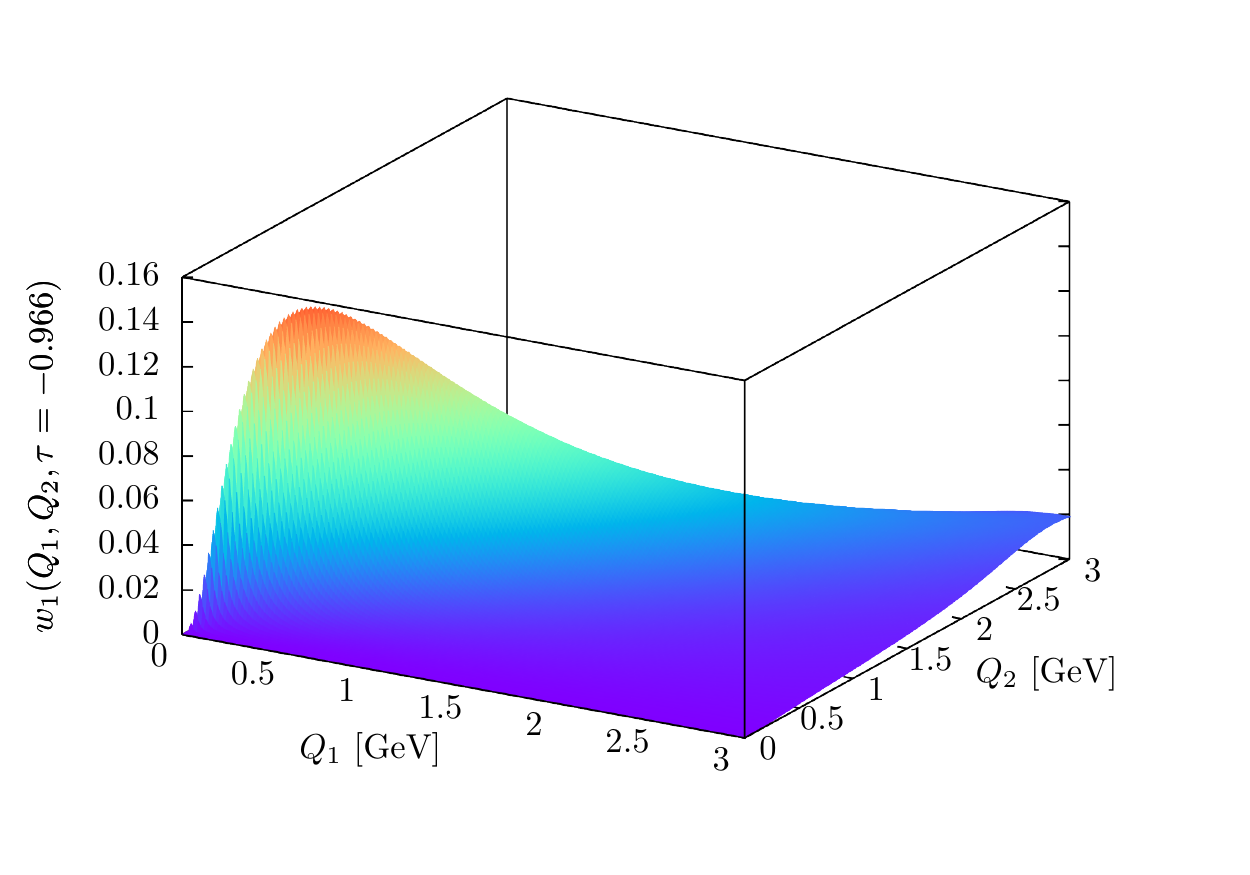}
\includegraphics[width=0.5\textwidth]{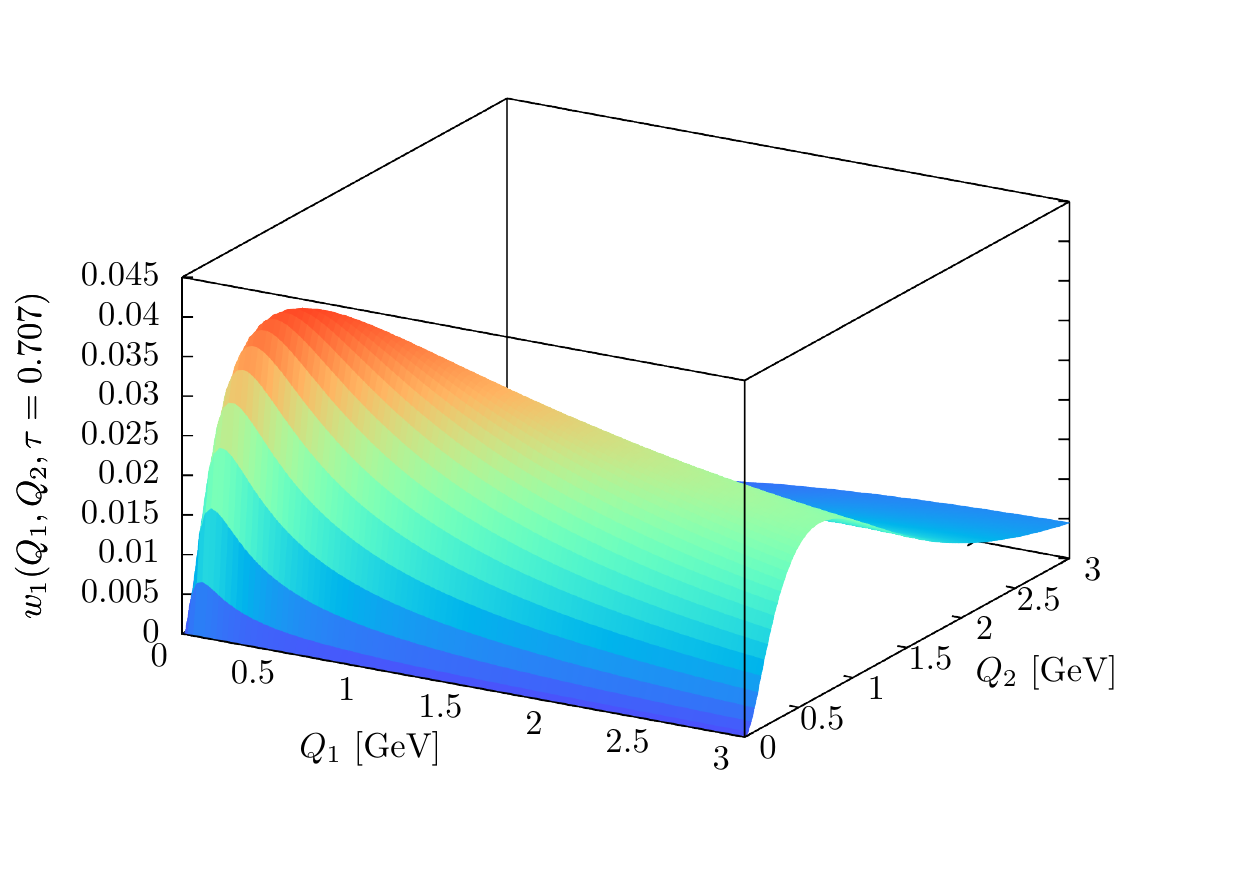}}

\caption{Weight functions $w_1(Q_1,Q_2,\tau)$ and $w_2(Q_1,Q_2,\tau)$
  for the $\eta$ and $\eta^\prime$ as function of the Euclidean
  momenta $Q_1$ and $Q_2$, each for two values of $\tau =
  \cos\theta$. Top left: $w_1$ for $\eta$ with $\theta =
  165^\circ~(\tau = -0.966)$, top right: $w_1$ for $\eta$ with $\theta
  = 45^\circ~(\tau = 0.707)$, middle left: $w_2$ for $\eta$ with
  $\theta = 165^\circ~(\tau = -0.966)$, middle right: $w_2$ for $\eta$
  with $\theta = 45^\circ~(\tau = 0.707)$. Note the different range in
  $Q_{1,2}$ for $w_2$. Bottom left: $w_1$ for $\eta^\prime$ with
  $\theta = 165^\circ~(\tau = -0.966)$, bottom right: $w_1$ for
  $\eta^\prime$ with $\theta = 45^\circ~(\tau = 0.707)$. The plots of
  $w_2$ for $\eta^\prime$ look similar to those for $\eta$ with the
  same $\theta$.}

\label{Fig:w_i_eta_etaprime}

\end{figure}

In Fig.~\ref{Fig:w_i_eta_etaprime} we have plotted the
model-independent weight functions $w_{1,2}(Q_1,Q_2,\tau)$ for the
$\eta$ and the $\eta^\prime$ as a function of $Q_1$ and $Q_2$ for the
two angles $\theta = 165^\circ$ and $45^\circ$. The shape of the plots
for $\theta = 120^\circ$ and $90^\circ$ is similar to the ones shown
for $\theta = 45^\circ$.

The only dependence on the pseudoscalars appears in the weight
functions $w_{1,2}(Q_1,Q_2,\tau)$ through the pseudoscalar
propagators, i.e.\ a factor $1/(Q_2^2 + m_{\rm P}^2)$ in $w_1$ and a
factor $1/((Q_1 + Q_2)^2 + m_{\rm P}^2) = 1/(Q_1^2 + 2 Q_1 Q_2 \tau +
Q_2^2 + m_{\rm P}^2)$ in $w_2$, see Eqs.~(\ref{w1}) and (\ref{w2}). As
can be seen from the plots in Fig.~\ref{Fig:w_i_eta_etaprime}, this
shifts the relevant momentum regions (peaks, ridges) in the weight
functions to higher momenta for $\eta$ compared to $\pi^0$ and even
higher for $\eta^\prime$. Note that for $w_1$ the momentum range in
the plots for $\eta$ and $\eta^\prime$ is now $0-3~\mbox{GeV}$,
whereas it was only $0-2~\mbox{GeV}$ for $\pi^0$ in
Fig.~\ref{Fig:w_i_pion}. It also leads to a suppression in the
absolute size of the weight functions due to the larger masses in the
propagators. This pattern will also be visible in the values for the
contributions to HLbL. For the bulk of the weight functions (maxima,
ridges) we observe the following approximate relations (for the same
angle $\theta$, but not necessarily at the same values of the momenta)
\bea 
\left. w_1 \right|_{\eta} & \approx & \frac{1}{6} \left. w_1
\right|_{\pi^0},  \label{wi_eta_vs_pi0} \\ 
\left. w_1 \right|_{\eta^\prime} & \approx & \frac{1}{2.5} \left. w_1
\right|_{\eta}.  \label{wi_etaprime_vs_eta} 
\eea 
Of course, the ratio of the weight functions is given by the ratio of
the propagators and is maximal at zero momenta and at that point equal
to the ratio of the squares of the masses, but at zero momenta the
weight functions themselves vanish. The combined effect is well
described by the relations in Eqs.~(\ref{wi_eta_vs_pi0}) and
(\ref{wi_etaprime_vs_eta}). Furthermore, for both $\eta$ and
$\eta^\prime$, the weight function $w_2$ is about a factor 20 smaller
than the corresponding weight function $w_1$.

The peaks for the weight function $w_1(Q_1,Q_2,\tau)$ for $\eta$ and
$\eta^\prime$ are less steep, compared to $\pi^0$, and the ridge is
quite broad in the $Q_2$-direction. Furthermore, the ridge falls off
only slowly in the $Q_1$-direction and for $\theta \leq 75^\circ$ is
still about half of the maximum out to values of $Q_1 =
3~\mbox{GeV}$. For the weight function $w_2(Q_1,Q_2,\tau)$ the peaks
are again less steep and larger values of the momenta contribute,
compared to the pion.

\begin{table}[h!] 

  \caption{Values of the maxima of the weight functions
    $w_{1,2}(Q_1,Q_2,\tau)$ and locations of the maxima in the
    $(Q_1,Q_2)$-plane for a selection of angles $\theta$. Top part:
    $\eta$-meson. Bottom part: $\eta^\prime$-meson.}     

\label{Tab:Maxima_weightfunctions_eta_etaprime}

\begin{center} 
\renewcommand{\arraystretch}{1.25}
\begin{tabular}{|r@{$^\circ$}l|r@{.}l|c|c|r@{.}l|c|}
\hline 
\multicolumn{2}{|c|}{$\theta~(\tau = \cos\theta)$} &
\multicolumn{2}{|c|}{Max.\ $w_1$} & $Q_1$ [GeV] & $Q_2$ [GeV] &
\multicolumn{2}{|c|}{Max.\ $w_2$} & $Q_1 = Q_2$ [GeV] \\ 
\hline 
$175$ & ~(-0.996) & ~0 & 117   & 0.328 & 0.328 & ~0 & 0061  & 0.143 \\
$165$ & ~(-0.966) &  0 & 341   & 0.327 & 0.327 &  0 & 018   & 0.142 \\
$150$ & ~(-0.866) &  0 & 616   & 0.325 & 0.323 &  0 & 032   & 0.137 \\
$135$ & ~(-0.707) &  0 & 778   & 0.323 & 0.317 &  0 & 041   & 0.131 \\
$120$ & ~(-0.5)   &  0 & 809   & 0.322 & 0.308 &  0 & 044   & 0.123 \\
$105$ & ~(-0.259) &  0 & 729   & 0.323 & 0.296 &  0 & 040   & 0.114 \\
$90$  & ~(0.0)    &  0 & 575   & 0.328 & 0.282 &  0 & 032   & 0.106 \\
$75$  & ~(0.259)  &  0 & 395   & 0.336 & 0.267 &  0 & 023   & 0.096 \\
$60$  & ~(0.5)    &  0 & 231   & 0.346 & 0.253 &  0 & 014   & 0.087 \\
$45$  & ~(0.707)  &  0 & 107   & 0.356 & 0.241 &  0 & 0063  & 0.077 \\
$30$  & ~(0.866)  &  0 & 034   & 0.363 & 0.231 &  0 & 0019  & 0.067 \\ 
$15$  & ~(0.966)  &  0 & 0044  & 0.367 & 0.225 &  0 & 00023 & 0.063 \\ 
$5$   & ~(0.996)  &  0 & 00017 & 0.368 & 0.224 &
\multicolumn{2}{|c|}{$8 \times 10^{-6}$} & 0.065 \\  
\hline \hline 
$175$ & ~(-0.996) &  0 & 049   & 0.434 & 0.434 &  0 & 0020   & 0.143 \\
$165$ & ~(-0.966) &  0 & 142   & 0.432 & 0.433 &  0 & 0059   & 0.142 \\
$150$ & ~(-0.866) &  0 & 255   & 0.427 & 0.430 &  0 & 011    & 0.139 \\
$135$ & ~(-0.707) &  0 & 320   & 0.419 & 0.423 &  0 & 014    & 0.134 \\
$120$ & ~(-0.5)   &  0 & 330   & 0.413 & 0.412 &  0 & 015    & 0.128 \\
$105$ & ~(-0.259) &  0 & 293   & 0.412 & 0.397 &  0 & 014    & 0.120 \\
$90$  & ~(0.0)    &  0 & 227   & 0.418 & 0.378 &  0 & 011    & 0.112 \\
$75$  & ~(0.259)  &  0 & 154   & 0.431 & 0.358 &  0 & 0079   & 0.102 \\
$60$  & ~(0.5)    &  0 & 088   & 0.451 & 0.340 &  0 & 0047   & 0.092 \\
$45$  & ~(0.707)  &  0 & 041   & 0.472 & 0.326 &  0 & 0022   & 0.082 \\
$30$  & ~(0.866)  &  0 & 013   & 0.491 & 0.315 &  0 & 00066  & 0.072 \\ 
$15$  & ~(0.966)  &  0 & 0017  & 0.504 & 0.309 &  0 & 000079 & 0.067 \\ 
$5$   & ~(0.996)  &  0 & 000062 & 0.508 & 0.307 &
\multicolumn{2}{|c|}{$3 \times 10^{-6}$} & 0.070 \\  
\hline 
\end{tabular}
\end{center} 

\end{table}

Table~\ref{Tab:Maxima_weightfunctions_eta_etaprime} shows for $\eta$
and $\eta^\prime$ the maxima of the weight functions
$w_{1,2}(Q_1,Q_2,\tau)$ and the locations of the maxima in the
$(Q_1,Q_2)$-plane for a selection of angles $\theta$. 
For the $\eta$-meson, the peak in the weight function
$w_1(Q_1,Q_2,\tau)$ is around $Q_1 \sim 0.32 - 0.37~\mbox{GeV}$, $Q_2
\sim 0.22 - 0.33~\mbox{GeV}$.  For $\theta \leq 120^\circ~(\tau \geq
-0.5)$, again the broad ridge along the $Q_1$ direction is visible
with $Q_2 \sim 0.4 - 0.7~\mbox{GeV}$ (maximum along the line $Q_1 =
2~\mbox{GeV}$).  The peak for the weight function $w_2(Q_1,Q_2,\tau)$
is around $Q_1 = Q_2 \sim 0.14~\mbox{GeV}$ for $\tau$ near $-1$ as for
the pion. The location of the peak moves down to $Q_1 = Q_2 \sim
0.06~\mbox{GeV}$ when $\tau$ is near $+1$. The global maximum and
minimum of $w_1(Q_1, Q_2, \tau)$ and the global maximum of $w_2(Q_1,
Q_2, \tau)$ are shown in Table~\ref{Tab:global_Max_Min}.

For the $\eta^\prime$, the peak in $w_1(Q_1,Q_2,\tau)$ now occurs for
even higher momenta, $Q_1 \sim 0.41 - 0.51~\mbox{GeV}$, $Q_2 \sim 0.31
- 0.43~\mbox{GeV}$. Here the broad ridge along the $Q_1$ direction
arises for $\theta \leq 120^\circ~(\tau \geq -0.5)$ with $Q_2 \sim 0.6
- 1~\mbox{GeV}$.  The locations of the peaks of $w_2(Q_1,Q_2,\tau)$ in
the $(Q_1,Q_2)$-plane follow a similar pattern as for the $\eta$
meson, see Table~\ref{Tab:Maxima_weightfunctions_eta_etaprime}. Again
the global maximum and minimum of $w_1(Q_1, Q_2, \tau)$ and the global
maximum of $w_2(Q_1, Q_2, \tau)$ have been collected in 
Table~\ref{Tab:global_Max_Min}.

\section{Relevant momentum regions in $a_\mu^{\HLbLP}$}
\label{Sec:momentum_regions}

In order to study the impact of different momentum regions on the
pseudoscalar-pole contribution, we need, at least for the integral
with the weight function $w_1(Q_1, Q_2, \tau)$ in
Eq.~(\ref{amupi0_1}), some knowledge on the form factor
$\FFP(-Q_1^2,-Q_2^2)$, since the integral diverges for a constant form
factor.\footnote{As already observed in Refs.~\cite{HK,KN_02}, for
  the integral with the weight function $w_2(Q_1,Q_2,\tau)$ in
  Eq.~(\ref{amupi0_2}) one obtains even for a constant WZW form factor
  a finite and small result:
  $\left(\frac{\alpha}{\pi}\right)^3 a_{\mu; {\rm WZW}}^{\HLbLpitwo} =
  2.5 \times 10^{-11}$, $\left(\frac{\alpha}{\pi}\right)^3 a_{\mu;
    {\rm WZW}}^{{\rm HLbL};\eta(2)} = 0.78 \times 10^{-11}$ and
  $\left(\frac{\alpha}{\pi}\right)^3 a_{\mu; {\rm WZW}}^{{\rm
      HLbL};\eta^\prime(2)} = 0.65 \times 10^{-11}$.}  For
illustration we take for the pion two simple models to perform the
integrals: Lowest Meson Dominance with an additional vector multiplet,
LMD+V model~\cite{KN_EPJC_01,KN_02}, based on the Minimal Hadronic
Approximation~\cite{MHA_LMD, Knecht_et_al_99} to large-$N_c$ QCD
matched to certain QCD short-distance constraints from the operator
product expansion (OPE)~\cite{OPE}, and the well-known and often used
VMD model. Of course, in the end, the models have to be replaced as
much as possible by experimental data on the single- and
double-virtual TFF or one can use a DR for the form factor
itself~\cite{DR_pion_TFF,DR_eta_etaprime_TFF, DR_eta_TFF_a2,
  DR_eta_TFF_double_virtual}.

Some details and properties of these two form factor models can be
found in Appendix~\ref{App:TFF_models}. There are two main differences
between the models. First, the LMD+V model does not factorize
$\FFa^{\rm LMD+V}(-Q_1^2,-Q_2^2) \neq f(Q_1^2) \times f(Q_2^2)$. Such
a factorization is also not expected in QCD. Second, the two models
have a different behavior of the double-virtual form factor for large
and equal momenta. The LMD+V model reproduces by construction the
OPE~\cite{OPE_TFF_1, OPE_TFF_2}, whereas the VMD form factor falls off
too fast (see Eq.~(\ref{OPE_condition}) for the exact OPE behavior):
\bea 
\FFa^{\rm LMD+V}(-Q^2,-Q^2) & \sim & \FFa^{\rm OPE}(-Q^2,-Q^2) \sim 
\frac{1}{Q^2} \, , \qquad \mbox{for large~}Q^2 \,
,  \label{large_Q_LMD+V} \\  
\FFa^{\rm VMD}(-Q^2,-Q^2) & \sim & \frac{1}{Q^4} \, , \qquad \mbox{for
  large~}Q^2 \, . \label{large_Q_VMD}  
\eea 
Nevertheless, as can be seen from Fig.~\ref{Fig:LMD+V_vs_VMD} and
Table~\ref{Tab:LMD+V_vs_VMD} in Appendix~\ref{App:TFF_models}, for not too
large momenta, $Q_1 = Q_2 = Q = 0.5~[0.75]~\mbox{GeV}$, the form
factors $\FFa(-Q^2, -Q^2)$ in the two models differ by only
$3\%~[10\%]$. Furthermore, both models give an equally good
description of the single-virtual TFF
$\FFa(-Q^2,0)$~\cite{KN_EPJC_01,KLOE2_impact}.

The LMD+V model was developed in Ref.~\cite{KN_EPJC_01} in the chiral
limit and assuming octet symmetry. This is certainly not a good
approximation for the more massive $\eta$ and $\eta^\prime$ mesons,
where also the nonet symmetry, the effect of the $U(1)_A$ anomaly and
the $\eta-\eta^\prime$-mixing have to be taken into account. Since we
are interested here in the determination of the relevant momentum
regions of the pseudoscalar-pole contributions to HLbL and the impact
of experimental measurement errors of the form factors, we will take
for the $\eta$ and $\eta^\prime$ mesons the usual VMD model for the
TFF, as already done in Refs.~\cite{KN_02,N_09}. See
Appendix~\ref{App:TFF_models} for more details about the VMD model
parameters for $\eta$ and $\eta^\prime$.

The two models yield the following results for the pole-contributions 
of the light pseudoscalars to HLbL (we list here only the central
values)\footnote{ Although the weight functions
  $w_{1,2}(Q_1,Q_2,\tau)$ vanish for $Q_{1,2} \to 0$ and $\tau \to \pm
  1$, for the numerical integration with VEGAS~\cite{VEGAS} we
  introduced a small infrared / collinear cutoff as follows: $Q_{1,2}
  \geq \epsilon~\mbox{GeV}$, $\tau \leq 1 - \epsilon$ and $\tau \geq
  -1 + \epsilon^2$ with $\epsilon = 10^{-6}$, where the latter
  condition takes into account the steeper slope for $\tau \to -1$,
  when $Q_1 = Q_2$, see Fig.~\ref{Fig:weightfunctions_1Dplots} and
  Eqs.~(\ref{w1_tau_to_-1_diagonal}) and (\ref{w2_tau_to_-1_diagonal})
  in Appendix~\ref{App:weight_functions}.}
\bea 
a_{\mu; {\rm LMD+V}}^{\HLbLpi} & = & 62.9 \times
10^{-11} \, , \label{amuLMD+V} \\ 
a_{\mu; {\rm VMD}}^{\HLbLpi} & = & 57.0 \times
10^{-11} \, , \label{amuVMD} \\
a_{\mu; {\rm VMD}}^{{\rm HLbL};\eta} & = & 14.5 \times
10^{-11}, \label{amueta} \\ 
a_{\mu; {\rm VMD}}^{{\rm HLbL};\eta^\prime} & = & 12.5 \times
10^{-11}.  \label{amuetaprime}   
\eea

The results (\ref{amuLMD+V}) and (\ref{amuVMD}) for the pion-pole
contribution in the two models are in the ballpark of many other
estimates, see Eq.~(\ref{range_HLbLpi0}), but they also differ by
$9.4\%$, relative to the LMD+V result, due to the different
high-energy behavior for the double-virtual TFF in
Eqs.~(\ref{large_Q_LMD+V}) and (\ref{large_Q_VMD}) for $Q_{1,2} \geq
1~\mbox{GeV}$. In fact, the pattern of the contributions to
$a_\mu^{\HLbLpi}$ is to a large extent determined by the
model-independent weight functions $w_{1,2}(Q_1,Q_2,\tau)$, which are
concentrated below about $0.5~\mbox{GeV}$, up to that ridge in $w_1$
along the $Q_1$ direction. As long as realistic form factor models for
the double-virtual case fall off at large momenta and do not differ
too much at low momenta (with the normalization from $\Gamma(\pi^0 \to
\gamma\gamma)$ and constraints from the single-virtual transition form
factor), we expect similar results for the pion-pole contribution at
the level of $15\%$ which is in fact what is seen in the
literature~\cite{HLbL_P_literature}. Nevertheless, due to the
ridge-like structure in the weight function $w_1$ along the $Q_1$
direction, the high-energy behavior of the form factors is relevant at
the precision of $10\%$ one is aiming for.

For the $\eta$ and $\eta^\prime$, the results in Eqs.~(\ref{amueta})
and (\ref{amuetaprime}) are as expected from the discussion of the
relative size of the weight functions in Eqs.~(\ref{wi_eta_vs_pi0})
and (\ref{wi_etaprime_vs_eta}). The result for $\eta$ is about a
factor 4 smaller than for the pion with VMD. The result for
$\eta^\prime$ is only slighly smaller than for $\eta$. Note that the
normalization of the form factors related to the decay $\Gamma({\rm P}
\to \gamma\gamma)$ and the momentum dependence due to different values
of $M_V$ for $\eta$ and $\eta^\prime$ also play an important role for
the results in Eqs.~(\ref{amueta}) and (\ref{amuetaprime}).

\begin{figure}[h!]

\centerline{\includegraphics[width=0.5\textwidth]{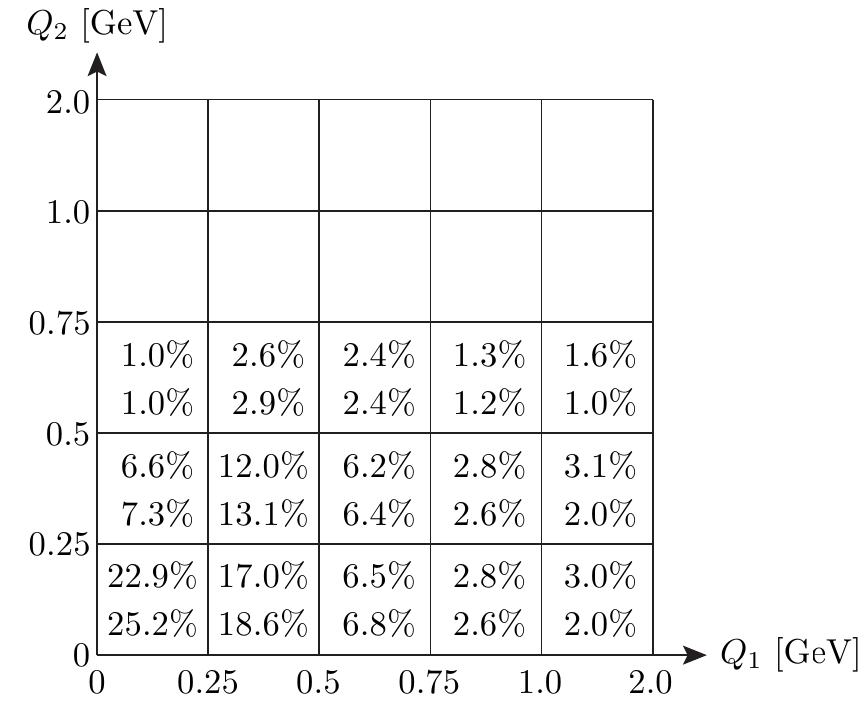}
\includegraphics[width=0.5\textwidth]{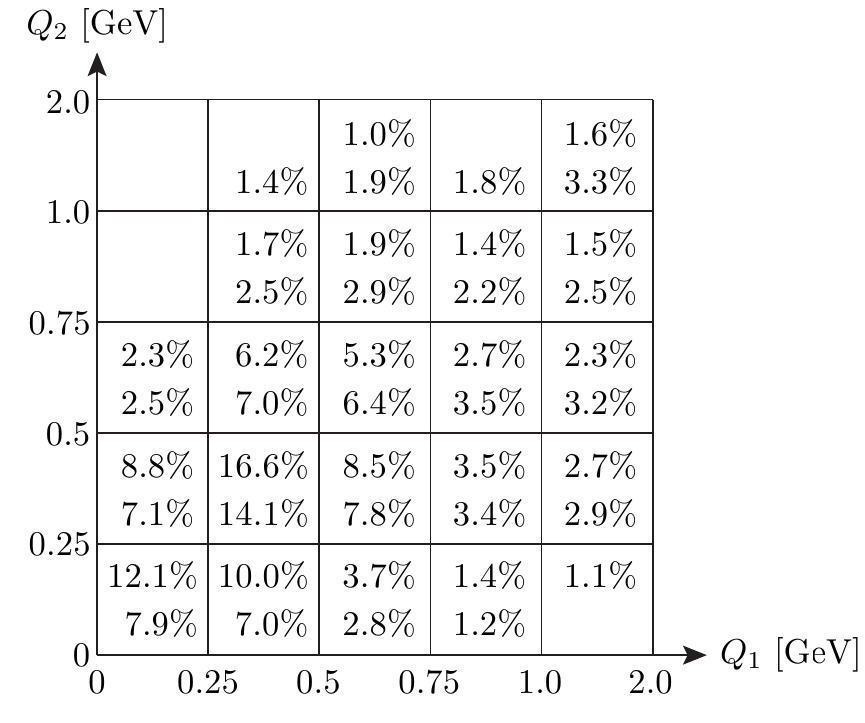}} 

\caption{Left panel: Relative contributions to the total
  $a_\mu^{\HLbLpi}$ from individual bins in the $(Q_1,Q_2)$-plane,
  integrated over all angles according to Eq.~(\ref{bins}). Note the
  larger size of the bins with $Q_{1,2} \geq 1~\mbox{GeV}$. Top line
  in each bin: LMD+V model, bottom line: VMD model. Contributions
  smaller than 1\% have not been displayed. For the LMD+V model there
  are further contributions bigger than 1\% along the $Q_1$-axis. For
  $2~\mbox{GeV} \leq Q_1 \leq 20~\mbox{GeV}$: $1.1\%$ in the bin $0
  \leq Q_2 \leq 0.25~\mbox{GeV}$ and $1.2\%$ for $0.25~\mbox{GeV} \leq
  Q_2 \leq 0.5~\mbox{GeV}$. Right panel: Relative contributions to the
  total of $a_\mu^{{\rm HLbL};\eta}$ and $a_\mu^{{\rm
      HLbL};\eta^\prime}$ with the VMD model from individual bins in
  the $(Q_1,Q_2)$-plane, integrated over all angles. Top line in each
  bin: $\eta$-meson, bottom line:
  $\eta^\prime$-meson.}

\label{Fig:PS_bins}

\end{figure}

For a more detailed analysis, we integrate in Eqs.~(\ref{amupi0_1})
and (\ref{amupi0_2}) over individual momentum bins and all angles
$\theta$ as follows
\be \label{bins} 
\int_{Q_{1,{\rm min}}}^{Q_{1,{\rm max}}} dQ_1 \int_{Q_{2,{\rm
      min}}}^{Q_{2,{\rm max}}} dQ_2 \int_{-1}^{1} d\tau 
\ee 
and display the results, relative to the totals in
Eqs.~(\ref{amuLMD+V})-(\ref{amuetaprime}), in Fig.~\ref{Fig:PS_bins}.

Since the absolute size of the weight function $w_1(Q_1,Q_2,\tau)$ is
much larger than $w_2(Q_1,Q_2,\tau)$, the contribution from the
integral $a_\mu^{\HLbLPone}$ in Eq.~(\ref{amupi0_1}) dominates over
$a_\mu^{\HLbLPtwo}$ in Eq.~(\ref{amupi0_2}). Therefore the asymmetry
seen in the $(Q_1,Q_2)$-plane in Fig.~\ref{Fig:PS_bins}, with larger
contributions below the diagonal, reflects the ridge-like structure of
$w_1(Q_1,Q_2,\tau)$ in Figs.~\ref{Fig:w_i_pion} and
\ref{Fig:w_i_eta_etaprime}.  Note that in relating the contributions
to $a_\mu^{\HLbLP}$ from different momentum regions to the form
factor, one has to take into account that what enters in the dominant
part $a_\mu^{\HLbLPone}$ in Eq.~(\ref{amupi0_1}) is the
double-offshell form factor at momenta $\FFP(-Q_1^2, -(Q_1+Q_2)^2)$,
not $\FFP(-Q_1^2,-Q_2^2)$.

For the pion, the largest contribution comes from the lowest bin
$Q_{1,2} \leq 0.25~\mbox{GeV}$ since a large part of the peaks in the
weight functions (for different angles $\theta$) is contained in that
bin. More than half of the contribution comes from the four bins with
$Q_{1,2} \leq 0.5~\mbox{GeV}$.  In contrast, for the $\eta$ and
$\eta^\prime$, it is not the bin $Q_{1,2} \leq 0.25~\mbox{GeV}$ which
yields the largest contribution, since the maxima of the weight
functions are shifted to higher momenta, around
$0.3-0.5~\mbox{GeV}$. Furthermore, more bins up to $Q_2 =
2~\mbox{GeV}$ now contribute at least $1\%$ to the total. This is
different from the pattern seen for $\pi^0$.  The plots of the weight
functions for $\eta$ and $\eta^\prime$ in
Figs.~\ref{Fig:w_i_eta_etaprime} show that now the region $1.5 -
2.5~\mbox{GeV}$ is also important for the evaluation of the $\eta$-
and $\eta^\prime$-pole contributions. However, as mentioned before,
the VMD model is known to have a too fast fall-off at large momenta,
compared to the OPE.  Therefore the size of the contributions
$a_\mu^{{\rm HLbL};\eta}$ and $a_\mu^{{\rm HLbL};\eta^\prime}$ in
Eqs.~(\ref{amueta}) and (\ref{amuetaprime}) might be underestimated by
the VMD model, which could also affect the relative importance of the
higher momentum region in Fig.~\ref{Fig:PS_bins}.

Integrating both $Q_1$ and $Q_2$ from zero up to some upper momentum
cutoff $\Lambda$ (integration over a square) and integrating over all
angles $\theta$, one obtains the results shown in
Table~\ref{Tab:cutoffdependence}, see also
Ref.~\cite{CD12_Proceedings} for a similar analysis for the
pseudoscalar-exchange contribution. This amounts to summing up the
individual bins shown in Fig.~\ref{Fig:PS_bins}.

\begin{table} 

  \caption{Pseudoscalar-pole contribution $a_\mu^{\HLbLP} \times
    10^{11}, {\rm P} = \pi^0,\eta,\eta^\prime$, for different form
    factor models obtained with a momentum
    cutoff~$\Lambda$. In brackets, relative contribution of the total
    obtained with $\Lambda = 20~\mbox{GeV}$.}  

\label{Tab:cutoffdependence}

\begin{center} 
\renewcommand{\arraystretch}{1.1}
\begin{tabular}{|r@{.}l|r@{.}l|r@{.}l|r@{.}l|r@{.}l|}
\hline 
\multicolumn{2}{|c|}{$\Lambda$ [GeV]} & \multicolumn{2}{|c|}{$\pi^0$
  [LMD+V]}  & \multicolumn{2}{|c|}{$\pi^0$ [VMD]}  &
\multicolumn{2}{|c|}{$\eta$ [VMD]}  &  \multicolumn{2}{|c|}{$\eta^\prime$
  [VMD]}  \\ 
\hline 
0 & 25 & 14 & 4~(22.9\%) & 14 & 4~(25.2\%) &  1 & 8~(12.1\%) &  1 &
0~(7.9\%)  \\  
0 & 5  & 36 & 8~(58.5\%) & 36 & 6~(64.2\%) &  6 & 9~(47.5\%) &  4 &
5~(36.1\%) \\  
0 & 75 & 48 & 5~(77.1\%) & 47 & 7~(83.8\%) & 10 & 7~(73.4\%) &  7 &
8~(62.5\%) \\  
1 & 0  & 54 & 1~(86.0\%) & 52 & 6~(92.3\%) & 12 & 6~(86.6\%) &  9 &
9~(79.1\%) \\  
1 & 5  & 58 & 8~(93.4\%) & 55 & 8~(97.8\%) & 14 & 0~(96.1\%) & 11 &
7~(93.1\%) \\  
2 & 0  & 60 & 5~(96.2\%) & 56 & 5~(99.2\%) & 14 & 3~(98.6\%) & 12 &
2~(97.4\%) \\  
5 & 0  & 62 & 5~(99.4\%) & 56 & 9~(99.9\%) & 14 & 5~(100\%)  & 12 &
5~(99.9\%) \\  
~20 & 0 & 62 & 9~(100\%)  & 57 & 0~(100\%)  & 14 & 5~(100\%)  & 12 &
5~(100\%) \\  
\hline 
\end{tabular}
\end{center} 

\end{table}

As one can see in Table~\ref{Tab:cutoffdependence}, for the pion more
than half of the final result stems from the region below $\Lambda =
0.5~\mbox{GeV}$ (59\% for LMD+V, 64\% for VMD) and the region below
$\Lambda = 1~\mbox{GeV}$ gives the bulk of the total result (86\% for
LMD+V, 92\% for VMD). The small difference between the form factor
models for small momenta $Q_{1,2} \leq 0.5~\mbox{GeV}$ is reflected in
the small absolute difference for $a_\mu^{\HLbLpi}$ in the two models
for $\Lambda \leq 0.75~\mbox{GeV}$. For instance, for $\Lambda =
0.5~\mbox{GeV}$, the difference is only $0.2 \times 10^{-11}$, i.e.\
about $0.5\%$ and for $\Lambda = 0.75~\mbox{GeV}$, the difference is
only $0.8 \times 10^{-11}$, i.e.\ $1.6\%$. The faster fall-off of the
VMD model at larger momenta beyond $1~\mbox{GeV}$, compared to the
LMD+V model, leads to a smaller contribution from that region to the
total. Therefore we can see in Fig.~\ref{Fig:PS_bins} that the main
contributions in the VMD model, relative to the total, are
concentrated at lower momenta, compared to the LMD+V model, in
particular below $0.75~\mbox{GeV}$.

On the other hand, Table~\ref{Tab:cutoffdependence} shows that for
$\eta$ and $\eta^\prime$, the region below $\Lambda = 0.25~\mbox{GeV}$
only gives a small contribution to the total (12\% for $\eta$, 8\% for
$\eta^\prime$). Up to $\Lambda = 0.5~\mbox{GeV}$, we get about half
(one third) for $\eta$ ($\eta^\prime$) and the bulk of the results
comes from the region below $\Lambda = 1.5~\mbox{GeV}$, 96\% for the
$\eta$ and 93\% for the $\eta^\prime$-meson. But note again, that the
VMD model might have a too strong fall-off at large momenta. A less
model-dependent evaluation of $a_\mu^{{\rm HLbL};\eta}$ and
$a_\mu^{{\rm HLbL};\eta^\prime}$, largely based on experimental data,
would therefore be highly preferable.

\section{Experimental status of transition form factors} 
\label{Sec:TFF_experiment}

For the calculation of the pseudoscalar-pole contribution
$a_\mu^{\HLbLP}$ with ${\rm P} = \pi^0, \eta, \eta^\prime$ in
Eqs.~(\ref{amupi0_1}) and (\ref{amupi0_2}) the single-virtual form
factor $\FFP(-Q^2,0)$ and the double-virtual form factor $\FFP(-Q_1^2,
-Q_2^2)$, both in the spacelike region, enter.  We are interested here
in the impact of uncertainties of experimental measurements of these
form factors on the precision of $a_\mu^{\HLbLP}$. In the following,
we will summarize the experimental information which is currently
available for these form factors or which should be available in the
near future. We will mostly just quote the uncertainties of several
experiments in different momentum regions, concentrating on the most
precise and most recent ones. More details about the actual
measurements can be found in the given references. See
Ref.~\cite{TFF_mini_review} for a brief overview of the various
experimental processes where information on the transition form
factors can be obtained. More details can be found in
Ref.~\cite{Landsberg}.

For the single-virtual form factor we parametrize the measurement
errors in Eqs.~(\ref{amupi0_1}) and (\ref{amupi0_2}) as follows:
\be \label{delta1}
\FFP(-Q^2,0) \to \FFP(-Q^2,0) \, \left(1 \pm \delta_{1,{\rm P}}(Q)
  \right).  
\ee 
The momentum dependent errors $\delta_{1, {\rm P}}(Q)$ in different
bins are displayed in Table~\ref{Tab:delta1} in
Section~\ref{Sec:impact}. Ideally, one should combine all the
experimental data to obtain $\FFP(-Q^2, 0)$ and $\delta_{1, {\rm
    P}}(Q)$ as function for all momenta. However, there are currently
still regions where there are no or rather unprecise data
available. We therefore need to make some assumptions and will employ
a simplified approach with bin-wise constant errors. Also, no
correlations between different bins are taken into account.

Similarly, for the double-virtual form factor we parametrize the
measurement errors in the following way
\be \label{delta2}
\FFP(-Q_1^2,-Q_2^2) \to \FFP(-Q_1^2,-Q_2^2) \, \left(1 \pm 
  \delta_{2, {\rm P}}(Q_1,Q_2) \right), 
\ee
where the assumed momentum dependent errors $\delta_{2, {\rm
    P}}(Q_1,Q_2)$ in different bins are shown in
Fig.~\ref{Fig:FF_errors_bins} in Section~\ref{Sec:impact}. Since there
are currently no data available for the double-offshell form factor,
we will use the results of a MC simulation~\cite{BESIII_private} for
BESIII, as described below, to estimate the errors $\delta_{2, {\rm
    P}}(Q_1,Q_2)$.

\subsection{Pion transition form factor}

For the single-virtual TFF of the pion $\FFa(-Q^2,0)$ the following
experimental information is available. The normalization of the form
factor can be obtained from the decay width $\Gamma(\pi^0 \to
\gamma\gamma) = (\pi \alpha^2 m_\pi^3/4) \FFa^2(0,0)$. For a detailed
overview of theory and experiment we refer to
Ref.~\cite{Bernstein_Holstein_13}. See also
Ref.~\cite{Kampf_review_11} which briefly reviews the theory for other
decay modes of the neutral pion that will be discussed below. From the
Particle Data Group (PDG) average~\cite{PDG_2015} for the decay width
$\Gamma(\pi^0 \to \gamma\gamma) = (7.63 \pm 0.16)~\mbox{eV}$, one
obtains a $1.1\%$ precision on the form factor normalization. The
error is largely driven by the single most precise measurement of the
decay width by the PrimEx collaboration $\Gamma(\pi^0 \to
\gamma\gamma) = (7.82 \pm 0.22)~\mbox{eV}$~\cite{PrimEx}. We will take
the corresponding precision of $1.4\%$ for the form factor at
vanishing momenta as conservative estimate on the normalization. An
improvement from PrimEx-II is expected soon at the level of $0.7\%$
for the form factor~\cite{PrimEx2}. There are also plans to measure
this decay at KLOE-2 with $0.5\%$ statistical precision for the
normalization of the form factor~\cite{KLOE2_impact}. Note that at the
level of $1-2\%$, quark mass corrections and radiative corrections to
the decay width need to be considered in order to compare
theory~\cite{pi0_gamma_gamma_theory} and experiment.

Another important experimental information is the slope of the form
factor at the origin. Following Ref.~\cite{Landsberg}, one
defines\footnote{The PDG~\cite{PDG_2015} parametrizes the form factor
  in conversion or Dalitz decays for small momenta in the linearized
  form $\FFa(q^2,0) = \FFa(0,0) \left(1 + a_{\pi^0} q^2 / m_{\pi^0}^2
  \right)$, therefore $b_{\pi^0} = a_{\pi^0} / m_{\pi^0}^2$.}
\be \label{TFF_slope} 
b_{\pi^0} = \left. \frac{1}{\FFa(0,0)} \frac{d \FFa(q^2,0)}{dq^2}
\right|_{q^2 = 0}. 
\ee
The PDG~\cite{PDG_2015} uses essentially the determination of the
slope by CELLO~\cite{CELLO} as their average $b_{\pi^0} = (1.76 \pm
0.22)~\mbox{GeV}^{-2}$, with a $12.5\%$ precision (assuming that the
systematic error is of the same size as the statistical error as
stated in Ref.~\cite{CELLO}). The CELLO collaboration was the first to
measure the single-virtual form factor $\FFa(-Q^2, 0)$ for spacelike
momenta in the momentum range $0.7 - 1.6~\mbox{GeV}$ in the process
$e^+ e^- \to e^+ e^- \gamma^* \gamma^* \to e^+ e^- \pi^0$. CELLO
determines the slope of the form factor using a simple VMD ansatz with
vector meson mass $\Lambda_{\pi^0} = (748 \pm 42)~\mbox{MeV}$ to fit
the data and evaluates the slope at zero momentum from the fitted
function as $b_{\pi^0} = 1/\Lambda_{\pi^0}^2$.  As pointed out in
Ref.~\cite{KN_EPJC_01}, this large extrapolation from $0.7~\mbox{GeV}$
to the origin might induce a model dependence or bias that is not
covered by the uncertainty from the fit to the data at higher
energies. In Ref.~\cite{Masjuan_12} it was argued, based on Pad\'e
approximants to form factor measurements in the spacelike region by
CELLO, CLEO~\cite{CLEO} and BABAR~\cite{BABAR}, that there might be a
$45\%$ systematic uncertainty in the slope from the modelling and the
large extrapolation. From a sequence of Pad\'e approximants,
Ref.~\cite{Masjuan_12} obtained the result $b_{\pi^0}[\mbox{Pad\'e}] =
(1.78 \pm 0.12)~\mbox{GeV}^{-2}$ with $6.9\%$ precision.  Other
experimental determinations of the slope in the timelike region from
the single Dalitz decay $\pi^0 \to e^+ e^- \gamma$ in
Ref.~\cite{pi0_TFF_slope_timelike}, published around the same time as
CELLO and used in the PDG average, yield mostly similar central
values, close to the CELLO result and the VMD prediction
$b_\pi[\mbox{VMD}] = 1/M_\rho^2 = 1.66~\mbox{GeV}^{-2}$, but with
uncertainties of $100\%$ or more (one experiment even obtained a
negative central value). Note that a proper treatment of radiative
corrections needs to be done in order to extract information on the
form factor or the branching ratio $\mbox{BR}(\pi^0 \to e^+ e^-
\gamma)$ from the data, as discussed in
Ref.~\cite{pi0_Dalitz_rad_corr}.

The single-virtual form factor $\FFa(-Q^2, 0)$ for spacelike momenta
has been measured by a series of experiments.  The CELLO~\cite{CELLO}
measurement was done in the momentum range $0.7 \leq Q \leq
1.6~\mbox{GeV}~(0.5 \leq Q^2 \leq 2.7~\mbox{GeV}^2)$. It has a
precision of about $8-9\%$ in the two bins from
$0.7-1~\mbox{GeV}$. Between $1-1.4~\mbox{GeV}$, the precision is about
$11-12\%$ and for the highest bin $20\%$. Later, CLEO~\cite{CLEO}
measured the form factor in the region $1.2 \leq Q \leq 3~\mbox{GeV}$
($1.5 \leq Q^2 \leq 9~\mbox{GeV}^2$). Between $1.2 - 1.5~\mbox{GeV}$
the precision is about $6-7\%$, between $1.5 - 2~\mbox{GeV}$ about
$8-11\%$ and above $2~\mbox{GeV}$ the uncertainty increases gradually
to $15\%$.  At higher energies, $2 \leq Q \leq 6.3~\mbox{GeV}~(4 \leq
Q^2 \leq 40~\mbox{GeV}^2)$, there are measurements of the form factor
available from BABAR~\cite{BABAR} and Belle~\cite{Belle}. Between
$2-3~\mbox{GeV}$ the precision of BABAR is about $3-4\%$. For Belle,
the form factor measurements in the two lowest bins with $2 \leq Q
\leq 2.4~\mbox{GeV}$ have errors of $16\%$ and $13\%$,
respectively. The error for the three next bins with $2.4 \leq Q \leq
3~\mbox{GeV}$ then drops to $5-6\%$. There is a disagreement between
the data of BABAR and Belle above about $3~\mbox{GeV}$, with the BABAR
data for the form factor not showing the $1/Q^2$ behavior expected
from QCD~\cite{Brodsky-Lepage}. While this puzzle needs to be
clarified, the form factor data above $3~\mbox{GeV}$ is not very
relevant for the muon $g-2$, as already discussed in
Ref.~\cite{CD09_Proceedings}. This can be seen from the plots of the
weight functions in Fig.~\ref{Fig:w_i_pion} and the results in
Table~\ref{Tab:cutoffdependence}.

An analysis is ongoing by BESIII~\cite{BESIII_single_virtual} to
measure the form factor in the region $0.5 \leq Q \leq 1.8~\mbox{GeV}$
($0.3 \leq Q^2 \leq 3.1~\mbox{GeV}^2$) with a precision of
$5-10\%$. Finally, according to the simulations performed in
Ref.~\cite{KLOE2_impact}, KLOE-2 should be able to measure the form
factor with 6\% statistical precision at even lower energies: $0.1
\leq Q \leq 0.3~\mbox{GeV}$ ($0.01 \leq Q^2 \leq 0.09~\mbox{GeV}^2$).

In principle, one can obtain information about the single-virtual form
factor $|\FFa(q^2,0)|$ at small timelike momenta from the single
Dalitz decay $\pi^0 \to \gamma^* \gamma \to e^+ e^- \gamma$, see
Ref.~\cite{pi0_TFF_slope_timelike}. However, due to the small pion
mass, the kinematical reach in $|q|$ is very small and form factor
effects are not clearly visible. This explains the difficulty to
extract for instance the slope parameter with a reasonable
precision. At higher momenta, the form factor enters in the process
$e^+ e^- \to \gamma^* \to \pi^0 \gamma$ and has been measured at SND
and CMD-2~\cite{pi0_gamma_SND_CMD-2}. Of course, one cannot simply
translate the form factor values $|\FFeta(q^2,0)|$ measured at
timelike momenta $q^2$ into the spacelike region $Q^2$. A proper
analytical continuation needs to be performed, preferably without
introducing too much model dependence.

Finally, recently a dispersion relation has been proposed in
Ref.~\cite{DR_pion_TFF} to determine the single and double-virtual
form factor. So far, only the single-virtual form factor has been
evaluated in this dispersive framework. The dispersion relation
yields, within uncertainties, a perfect description of the
experimental data for $e^+ e^- \to \pi^0\gamma$ from
Ref.~\cite{pi0_gamma_SND_CMD-2}. At small spacelike momenta, very
small errors are obtained, e.g.\ for $Q = 0.25,~0.5,~1~\mbox{GeV}$,
the uncertainties for $\FFa(-Q^2,0)$ are
$0.24\%,~0.9\%,~3.8\%$~\cite{DR_pion_TFF_private}. Note that to these
uncertainties an error of about $1.4\%$ for the normalization at $Q^2
= 0$ has to be added in quadrature. The normalization error might go
down further soon, thanks to PrimEx-II~\cite{PrimEx2}. The dispersion
relation predicts the slope with $2\%$ precision: $b_\pi[\mbox{DR}] =
(1.69 \pm 0.03)~\mbox{GeV}^{-2}$. This agrees quite well with the
naive VMD estimate, but this might be a coincidence, see the
discussion in Ref.~\cite{DR_pion_TFF}.

Based on the current experimental situation and the progress expected
in the next few years, we obtain the momentum dependent errors
$\delta_{1,\pi^0}(Q)$ in different bins as shown in
Table~\ref{Tab:delta1} in Section~\ref{Sec:impact}. Currently, no
experimental data on the pion TFF is available below the point $Q =
0.7~\mbox{GeV}$ from CELLO~\cite{CELLO}, except for the normalization
from $\Gamma(\pi^0 \to \gamma\gamma)$ and estimates of the
slope. BESIII~\cite{BESIII_single_virtual} should publish their
results down to $0.5~\mbox{GeV}$ soon and also
KLOE-2~\cite{KLOE2_impact} might perform measurements for $0.1 \leq Q
\leq 0.3~\mbox{GeV}$. For the lowest bin $0 \leq Q < 0.5~\mbox{GeV}$,
we therefore assume an error, based on ``extrapolating'' the current
data sets and the data that will be available in a few years. In the
bin $0.5 - 1~\mbox{GeV}$ mostly the data from CELLO and future BESIII
data lead to the assumed precision, while in the region
$1-2~\mbox{GeV}$ these are data from CLEO~\cite{CLEO} and future
BESIII data. Finally, for $2-3~\mbox{GeV}$, the data from
BABAR~\cite{BABAR} are the most precise. If one uses the dispersion
relation from Ref.~\cite{DR_pion_TFF} below $1~\mbox{GeV}$ and the
current error on the normalization from the decay width, one obtains
(conservatively) the uncertainties in the lowest two bins given in
brackets in Table~\ref{Tab:delta1}.

The second ingredient in Eqs.~(\ref{amupi0_1}) and (\ref{amupi0_2}) is
the double-virtual form factor $\FFa(-Q_1^2, -Q_2^2)$. Currently,
there are no direct experimental measurements available for this form
factor at spacelike momenta. From the double Dalitz decay $\pi^0 \to
\gamma^* \gamma^* \to e^+ e^- e^+ e^-$ one can obtain the double
off-shell form factor $|\FFa(q_1^2, q_2^2)|$ at small invariant
momenta in the timelike region~\cite{KTeV_pi0_to_4e}, but the results
are inconclusive, even translating into a negative value for the slope
of the form factor $b_{\pi^0} = (-2.2 \pm 2.2)~\mbox{GeV}^{-2}$ (with
100\% uncertainty). There might again be some issues with potentially
large radiative corrections with respect to the extraction of the form
factor and its slope from the data, see
Refs.~\cite{pi0_double_Dalitz_rad_corr, Kampf_review_11}.

From processes in the timelike region like $\omega \to \pi^0 \gamma^*
\to \pi^0 \ell^+ \ell^-~(\ell = e,\mu)$, $\phi \to \pi^0 \gamma^* \to
\pi^0 \ell^+ \ell^-$ and $e^+ e^- \to \gamma^* \to \omega\pi^0$ there
is information available for $|\FFa(m_V^2, q^2)|$, i.e. along certain
lines in the two dimensional plane~$(q_1^2,
q_2^2)$~\cite{TFF_mini_review}. Again, it is not so straightforward to
use this information measured far in the timelike region to obtain the
form factor with spacelike momenta.

There is also indirect information available on the double off-shell
form factor $\FFa(q_1^2, q_2^2)$ from the loop-induced decay $\pi^0
\to e^+ e^-$~\cite{PDG_2015, pi0_to_ee}. Without a form factor at the
$\pi^0-\gamma^*-\gamma^*$-vertex, the loop integral in the this decay
is ultraviolet divergent. One therefore obtains short-distance
constraints on the form factor $\FFa(-Q_1^2, -Q_2^2)$ in the spacelike
region, see Ref.~\cite{Knecht_et_al_99}. The connection between the
pion (pseudoscalar) decay into a lepton pair and the pion
(pseudoscalar) pole contribution to HLbL was already pointed out in
Refs.~\cite{KN_02, Knecht_et_al_PRL_02}.  It was later taken up in
Ref.~\cite{HLbL_PS_vs_lepton_pair_decay} and problems to explain both
processes simultaneously with the same model have been
stressed. Again, there are potential issues with radiative corrections
to extract the decay rate $\pi^0 \to e^+ e^-$ from the measured data,
see Ref.~\cite{pi0_lepton_pair_rad_corr}.

The decay $\pi^0 \to \gamma\gamma$ and the double-virtual form factor
$\FFa(-Q_1^2, -Q_2^2)$ has also been studied in Lattice
QCD~\cite{TFF_Lattice_1, TFF_Lattice_2} for spacelike and timelike
momenta. While at low momenta a description of the form factor by the
VMD model seems to work, at higher spacelike momenta, i.e.\ above
$0.5-1~\mbox{GeV}$, deviations are seen, in particular the impact of
excited states in the vector channel. For the pion decay width the
result $\Gamma(\pi^0 \to \gamma\gamma) = 7.83(31)(49)~\mbox{eV}$ has
been quoted in the most recent paper in Ref.~\cite{TFF_Lattice_2},
rather close to the experimental result from the PDG and PrimEx, but
the precision of $7.4\%$ is not yet competitive with the experimental
uncertainty. The result is the extrapolation to the physical pion
mass, however, only one lattice spacing has been used. The
double-virtual form factor with off-shell momenta has only been
studied in Refs.~\cite{TFF_Lattice_1, TFF_Lattice_2} for unphysical
pion masses ($M_{\pi} \geq 300~\mbox{MeV}$), which lead to
vector-meson masses of $1~\mbox{GeV}$ or more. Therefore more studies
are needed for firm conclusions about the applicability and
generalization of VMD.

Because of the lack of direct experimental measurements of the
double-virtual form factor $\FFa(-Q_1^2, -Q_2^2)$, models have been
used to describe the form factor in the spacelike region and thus all
current evaluations of $a_\mu^{{\rm HLbL;}\pi^0}$ are model
dependent. Often the assumption is made that the form factor
factorizes $\FFa(-Q_1^2, -Q_2^2) = f(Q_1^2) \times f(Q_2^2)$, like in
the popular VMD model, but, as already mentioned, this factorization
is not expected in QCD. With this assumption, the double-virtual form
factor is then completely determined by the single-virtual form factor
where a lot of experimental information is available, although not yet
in the low-momentum region $Q \leq 0.7~\mbox{GeV}$, which is very
important for the pion-pole contribution, as we have seen
before. Nevertheless, at least for low momenta, the assumption of
factorization might work well numerically.  For instance, the LMD+V
model, which does not factorize, and the VMD model differ for $Q_1 =
Q_2 = 0.5~\mbox{GeV}$ by only $3\%$.

Of course, it would be preferable to replace these model assumptions
as much as possible by experimental data. In fact, it is planned to
determine the double-virtual form factor at BESIII for momenta $0.5
\leq Q_{1,2} \leq 1.5~\mbox{GeV}$ and a first analysis is already in
progress~\cite{BESIII_private}, based on existing data. More data will
be collected in the coming years.  Maybe at very low momenta $Q_{1,2}
\leq 0.5~\mbox{GeV}$ also KLOE-2 could measure this double-virtual
form factor and Belle 2 could maybe measure it at higher momenta
$Q_{1,2} \geq 1~\mbox{GeV}$. It would be highly welcome, if some
simulations and data analysis would be performed.

The dispersion relation derived in Ref.~\cite{DR_pion_TFF} allows to
determine the double-virtual form factor, but has not yet been
evaluated. This might allow to check the factorization property,
although at momenta approaching $1~\mbox{GeV}$, where the LMD+V and
the VMD model differ already by $23\%$, it remains to be seen whether
the dispersive approach will give reliable results. If factorization
works below $1~\mbox{GeV}$, the precision of the dispersive approach
for the form factor could be $2 \times 4\% = 8\%$ (or better), see
Table~\ref{Tab:delta1}.

In view of the current absence of direct experimental information on
the double-virtual form factor $\FFa(-Q_1^2, -Q_2^2)$, we use as
estimate of the measurement errors $\delta_{2,\pi^0}(Q_1, Q_2)$ the
results of a Monte Carlo simulation~\cite{BESIII_private} for the
BESIII detector using the LMD+V model in the EKHARA event
generator~\cite{EKHARA} for the signal process $e^+ e^- \to e^+ e^-
\gamma^*\gamma^* \to e^+ e^- \pi^0$. The results for $\delta_{2,
  \pi^0}(Q_1,Q_2)$ in different momentum bins are shown in
Fig.~\ref{Fig:FF_errors_bins} in Section~\ref{Sec:impact}.

Since the number of MC events $N_i$ in bin number $i$ is proportional
to the cross-section $\sigma_i$ (in that bin) and since for the
calculation of the cross-section the form factor enters squared, the
statistical error on the form factor measurement is given according to
Poisson statistics by
\be \label{stat_error}
\sigma_i \sim \FFa^2 \Rightarrow \frac{\delta \FFa}{\FFa} =
\frac{\sqrt{N_i}}{2 N_i}.  
\ee
To simplify the appearance of Fig.~\ref{Fig:FF_errors_bins}, we have
rounded the number of events from the Monte Carlo simulation to
integer values (similarly for the percentage errors) and symmetrized
the entries off the diagonal. In total there are 605 events in the
displayed momentum region.

In the lowest momentum bin $Q_{1,2} \leq 0.5~\mbox{GeV}$, there are no
events in the simulation, because of the acceptance of the
detector. When both $Q_{1,2}^2$ are small, both photons are almost
real and the scattered electrons and positrons escape detection along
the beam pipe. As a further assumption, we have therefore taken the
average of the uncertainties in the three neighboring bins as estimate
for the error in that lowest bin. This ``extrapolation'' from the
neighboring bins seems justified, since information along the two axis
is (or will soon be) available and the value at the origin is known
quite precisely from the decay width. Note that although the form
factor for spacelike (Euclidian) momenta is rather smooth in the two
models considered in this paper (and in other models as well), see
Fig.~\ref{Fig:LMD+V_vs_VMD} in Appendix~\ref{App:TFF_models}, it is
far from being a constant and some nontrivial extrapolation is
needed. For instance for $Q^2 = (0.5~\mbox{GeV})^2$ we get for both the
LMD+V and the VMD model $\FFa(-Q^2,0)/\FFa(0,0) = 0.7$ and
$\FFa(-Q^2,-Q^2)/\FFa(0,0) = 0.5$. Of course, a direct experimental
measurement in that lowest bin would be helpful. In the meantime, the
dispersive approach from Ref.~\cite{DR_pion_TFF} will hopefully give
reliable results at these low momenta.

The Monte Carlo simulation~\cite{BESIII_private} corresponds to a data
sample with an integrated luminosity of $2.9~\mbox{fb}^{-1}$,
collected at BESIII at an energy of $\sqrt{s} =
3.773~\mbox{GeV}$. This is approximately half of the data set
collected at BESIII so far. We should point out that the simulation
only included signal events without any decay of the $\pi^0$ and
assumed $100\%$ detection efficiency and acceptance. There is a large
background from Bhabha events with additional radiated photons, which
has to be removed by cuts or more sophisticated analysis techniques,
like neural networks.  Based on a first preliminary analysis of the
actual BESIII data~\cite{BESIII_private} with strong cuts to reduce
the background, it seems possible that the number of events and the
corresponding precision for $\FFa(-Q_1^2,-Q_2^2)$ shown in
Fig.~\ref{Fig:FF_errors_bins} could be achievable with the current
data set plus a few more years of data taking at BESIII.

On the other hand, once experimental data will be available, e.g.\
event rates in the different momentum bins, there will still be the
task to unfold the data to reconstruct the form factor $\FFa(-Q_1^2,
-Q_2^2)$. This should be done without introducing too much model
dependence, i.e.\ more sophisticated approaches are needed than what
is done for the single-virtual form factor where often a simple VMD
form factor is used as fitting function. As a first approximation, one
could maybe even use a constant form factor in each momentum bin. In
this sense the situation is somewhat different from the hadronic
vacuum polarization contribution to the muon $g-2$, where one only
needs to insert the hadronic cross-section into the dispersion
integral with a known kernel function and can then use a simple
trapezoidal rule to perform the integral.

\subsection{$\eta$ transition form factor}

For the $\eta$-meson, the following experimental information is
available about the single-virtual form factor $\FFeta(-Q^2,0)$. From the
PDG average~\cite{PDG_2015} for the decay width $\Gamma(\eta \to
\gamma\gamma) = (0.516 \pm 0.018)~\mbox{keV}$ one obtains a $1.7\%$
precision on the form factor normalization. The error is driven by the
recent measurement $\Gamma(\eta \to \gamma\gamma) = (0.520 \pm
0.024)~\mbox{keV}$ by the KLOE-2
collaboration~\cite{KLOE-2_eta_gamma_gamma}, which gives a $2.3\%$
precision for the form factor at vanishing momenta.

In contrast to the pion, the single Dalitz decay $\eta \to \ell^+
\ell^- \gamma$, with $\ell = e, \mu$ now has enough phase-space, so
that the slope and the form factor can be measured in the timelike
region.  The slope of the form factor at zero momentum has been
extracted with a precision of $9.1\%$ from measurements of the form
factor by NA60~\cite{NA60}\footnote{In the conference
  proceedings~\cite{NA60_Proceedings} by the NA60 collaboration a
  further analysis even claimed a precision of $3.7\%$ for the slope.}
and by A2~\cite{A2}: $b_\eta[\mbox{NA60, A2}] = (1.95 \pm
0.18)~\mbox{GeV}^{-2}$. The NA60 collaboration measured the Dalitz
decay $\eta \to \mu^+\mu^-\gamma$ approximately in the range $220 \leq
|q| \leq 480~\mbox{MeV}$, whereas the A2 collaboration measured the
decay $\eta \to e^+ e^- \gamma$ down to very small momenta, $45 \leq
|q| \leq 450~\mbox{MeV}$. As in earlier experiments, a VMD ansatz was
fitted to the data and from this fit function the slope was
calculated. In Ref.~\cite{Escribano_et_al_15} it was argued, based on
an analysis using Pad\'e approximants for the form factor, that the
modelling of the data with a simple VMD ansatz and the extrapolation
to zero might induce an additional error on the slope of about $5\%$
for experimental determinations in the timelike region. Combining
various recent timelike and spacelike data,
Ref.~\cite{Escribano_et_al_15} obtained a $2\%$ determination of the
slope: $b_\eta[\mbox{Pad\'e}] = (1.919 \pm 0.037)~\mbox{GeV}^{-2}$,
improving on an earlier $11\%$ determination based on spacelike data
in Ref.~\cite{Escribano_et_al_14}: $b_\eta[\mbox{Pad\'e, spacelike}] =
(2.00 \pm 0.22)~\mbox{GeV}^{-2}$.

The slope was also determined from measurements of the form factor
$\FFeta(-Q^2,0)$ in the spacelike region. The CELLO
collaboration~\cite{CELLO} measured in the region $0.5 -
1.8~\mbox{GeV}$ and the slope was extracted with about $21\%$
precision (assuming, as for the pion, that the systematic error is of
the same size as the statistical error). However, since a simple VMD
ansatz was fitted to the data above $0.5~\mbox{GeV}$, there could be
some large bias and uncertainty from the extrapolation to zero
momentum, as already discussed in the context of determinations of the
slope of the form factor for the pion. In fact, the fitted vector
meson mass $\Lambda_\eta[\mbox{CELLO}] = (839 \pm 63)~\mbox{MeV}$
differs quite substantially from the value $\Lambda_\eta[\mbox{NA60,
  A2}] = (716 \pm 33)~\mbox{MeV}$ obtained by NA60 and A2. Shortly
before CELLO, the form factor in the spacelike region has also been
measured by the TPC/2$\gamma$ collaboration~\cite{TPC_2gamma} in the
region $0.3-2.6~\mbox{GeV}$. The fitted vector meson mass
$\Lambda_\eta[\mbox{TPC/2$\gamma$}] = (700 \pm 80)~\mbox{MeV}$ is
rather close to the value extracted in the timelike region. The
TPC/2$\gamma$ collaboration did not evaluate the slope, their
measurement would translate into a precision of $23\%$ for the
slope. The fit by the CLEO collaboration~\cite{CLEO} of their form
factor data in the spacelike region from $1.2 - 4.5~\mbox{GeV}$ with a
VMD model yielded an even better determination of the vector meson
mass $\Lambda_\eta(\mbox{CLEO}) = (774 \pm 29)~\mbox{MeV}$. This would
formally translate into an $8\%$ determination of the slope, but the
extrapolation to zero seems even more questionable than for CELLO and
therefore CLEO did not quote a value for the slope. The value of
$\Lambda_\eta$ by CLEO differs by two standard deviations from the
results by NA60 and A2. There is, however, no reason why the fitted
values of $\Lambda_\eta$ for large spacelike and small timelike
momenta should be the same. Note that we used the value of
$\Lambda_\eta$ from the CLEO fit to fix the vector meson mass in our
VMD model to obtain the result in Eq.~(\ref{amueta}).

The form factor $\FFeta(-Q^2,0)$ has been measured in the spacelike
region for the first time by the TPC/$2\gamma$
collaboration~\cite{TPC_2gamma} for $0.3 \leq Q \leq
2.6~\mbox{GeV}~(0.1 \leq Q^2 \leq 7~\mbox{GeV}^2)$. Since the data
points are not tabulated, the measurement precision is difficult to
estimate from the logarithmic plot in their paper.  The
CELLO~\cite{CELLO} collaboration measured the form factor for $0.5
\leq Q \leq 1.8~\mbox{GeV}~(0.3 \leq Q^2 \leq 3.4~\mbox{GeV}^2)$. The
precision for $0.5 \leq Q \leq 0.9~\mbox{GeV}$ is $14\%$ and for $0.9
\leq Q \leq 1.1~\mbox{GeV}$ about $19\%$. In the two bins above
$1.1~\mbox{GeV}$, the precision is $23\%$ and $18\%$,
respectively. The CLEO collaboration~\cite{CLEO} measured the form
factor in the range $1.2 \leq Q \leq 4.5~\mbox{GeV}~(1.5 \leq Q^2 \leq
20~\mbox{GeV}^2)$. For $1.2 - 1.6~\mbox{GeV}$ the best precision in
some decay channels is about $8-9\%$, between $1.6 - 2~\mbox{GeV}$
about $8-10\%$ and above $2~\mbox{GeV}$ about $10-14\%$ (or even more
than $20\%$ in the highest bin in some decay channels).  More
recently, BABAR~\cite{BABAR_TFF_eta_etaprime} measured the form at
higher momenta, $2 \leq Q \leq 6.3~\mbox{GeV}~(4 \leq Q^2 \leq
40~\mbox{GeV}^2)$. Between $2-3~\mbox{GeV}$ the precision is about
$4-5\%$. There is also one earlier measurement by BABAR at very large
timelike momenta $q^2 = 112~\mbox{GeV}^2$~\cite{BABAR_TFF_timelike},
which is compatible with the results for large spacelike momenta.

It is difficult to extract the precision of the measurement of the
transition form factor $|\FFeta(q^2,0)|$ in the timelike region by the
NA60 collaboration from the logarithmic plot in Ref.~\cite{NA60}. The
precision of the form factor measurements by the A2
collaboration~\cite{A2} between $45-150~\mbox{MeV}$ is about
$1.5-2\%$, between $150-300~\mbox{MeV}$ about $2-5\%$, between
$300-400~\mbox{MeV}$ about $5-11\%$ and above $400~\mbox{MeV}$ around
$15-28\%$. In the region $220-400~\mbox{MeV}$, the precision of NA60
is about the same as A2, while above $400~\mbox{MeV}$, the precision
of NA60 seems a bit better. As for the pion, one then needs to
properly map the values measured in the timelike into the spacelike
region via an analytical continuation, without introducing new model
dependence.

In Ref.~\cite{DR_eta_etaprime_TFF} a dispersion relation was proposed
and evaluated for the single-virtual TFF $\FFeta(q^2,0)$. It yields a
rather precise prediction for the form factor and agrees well with the
most precise measurements by NA60~\cite{NA60, NA60_Proceedings} and
A2~\cite{A2} up to $0.25~\mbox{GeV}^2$ in the timelike region. A
prediction for the slope with a precision of $11\%$ was given in
Ref.~\cite{DR_eta_etaprime_TFF}, but it was later pointed out in
Ref.~\cite{DR_eta_TFF_a2} that there are effects of the $a_2$-tensor
meson in the left-hand cut of the form factor, which shift the slope
down by about $7\%$. A recent
reanalysis~\cite{DR_eta_TFF_double_virtual} then leads to the result
$b_\eta[\mbox{DR}] = 1.9^{+0.2}_{-0.1}~\mbox{GeV}^{-2}$, i.e.\ with
about $11\%$ precision, neglecting the tiny effects
$-0.04~\mbox{GeV}^{-2}$ from the isoscalar contribution. This value is
compatible with the experimental results by NA60 and A2 and with the
recent analysis using Pad\'e approximants~\cite{Escribano_et_al_14,
  Escribano_et_al_15}, but bigger than older theoretical
estimates~\cite{Ametller_et_al_92}, using ChPT (1-loop with resonance
saturation for the low-energy constants), VMD, quark-loop models and
the Brodsky-Lepage interpolation formula for the
TFF.

From these observations, we deduce the values for the measurement
errors $\delta_{1,\eta}(Q)$ for the single-virtual TFF
$\FFeta(-Q^2,0)$ shown in Table~\ref{Tab:delta1}. We assume an error
in the lowest bin, $0 \leq Q \leq 0.5~\mbox{GeV}$, as done already for
the pion. There is one data point available from a measurement by the
TPC/2$\gamma$ collaboration~\cite{TPC_2gamma}, with a bin starting at
$0.3~\mbox{GeV}$, but it is difficult to estimate the uncertainty. We
hope that the information on the normalization, the slope at zero and
the data in the timelike region can be used to obtain some reliable
estimate in that low momentum region. In the bin $0.5 \leq Q \leq
1~\mbox{GeV}$ there is data available from TPC/2$\gamma$ and from
CELLO~\cite{CELLO}, but they are not very precise. It would be very
helpful, if BESIII could measure the $\eta$-TFF in that region in the
near future. In the bin $1-2~\mbox{GeV}$ there is data available from
CLEO~\cite{CLEO} with about the given precision and for
$2-3~\mbox{GeV}$ the data by BABAR~\cite{BABAR_TFF_eta_etaprime} are
the most precise.

Concerning the double-virtual form factor $\FFeta(-Q_1^2, -Q_2^2)$,
there are no measurements in the spacelike region yet. The branching
ratio of the double Dalitz decay $\eta \to e^+ e^- e^+ e^-$ has been
measured~\cite{eta_to_4e}, but no attempt was made to extract the TFF
in the timelike region. Again, there are some indirect constraints
from the loop-induced decay $\eta \to \mu^+ \mu^-$~\cite{PDG_2015,
  HLbL_PS_vs_lepton_pair_decay}, see also the recent analysis in
Ref.~\cite{PS_to_lepton_pairs_Pade}. The dispersive approach was recently
extended to the double-virtual TFF $\FFeta(q_1^2, q_2^2)$ in
Ref.~\cite{DR_eta_TFF_double_virtual}. Because of limited input data
for the DR, it could not be evaluated for general momenta. It was
shown, however, that for $q_1^2 \ll 1~\mbox{GeV}^2$ and
$1~\mbox{GeV}^2 \leq q_2^2 \leq (4.5~\mbox{GeV})^2$ the form factor is
compatible with the factorization ansatz, if the effects of the $a_2$
meson are taken into account.

We therefore use again the results of the MC simulation
~\cite{BESIII_private} for BESIII, but now with the VMD model in
EKHARA~\cite{EKHARA} and display the corresponding estimates for the
errors $\delta_{2,\eta}(Q_1,Q_2)$ in
Fig.~\ref{Fig:FF_errors_bins}. For this simulation, there are in total
345 events in the given momentum region.

\subsection{$\eta^\prime$ transition form factor}

For the $\eta^\prime$ meson, one has the following experimental
information about the single-virtual TFF $\FFetaprime(-Q^2,0)$. The
PDG average~\cite{PDG_2015} $\Gamma(\eta^\prime \to \gamma\gamma)
= (4.28 \pm 0.19)~\mbox{keV}$ leads to a $2.2\%$ determination on the
form factor normalization at zero momentum. As for $\pi^0$ and $\eta$,
the uncertainty is driven mostly by one experiment, this time by the
L3 collaboration~\cite{L3}. Their value $\Gamma(\eta^\prime \to
\gamma\gamma) = (4.17 \pm 0.29)~\mbox{keV}$ leads to a $3.5\%$
precision for the form factor.

As for the $\eta$-meson, now the single Dalitz decay $\eta^\prime \to
\ell^+ \ell^- \gamma, \ell = e,\mu$, has enough phase-space, so that
the slope and the form factor can be measured. The slope of the form
factor has been determined very recently with a precision of $11.7\%$
from a measurement of the form factor $|\FFetaprime(q^2,0)|$ in the
timelike region by the BESIII collaboration~\cite{BESIII_timelike}:
$b_{\eta^\prime}[\mbox{BESIII}] = (1.60 \pm
0.19)~\mbox{GeV}^{-2}$. The Dalitz decay $\eta^\prime \to e^+ e^-
\gamma$ was measured in the momentum range $0 \leq |q| \leq
0.8~\mbox{GeV}$. If one uses a simple VMD ansatz as in
Eq.~(\ref{VMD}), a pole appears in that momentum region at the vector
meson mass. Therefore a Breit-Wigner ansatz with a pole mass $\Lambda$
and a width $\gamma$ has to be fitted (or a sum of such terms with
several vector mesons like $\rho, \omega,\phi$, see also
Ref.~\cite{Landsberg}). The BESIII data was fitted with a single
vector meson and from this fit function the vector meson mass
$\Lambda_{\eta^\prime}[\mbox{BESIII}] = 790 \pm 40~\mbox{MeV}$ and
the corresponding slope was calculated. Using instead for the bins
below $0.5~\mbox{GeV}$, i.e.\ below the $\rho$-pole, a VMD ansatz
without a width parameter leads to a value for the vector meson mass
and the slope which is consistent with the first method. Previously,
the slope had only been determined from a measurement of the form
factor $|\FFetaprime(q^2,0)|$ of the decay $\eta^\prime \to \mu^+
\mu^- \gamma$ by the Lepton-G collaboration~\cite{Lepton-G_etaprime}
in the timelike region $0.2 \leq |q| \leq 0.9~\mbox{GeV}$ with a
precision of about $24\%$ as quoted in Ref.~\cite{Landsberg},
corresponding to $\Lambda_{\eta^\prime}[\mbox{Lepton-G}] = 770 \pm
90~\mbox{MeV}$.

Again there are also determinations of the slope by measurements of
the form factor $\FFetaprime(-Q^2,0)$ in the spacelike region.  The
form factor was measured by the L3 collaboration~\cite{L3} in the
region $0.1 \leq Q \leq 3.2~\mbox{GeV}$. The data were fitted with a
VMD ansatz with $\Lambda_{\eta^\prime}[\mbox{L3}] = (900 \pm
51)~\mbox{MeV}$, which differs significantly from the value obtained
by BESIII. The L3 collaboration did not translate their result into a
determination of the slope, which would give a $11.3\%$
precision. Although there are three data points by L3 at rather low
momentum values (two of them below $0.5~\mbox{GeV}$), there are also
two data points with large momenta above $1~\mbox{GeV}$, which might
distort the fit and the result for $\Lambda_{\eta^\prime}$. The CELLO
collaboration~\cite{CELLO} has measured the form factor in the
spacelike region $0.5 - 4.5~\mbox{GeV}$ and a VMD fit yielded
$\Lambda_{\eta^\prime}[\mbox{CELLO}] = 794 \pm 44~\mbox{MeV}$, which
is rather close to the value obtained by BESIII. This gives a
$15.7\%$ determination of the slope (assuming, as for the pion, that
the systematic error is of the same size as the statistical
error). Shortly before CELLO, the TPC/$2\gamma$
collaboration~\cite{TPC_2gamma} measured the form factor at slightly
lower spacelike momenta $0.3 - 2.6~\mbox{GeV}$ and obtained
$\Lambda_{\eta^\prime}[\mbox{TPC/2$\gamma$}] = 850 \pm 70~\mbox{MeV}$,
i.e.\ a bit higher than CELLO, but with similar precision. The
TPC/2$\gamma$ collaboration did not evaluate the slope. As for the
pion and the $\eta$, the VMD fit of the data measured by the CLEO
collaboration~\cite{CLEO} in the spacelike region $1.2 -
5.5~\mbox{GeV}$ yields the most precise determination of the vector
meson mass parameter $\Lambda_{\eta^\prime}[\mbox{CLEO}] = 859 \pm
28~\mbox{MeV}$. This result deviates by about two standard deviations
from the values obtained by BESIII in the timelike region and by
CELLO in the spacelike region. Again, because of the large
extrapolation to zero momentum, CLEO did not quote a result for the
slope. Formally, their result would be a determination of the slope
with $6.6\%$ precision. Note that we used the value of
$\Lambda_{\eta^\prime}$ from the CLEO fit to fix the vector meson mass
in our VMD model to obtain the result in Eq.~(\ref{amuetaprime}).

The form factor $\FFetaprime(-Q^2,0)$ has been measured in the
spacelike region for the first time by the PLUTO
collaboration~\cite{PLUTO} for $0.4 \leq Q \leq 1.0~\mbox{GeV}~(0.2
\leq Q^2 \leq 1~\mbox{GeV}^2)$. For the lowest bin $0.4 -
0.6~\mbox{GeV}$ the precision was about $20\%$, for the two higher
bins only about $40\%$. Then there was the measurement by the
TPC/2$\gamma$ collaboration~\cite{TPC_2gamma} in the region $0.3 \leq
Q \leq 2.6~\mbox{GeV}~(0.1 \leq Q^2 \leq 7~\mbox{GeV}^2)$.  The
measurement precision is difficult to estimate from the logarithmic
plot in their paper. The CELLO~\cite{CELLO} collaboration measured the
form factor for $0.5 \leq Q \leq 4.5~\mbox{GeV}~(0.3 \leq Q^2 \leq
20~\mbox{GeV}^2)$. The precision for $0.5 \leq Q \leq 0.9~\mbox{GeV}$
is about $11\%$ and for $0.9 \leq Q \leq 1.1~\mbox{GeV}$ about
$13\%$. In the two bins $1.1-1.8~\mbox{GeV}$, the precision is $14\%$
and $17\%$, respectively, and above $1.8~\mbox{GeV}$ it is $30\%$. The
CLEO collaboration~\cite{CLEO} measured the form factor in the range
$1.2 \leq Q \leq 4.5~\mbox{GeV}~(1.5 \leq Q^2 \leq
30~\mbox{GeV}^2)$. For $1.2 - 1.6~\mbox{GeV}$ the best precision in
some decay channels is about $7\%$, between $1.6 - 2~\mbox{GeV}$ about
$8\%$ and above $2~\mbox{GeV}$ about $8-12\%$ (or much more in some
decay channels). The L3 collaboration~\cite{L3} measured the form
factor in the region $0.1 \leq Q \leq 3.2~\mbox{GeV}~(0.01 \leq Q^2
\leq 10~\mbox{GeV}^2)$. There are three untagged measurement points in
the bins $(0.1 - 0.4), (0.4 - 0.5), (0.5 - 0.9)~\mbox{GeV}$ with
precisions of $5\%, 8\%, 11\%$. The two bins with tagged events from
$1.2 - 3.2~\mbox{GeV}$ have a precision of about $15\%$. More
recently, BABAR~\cite{BABAR_TFF_eta_etaprime} measured the form at
higher momenta, $2 \leq Q \leq 6.3~\mbox{GeV}~(4 \leq Q^2 \leq
40~\mbox{GeV}^2)$. Between $2-3~\mbox{GeV}$ the precision is about
$4\%$. The measurement by BABAR at very large timelike momenta $q^2 =
112~\mbox{GeV}^2$~\cite{BABAR_TFF_timelike} is compatible with the
results for large spacelike momenta.

The precision of the measurement of the form factor
$|\FFetaprime(q^2,0)|$ in the timelike region by
BESIII~\cite{BESIII_timelike} is about $2.8\%$ in the lowest bin $|q|
\leq 0.1~\mbox{GeV}$. For $0.1-0.3~\mbox{GeV}$ it is about $7\%$ and
for $0.3-0.5~\mbox{GeV}$ it is $11\%$. Just below the peak region
$0.5-0.8~\mbox{GeV}$ it is about $14\%$. Since for the Lepton-G
experiment~\cite{Lepton-G_etaprime} there is only a logarithmic plot
available in Ref.~\cite{Landsberg}, it is not really possible to give
an estimate on the precision. Again, one needs to perform an
analytical continuation to obtain the form factor
$\FFetaprime(-Q^2,0)$ in the spacelike region.

The dispersion relation for the single-virtual $\eta$-TFF proposed in
Ref.~\cite{DR_eta_etaprime_TFF} can also be used for the
$\eta^\prime$-meson, under the additional assumption that the slopes
of the pion spectra in $\eta \to \pi\pi\gamma$ and $\eta^\prime \to
\pi\pi\gamma$ are identical. This then yields
$b_{\eta^\prime}[\mbox{DR}] = 1.53^{+0.15}_{-0.08}~\mbox{GeV}^{-2}$
with about $10\%$ precision. As can be seen from the erratum of
Ref.~\cite{DR_eta_etaprime_TFF}, the assumption of an identical slope
in the spectral shape is compatible with the data, but it might not be
fulfilled completely. Furthermore, as pointed out in
Ref.~\cite{DR_eta_TFF_a2}, there could be the effect of the
$a_2$-tensor meson in the left-hand cut of the form factor, which
could be larger than for the $\eta$-TFF.  Therefore, this dispersive
evaluation of the slope needs to be scrutinized further, when more
precise data become available in the future. The value for the slope
itself agrees well with the new result by
BESIII~\cite{BESIII_timelike}. A recent determination of the slope
using Pad\'e approximants to spacelike and timelike TFF data yields a
somewhat smaller central value: $b_{\eta^\prime}[\mbox{Pad\'e}] =
(1.43 \pm 0.04)~\mbox{GeV}^{-2}$~\cite{Escribano_et_al_15_etaprime}
with about $3\%$ precision, improving considerably on an earlier
estimate that used only spacelike data: $b_{\eta^\prime}[\mbox{Pad\'e;
  spacelike}] = (1.42 \pm
0.18)~\mbox{GeV}^{-2}$~\cite{Escribano_et_al_14}. Older estimates
based on ChPT, VMD and quark-loop models give here similar
results~\cite{Ametller_et_al_92}.

From all these experimental results, we obtain the values for the
measurement errors $\delta_{1,\eta^\prime}(Q)$ for the single-virtual
TFF $\FFetaprime(-Q^2,0)$ shown in Table~\ref{Tab:delta1}. For the
$\eta^\prime$-meson the experimental situation is a bit better than
for $\pi^0$ and $\eta$, since there are in principle quite precise
data available by the L3 collaboration~\cite{L3} for $0.1 \leq Q \leq
0.5~\mbox{GeV}$. Since these measurements of the form factor are based
on untagged events, it would be good to have a cross check by some
other experiment with lepton tagging in the future, like BESIII. In
the bin $0.5 \leq Q \leq 1~\mbox{GeV}$ there are measurements with
similar precision available from L3 and CELLO~\cite{CELLO}.  Again,
more precise data in this region from BESIII or Belle-2 would be very
useful. In the bin $1-2~\mbox{GeV}$ there are measurements with the
given precision from CLEO~\cite{CLEO} and for $2-3~\mbox{GeV}$ the
data of BABAR~\cite{BABAR_TFF_eta_etaprime} have the precision listed
in Table~\ref{Tab:delta1}.

There are no measurements of the double-virtual form factor
$\FFetaprime(-Q_1^2, -Q_2^2)$, neither in the spacelike nor the
timelike region. According to Ref.~\cite{PDG_2015} there are not even
measurements of the corresponding branching ratios of the double
Dalitz decays $\eta^\prime \to \ell^+ \ell^- \ell^+ \ell^-$, $\ell =
e,\mu$. Furthermore, there is only an upper bound on the branching
ratio of the loop-induced decay $\eta^\prime \to e^+
e^-$~\cite{PDG_2015}, see also the discussion in
Ref.~\cite{PS_to_lepton_pairs_Pade}. For the errors
$\delta_{2,\eta^\prime}(Q_1,Q_2)$ we use therefore again the results
of the MC simulation~\cite{BESIII_private} for BESIII with the VMD
model in EKHARA~\cite{EKHARA} and display them in
Fig.~\ref{Fig:FF_errors_bins}. There are 902 events in the given
momentum region.

\section{Impact of form factor uncertainties on $a_\mu^{{\rm HLbL; P}}$} 
\label{Sec:impact}

\begin{figure}[h!]

\centerline{\includegraphics[width=0.5\textwidth]{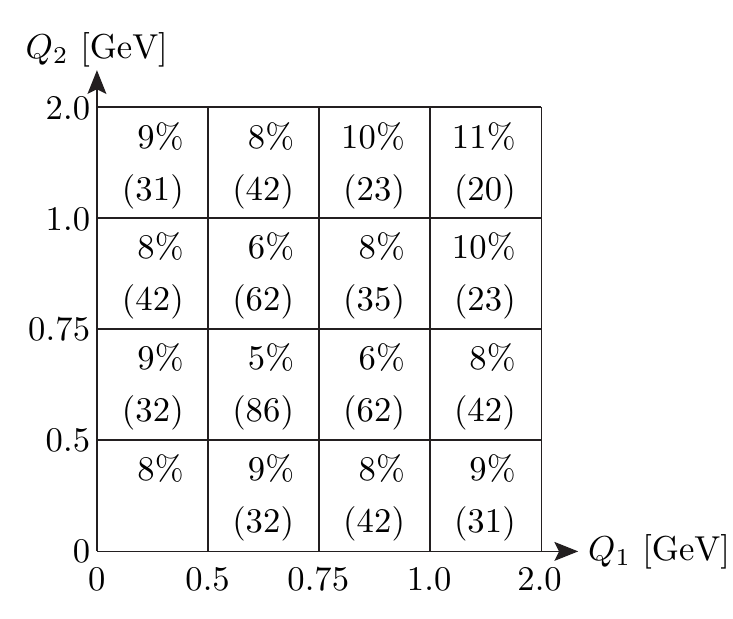}
\includegraphics[width=0.5\textwidth]{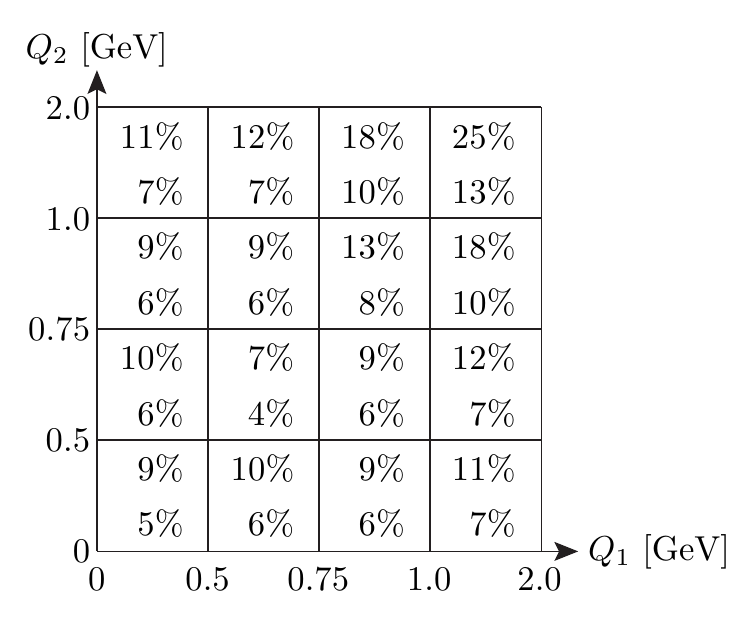}}

\caption{Left panel: Assumed relative error $\delta_{2,
    \pi^0}(Q_1,Q_2)$ of the pion TFF $\FFa(-Q_1^2,-Q_2^2)$ in
  different momentum bins. Note the unequal bin sizes. In brackets the
  number of MC events $N_i$ in each bin according to the simulation
  with the LMD+V model for BESIII. In total, there are 605
  events. For the lowest bin, $Q_{1,2} \leq 0.5~\mbox{GeV}$, there are
  no events in the simulation due to the detector acceptance. In that
  bin, we assume as error the average of the three neighboring
  bins. For $Q_{1,2} \geq 2~\mbox{GeV}$, we take a constant error of
  15\%. Right panel: Assumed relative error $\delta_{2, {\rm
      P}}(Q_1,Q_2)$ of the form factor $\FFP(-Q_1^2,-Q_2^2)$ for $P =
  \eta, \eta^\prime$ in different momentum bins according to the MC
  simulations with the VMD form factor for $\eta$ (top line) and
  $\eta^\prime$ (bottom line). The error in the lowest bin is obtained
  by averaging the neighboring bins. For $Q_{1,2} \geq 2~\mbox{GeV}$,
  we assume a constant error of 25\% for $\eta$ and 15\% for
  $\eta^\prime$.}

\label{Fig:FF_errors_bins}

\end{figure}

Based on the discussion in the previous Section, we summarize in
Table~\ref{Tab:delta1} the precision $\delta_{1, {\rm P}}(Q)$ that is
currently reached, or should soon be available, for the single-virtual
TFF $\FFP(-Q^2, 0)$ for all three light pseudoscalars. In
Fig.~\ref{Fig:FF_errors_bins} we show the estimated precision
$\delta_{2, {\rm P}}(Q_1, Q_2)$ for the double-virtual form factor
$\FFP(-Q_1^2, -Q_2^2)$ based on the MC simulation for
BESIII~\cite{BESIII_private}.

\begin{table}[h!] 

  \caption{Relative error $\delta_{1,{\rm P}}(Q)$ on the form factor
    $\FFP(-Q^2,0)$ for ${\rm P} = \pi^0, \eta, \eta^\prime$ in
    different momentum regions. The errors for $\delta_{1,\pi^0}(Q)$
    and $\delta_{1,\eta}(Q)$ below $0.5~\mbox{GeV}$ are based on
    assumptions, see discussions in the text. In brackets for $\pi^0$
    the uncertainties with a dispersion relation for the transition
    form factor.}    

\label{Tab:delta1}

\begin{center}
\renewcommand{\arraystretch}{1.1}
\begin{tabular}{|r@{$~Q~$}l|r@{\%}l|r@{\%}l|r@{\%}l|}
\hline 
\multicolumn{2}{|c|}{Region [GeV]} &
\multicolumn{2}{|c|}{$\delta_{1,\pi^0}(Q)$} &
\multicolumn{2}{|c|}{$\delta_{1,\eta}(Q)$}  &  
\multicolumn{2}{|c|}{$\delta_{1,\eta^\prime}(Q)$}  \\ 
\hline
$0 \leq $   & $ < 0.5$ & 5 & ~[2\%] & ~~10 & &    6 & \\ 
$0.5 \leq $ & $ < 1$   & 7 & ~[4\%] &   15 & & ~~11 & \\ 
$1 \leq $   & $ < 2$   & 8 &        &    8 & &    7 & \\  
$2 \leq $   &          & 4 &        &    4 & &    4 & \\ 
\hline 
\end{tabular}
\end{center} 

\end{table}

Taking again the LMD+V and VMD models for illustration, the assumed
momentum dependent errors from Table~\ref{Tab:delta1} and
Fig.~\ref{Fig:FF_errors_bins} impact the precision for the
pseudoscalar-pole contributions to HLbL as follows
\bea
a_{\mu; {\rm LMD+V}}^{\HLbLpi} & = & {62.9}^{+8.9}_{-8.2} \times
10^{-11} \quad \left(^{+14.1\%}_{-13.1\%}\right), \label{LMD+V_impact}
\\ 
a_{\mu; {\rm VMD}}^{\HLbLpi} & = & {57.0}^{+7.8}_{-7.3} \times 
10^{-11} \quad \left(^{+13.7\%}_{-12.7\%}\right), \label{VMD_impact}   
\\ 
a_{\mu; {\rm VMD}}^{{\rm HLbL};\eta} & = & 14.5^{+3.4}_{-3.0} \times
10^{-11} \quad \left(^{+23.4\%}_{-20.8\%}\right), \label{VMD_impact_eta}
\\ 
a_{\mu; {\rm VMD}}^{{\rm HLbL};\eta^\prime} & = & 12.5^{+1.9}_{-1.7}
\times 10^{-11} \quad
\left(^{+15.1\%}_{-13.9\%}\right). \label{VMD_impact_etaprime}  
\eea
While for the pion the absolute variations are different for the two
models, as are the central values, the relative uncertainty for both
models is around $14\%$. This will also be visible in the following
more detailed analysis. We therefore expect that using other form
factor models, and, eventually, using experimental data for the
single- and double-virtual form factors, will not substantially change
the following observations and conclusions.  Of course, a more
sophisticated error analysis will be needed, once experimental
informations on the double-virtual form factor will become available,
also taking into account correlations between data points in different
bins.

More details have been collected in Table~\ref{Tab:amuHLbLP_impact},
which contains the results from
Eqs.~(\ref{LMD+V_impact})-(\ref{VMD_impact_etaprime}) in the first
line. The impact of the uncertainties in different momentum regions,
below and above $0.5~\mbox{GeV}$, are shown in the lines $2-5$ in the
table and potential further improvements in the last five lines. Line
6 shows the impact of the use of the DR for the single-virtual TFF
$\FFa(-Q^2,0)$ for the pion in the two bins below $1~\mbox{GeV}$, as
indicated in Table~\ref{Tab:delta1}. The lines 7 and 8 show the effect
of an improvement of $\delta_{1,\eta}(Q)$ in the two bins below
$1~\mbox{GeV}$ from $10\%, 15\%$ to $8\%, 10\%$ and of
$\delta_{1,\eta^\prime}(Q)$ from $6\%, 11\%$ to $5\%, 8\%$. Such a
precision is similar to what has already been achieved in the region
$1-2~\mbox{GeV}$ and could for instance come from future measurements
at BESIII. Line number 9 shows the precision on $a_\mu^{{\rm HLbL};
  \pi^0, \eta, \eta^\prime}$ from an improvement in the lowest bin of
$\delta_2(Q_1, Q_2)$ from $8\%, 9\%, 5\%$ to $5\%, 7\%, 4\%$, i.e.\
what is obtained in the MC simulation for the second bin along the
diagonal in Fig.~\ref{Fig:FF_errors_bins}. Finally, line 10 shows what
would happen, if the second bins along the two axis in $\delta_2$
would have the same precision as well, i.e.\ all four bins below
$0.75~\mbox{GeV}$ would have the same precision as the second bin
along the diagonal.

\begin{table}[h!]

  \caption{Impact of assumed measurement errors $\delta_{1, {\rm
        P}}(Q)$ and $\delta_{2, {\rm P}}(Q_1,Q_2)$ in the form factors
    $\FFP(-Q^2,0)$ and $\FFP(-Q_1^2, -Q_2^2)$ on the relative
    precision of the pseudoscalar-pole contributions (first
    line). Lines $2-5$ show the effects of uncertainties in different
    momentum regions below and above $0.5~\mbox{GeV}$. Lines $6-10$
    show the impact of potential improvements of some of the assumed
    errors.}         

\label{Tab:amuHLbLP_impact}

{\small 

\vspace*{-0.3cm}
\begin{center}
\renewcommand{\arraystretch}{1.5}
\begin{tabular}{|c|c|c|c|l|}
\hline 
$\frac{\delta a_{\mu; {\rm LMD+V}}^{\HLbLpi}}{a_{\mu; {\rm LMD+V}}^{\HLbLpi}}$ & 
$\frac{\delta a_{\mu; {\rm VMD}}^{\HLbLpi}}{a_{\mu; {\rm VMD}}^{\HLbLpi}}$  & 
$\frac{\delta a_{\mu; {\rm VMD}}^{{\rm HLbL};\eta}}{a_{\mu; {\rm
      VMD}}^{{\rm HLbL};\eta}}$ &  
$\frac{\delta a_{\mu; {\rm VMD}}^{{\rm HLbL};\eta^\prime}}{a_{\mu;
    {\rm VMD}}^{{\rm HLbL};\eta^\prime}}$ &  
\multicolumn{1}{|c|}{Comment} \\  
\hline 
${}^{+14.1\%}_{-13.1\%}$ & ${}^{+13.7\%}_{-12.7\%}$ & 
${}^{+23.4\%}_{-20.8\%}$ & ${}^{+15.1\%}_{-13.9\%}$ & 
Given $\delta_1,\delta_2$ \\
\hline
${}^{+4.3\%}_{-4.2\%}$ & ${}^{+4.4\%}_{-4.3\%}$ & 
${}^{+6.9\%}_{-6.8\%}$ & ${}^{+3.4\%}_{-3.3\%}$ &  
Bin $Q < 0.5~\mbox{GeV}$ in $\delta_1$ as given, rest: $\delta_{1,2} = 0$ \\ 
\hline 
${}^{+1.1\%}_{-1.0\%}$ & ${}^{+1.0\%}_{-0.9\%}$ & 
${}^{+4.4\%}_{-4.3\%}$ & ${}^{+4.5\%}_{-4.4\%}$ &  
Bins $Q \geq 0.5~\mbox{GeV}$ in $\delta_1$ as given, rest:
$\delta_{1,2} = 0$ \\ 
\hline 
${}^{+4.5\%}_{-4.4\%}$ & ${}^{+4.9\%}_{-4.8\%}$ & 
${}^{+4.0\%}_{-4.0\%}$ & ${}^{+1.7\%}_{-1.7\%}$ &  
Bin $Q_{1,2} < 0.5~\mbox{GeV}$ in $\delta_2$ as given, rest:
$\delta_{1,2} = 0$ \\ 
\hline 
${}^{+3.9\%}_{-3.8\%}$ & ${}^{+3.2\%}_{-3.1\%}$ & 
${}^{+7.0\%}_{-6.8\%}$ & ${}^{+5.1\%}_{-5.0\%} $ &  
Bins $Q_{1,2} \geq 0.5~\mbox{GeV}$ in $\delta_2$ as given, rest:
$\delta_{1,2} = 0$ \\ 
\hline 
${}^{+10.9\%}_{-10.5\%}$ & ${}^{+10.6\%}_{-10.1\%}$ &  
$-$ & $-$ & 
Given $\delta_1,\delta_2$, lowest two bins in $\delta_{1,\pi^0}$: $2\%,
4\%$ \\   
\hline 
$-$ & $-$ & 
${}^{+20.4\%}_{-18.5\%}$ & $-$ &  
Given $\delta_1,\delta_2$, lowest two bins in $\delta_{1,\eta}$:
$8\%, 10\%$  \\  
\hline  
$-$ & $-$ & 
$-$ & ${}^{+13.4\%}_{-12.5\%}$ &  
Given $\delta_1,\delta_2$, lowest two bins in
$\delta_{1,\eta^\prime}$: $5\%, 8\%$  \\   
\hline  
${}^{+12.4\%}_{-11.6\%}$ & ${}^{+11.8\%}_{-11.0\%}$ & 
${}^{+22.4\%}_{-20.0\%}$ & ${}^{+14.8\%}_{-13.6\%}$ & 
$\pi^0, \eta, \eta^\prime$: given $\delta_1,\delta_2$, lowest
bin $\delta_2$: $5\%, 7\%, 4\%$ \\
\hline 
${}^{+12.0\%}_{-11.2\%}$ & ${}^{+11.4\%}_{-10.6\%}$ & 
${}^{+21.9\%}_{-19.6\%}$ & ${}^{+14.4\%}_{-13.4\%} $ & 
In addition, bins in $\delta_2$ close to lowest: $5\%, 7\%, 4\%$ \\ 
\hline   
\end{tabular}

\end{center} 
}

\end{table}

For the pion, the largest uncertainty of about $5\%$ comes from the
lowest bin $Q_{1,2} \leq 0.5~\mbox{GeV}$ in the $(Q_1,Q_2)$-plane for
$\delta_2$ in Fig.~\ref{Fig:FF_errors_bins} (fourth line in the
table).  Some improvement could be achieved, if the error in that
lowest bin and the neighboring bins ($Q_{1,2} \leq 0.75~\mbox{GeV}$)
could be reduced, to a total error of about $12\%$, see the lines 9
and 10 in Table~\ref{Tab:amuHLbLP_impact}. The second largest
uncertainty of $4.4\%$ for the pion stems from the lowest bin $Q <
0.5~\mbox{GeV}$ in $\delta_1$ (second line in the table). Here the use
of a dispersion relation for the single-virtual form factor
$\FFa(-Q^2, 0)$ for $Q < 1~\mbox{GeV}$ (see values in brackets in
Table~\ref{Tab:delta1}) could bring the total error of $14\%$ down to
$11\%$, see the sixth line. As seen from the plots of the weight
functions and the relative contribution to the total from different
momentum regions in Fig.~\ref{Fig:PS_bins}, for the pion more precise
data for the single- and double-virtual form factor in the region
below $0.5~\mbox{GeV}$ would be very important to reduce the error for
the pion-pole contribution to HLbL.

For the $\eta$ meson, the largest uncertainties of $7\%$ originate
from the region of $\delta_2$ above $0.5~\mbox{GeV}$ (5th line) and
from the lowest bin in $\delta_1$ (2nd line).  For the $\eta^\prime$,
the largest uncertainty of $5\%$ comes again from the region of
$\delta_2$ above $0.5~\mbox{GeV}$ (5th line). The second largest
uncertainty of $4.5\%$ comes from the bins in $\delta_1$ above
$0.5~\mbox{GeV}$ (3rd line). For $\eta$ and $\eta^\prime$ the errors
go down to $20\%$ and $13\%$, if the uncertainty in the two lowest
bins in $\delta_1$ could be reduced, see lines 7 and 8 in
Table~\ref{Tab:amuHLbLP_impact}. There is only a small reduction of
the uncertainty by one percentage point, if the errors in the lowest
few bins of $\delta_2$ with $Q_{1,2} \leq 0.75~\mbox{GeV}$ could be
reduced further, see lines 9 and 10. This is not unexpected, since for
the $\eta$ and $\eta^\prime$ that very-low momentum region is not as
important as for the $\pi^0$, see Fig.~\ref{Fig:PS_bins}. Here, more
precise data on the single- and double-virtual form factor in some
intermediate region $0.5-1.5~\mbox{GeV}$ would be very important to
reduce the uncertainty of the $\eta$- and $\eta^\prime$-pole
contributions to HLbL.

The size of the uncertainties can be understood approximately as
follows. If the errors would be independent of the momenta and small,
so that it is sufficient to take only the terms linear in $\delta_{1,
  {\rm P}}$ and $\delta_{2, {\rm P}}$, one would expect the following
uncertainty for the pseudoscalar-pole contribution
\be \label{delta_a_mu}
\delta a_\mu^{\HLbLP} \approx \left( \delta_{1, {\rm P}} + \delta_{2,
    {\rm P}} \right) \, a_\mu^{\HLbLP}.  
\ee
While the situation is more complicated with the given momentum
dependent errors, one can observe the following for the impact of the
uncertainties in the region below $0.5~\mbox{GeV}$. We only discuss
the case of the pion with the LMD+V form factor. The analysis is
similar for the pion with the VMD form factor and for $\eta$ and
$\eta^\prime$ with the VMD form factor. For the LMD+V model, the
region $Q_{1,2} < 0.5~\mbox{GeV}$ yields $59\%$ of the total result,
see Table~\ref{Tab:cutoffdependence}. Therefore, the effect of the
lowest bin of $\delta_{2, \pi^0}$ in the $(Q_1,Q_2)$-plane with an
assumed uncertainty of $8\%$ (see Fig.~\ref{Fig:FF_errors_bins})
translates with $\delta_{1, \pi^0} = 0$ in Eq.~(\ref{delta_a_mu}) to
about $8\% \times 0.59 \approx 4.7\%$ error in $a_\mu^{\HLbLpi}$. This
agrees quite well with the $4.5\%$ in line 4 of
Table~\ref{Tab:amuHLbLP_impact}. In doing so we neglected, however, in
the numerically dominating contribution in Eq.~(\ref{amupi0_1}), which
involves the weight function $w_1$ and the double-virtual form factor
$\FFa(-Q_1^2, -(Q_1 + Q_2)^2)$, the momentum dependence in
$\delta_{2,\pi^0}(|Q_1|, |Q_1 + Q_2|)$ with $|Q_1 + Q_2| = \sqrt{Q_1^2
  + 2 Q_1 Q_2 \tau + Q_2^2}$, see Eq.~(\ref{delta2}). This momentum
dependence in $\delta_{2,\pi^0}$ leads to an upper cutoff of
$0.5~\mbox{GeV}$ on $Q_1$ and to a cutoff on $Q_2$, which varies
between $0-1~\mbox{GeV}$ as a function of $Q_1$ and $\tau$. However,
as can be seen from Fig.~\ref{Fig:PS_bins}, summing up all the bins
with $Q_1 < 0.5~\mbox{GeV}$ and $Q_2 < 1~\mbox{GeV}$ adds at most a
few percent to the $59\%$.

On the other hand, looking at the impact of a non-zero
$\delta_{1,\pi^0}(Q)$ for $Q < 0.5~\mbox{GeV}$ in line 2 in
Table~\ref{Tab:amuHLbLP_impact} (with $\delta_{2,\pi^0} = 0$ in
Eq.~(\ref{delta_a_mu})), one has to take into account that in
Eq.~(\ref{amupi0_1}) the single-virtual form factor $\FFa(-Q_2^2, 0)$
enters. Therefore the corresponding non-zero $\delta_{1,\pi^0}(Q_2)$
restricts only the integration region $Q_2 < 0.5~\mbox{GeV}$, while
$Q_1$ runs up to infinity. This amounts to summing up the indiviual
bins in Fig.~\ref{Fig:PS_bins} in the direction of $Q_1$ with $Q_2 <
0.5~\mbox{GeV}$, i.e.\ the lowest two bins in the
$Q_2$-direction. This summation yields about $85\%$ of the total. With
a precision of $5\%$ in $\delta_{1,\pi^0}$ this gives $85\% \times
0.05 \approx 4.2\%$, to be compared with $4.3\%$ in
Table~\ref{Tab:amuHLbLP_impact}.

\section{Conclusions} 
\label{Sec:conclusions}

Recently, a dispersive approach to HLbL in the muon $g-2$ has been
proposed in Refs.~\cite{HLbL_DR_Bern_Bonn, HLbL_DR_Mainz}, which
connects the presumably numerically dominant contributions from the
light pseudoscalars $({\rm P} = \pi^0, \eta, \eta^\prime)$ and the
two-pion intermediate states to, in principle, measurable quantities,
like the single- and double-virtual pseudoscalar transition form
factor $\FFP(-Q_1^2, -Q_2^2)$ and the scattering of two off-shell
photons into two pions.

In this paper we studied in detail the pseudoscalar-pole contribution
to HLbL in this dispersive framework. The three-dimensional integral
representation for $a_\mu^{\HLbLP}$ from Ref.~\cite{JN_09}, shown in
Eqs.~(\ref{amupi0_1}) and (\ref{amupi0_2}), allows one to separate the
generic kinematics, described by model-independent weight functions
$w_{1,2}(Q_1, Q_2, \tau)$, from the double-virtual transition form
factors $\FFP(-Q_1^2, -Q_2^2)$, which can, in principle, be
measured. From the weight functions one can already identify which are
the most important momentum regions in the pseudoscalar-pole
contribution to HLbL. However, the weight-function $w_1$ has a slowly
decreasing ridge and the corresponding integral for HLbL diverges
without form factors. We therefore used two simple form factor models
(LMD+V, VMD) to evaluate $a_\mu^{\HLbLP}$. From this we deduced that
the relevant momentum region for $\pi^0$ is below about
$1~\mbox{GeV}$, irrespective of which of the two models is used.  For
$\eta$ and $\eta^\prime$, we only used a simple VMD model and showed
that the region below about $1.5~\mbox{GeV}$ gives the bulk of the
result. However, since the double-virtual VMD form factor falls off
too fast at high momenta compared to the predictions of the OPE, it
might be that the momentum region $1.5-2.5~\mbox{GeV}$ is cut off too
much. It would therefore be very useful, if experimental data on the
double-virtual form factor, e.g.\ from planned measurements at BESIII,
could reduce this model dependence.

If the assumed measurement errors $\delta_{1,{\rm P}}(Q)$ in
Table~\ref{Tab:delta1} and $\delta_{2, {\rm P}}(Q_1,Q_2)$ in
Fig.~\ref{Fig:FF_errors_bins} on the single- and double-virtual TFF
can be achieved in the coming years, one could obtain the following,
largely data driven, uncertainties for the pseudoscalar-pole
contributions to HLbL (see
Eqs.~(\ref{LMD+V_impact})-(\ref{VMD_impact_etaprime}) and
Table~\ref{Tab:amuHLbLP_impact})
\bea 
\frac{\delta a_\mu^{\HLbLpi}}{a_\mu^{\HLbLpi}} & = & 14\% \quad
[11\%],  \label{delta_a_pion} \\   
\frac{\delta a_\mu^{{\rm HLbL};\eta}}{a_\mu^{{\rm HLbL};\eta}} & = & 
23\%,  \\   
\frac{\delta a_\mu^{{\rm HLbL};\eta^\prime}}{a_\mu^{{\rm
      HLbL};\eta^\prime}} & = & 15\%.   
\eea 
The result in bracket for the pion in Eq.~(\ref{delta_a_pion}) uses
the DR~\cite{DR_pion_TFF} for the single-virtual TFF $\FFa(-Q^2, 0)$
below $1~\mbox{GeV}$. Compared to the range of estimates in the
literature in Eqs.~(\ref{range_HLbLpi0}) and (\ref{range_HLbLP}) this
would definitely be some progress, as it would be largely based on 
experimental input data only. More work is needed, however, to reach a
precision of $10\%$ for all three contributions which is envisioned in
the data-driven approach to HLbL~\cite{HLbL_DR_Bern_Bonn,
  HLbL_DR_Mainz}. We have shown in Table~\ref{Tab:amuHLbLP_impact},
what would be the impact of further potential improvements in the
measurement precision of the single-virtual TFF below $1~\mbox{GeV}$
and the double-virtual form factor below about $0.75~\mbox{GeV}$. 

One should keep in mind that the MC simulation from which the
estimates for $\delta_{2; {\rm P}}$ are derived, only considered the
signal process and no backgrounds. On the other hand, since the
simulation was based on about half of the data already collected at
BESIII, it seems not unreasonable that the assumed precision could be
reached with the full, almost doubled, data set in a few years, when
appropriate cuts are imposed or more sophisticated analysis tools are
used~\cite{BESIII_private}. It remains to be seen, how the unfolding
of event rates can be done to reconstruct the form factors, without
introducting too much model dependence. Once this is achieved, a much
more refined analysis can be performed, combining all existing data
for the single- and double-virtual TFF, also taking into account
correlations. One can then also combine this with the dispersive
framework for the TFF themselves~\cite{DR_pion_TFF,
  DR_eta_etaprime_TFF, DR_eta_TFF_a2, DR_eta_TFF_double_virtual}.

We hope that our analysis of the relevant momentum regions in HLbL
will be a useful guide for the experimental collaborations to look in
more details at the various processes where the single- and
double-virtual form factors can be measured. For instance,
measurements of the double-virtual form factors, e.g.\ by KLOE-2 for
$\pi^0$ in the low-momentum region $Q \leq 0.5~\mbox{GeV}$, where
BESIII cannot detect any events, or by Belle 2 for $\eta, \eta^\prime$
for higher momenta, $1 - 1.5~\mbox{GeV}$, would be very helpful.

We close by noting that similar three-dimensional integral
representations with corresponding weight functions have been derived
for the scalar-exchange contribution to HLbL in
Ref.~\cite{nonlocal_chiQM_P_S} and for all contributions to HLbL in
Refs.~\cite{Bijnens_MITP_14, HLbL_DR_Bern_Bonn}. They can be analyzed
along the same lines as shown here for $w_1$ and $w_2$ for the
pseudoscalars to identify the relevant momentum regions in a
model-independent way. This in turn can then help to plan future
measurement, e.g.\ for $\gamma^* \gamma^* \to \pi^+ \pi^-, \pi^0
\pi^0$, which is needed as input for the dispersive
framework~\cite{HLbL_DR_Bern_Bonn, HLbL_DR_Mainz}. In this way, a
concerted effort of theory and experiment can hopefully reduce and
better control the uncertainty in the HLbL contribution to the muon
$g-2$, so that one can fully profit from the upcoming future
experiments at Fermilab and J-PARC~\cite{future_g-2_exp}.

\section*{Acknowledgments} 

I am grateful to Achim Denig, Christoph Redmer and Pascal Wasser for
providing me with results of MC simulations for planned transition
form factor measurements at BESIII and to Martin Hoferichter and
Bastian Kubis for sharing information about the precision of the DR
approach to the pion TFF. I thank them and Hans Bijnens, Gilberto
Colangelo, Henryk Czy\. z, Simon Eidelman, Fred Jegerlehner, Marc
Knecht, Marc Vanderhaeghen and Graziano Venanzoni for discussions and
suggestions. This work was supported by Deutsche
Forschungsgemeinschaft (DFG) through the Collaborative Research Center
``The Low-Energy Frontier of the Standard Model'' (SFB 1044).

\appendix 

\renewcommand{\theequation}{\thesection\arabic{equation}}

\section{Weight functions in the integral
  representations for $a_\mu^{\HLbLpi}$}
\label{App:weight_functions}

\setcounter{equation}{0}

The kinematic functions in the two-loop integral for $a_\mu^{\HLbLpi}$
in Eqs.~(\ref{amupi0_start_1}) and (\ref{amupi0_start_2}) have been
evaluated in Ref.~\cite{KN_02} and read as follows
\bea
\tilde T_1(q_1,q_2;p) & = & (-64 \pi^6) \left[ \frac{16}{3}\, (p \cdot q_1)
  \, (p \cdot q_2) \, (q_1 \cdot q_2)
\,-\, \frac{16}{3}\, (p \cdot q_2)^2 \, q_1^2 \right. \nonumber \\
&& 
\qquad\qquad -\, \frac{8}{3}\, (p \cdot q_1) \, (q_1 \cdot q_2) \, q_2^2
\,+\, 8 (p \cdot q_2) \, q_1^2 \, q_2^2
\,-\,\frac{16}{3} (p \cdot q_2) \, (q_1 \cdot q_2)^2 \nonumber
\\
&& 
\left. \qquad\qquad +\, \frac{16}{3}\, m_\mu^2 \, q_1^2 \, q_2^2
\,-\, \frac{16}{3}\, m_\mu^2 \, (q_1 \cdot q_2)^2 \right],  \\
\tilde T_2(q_1,q_2;p) & = & (-64 \pi^6) \left[ \frac{16}{3}\, (p \cdot q_1)
  \, (p \cdot q_2) \, (q_1 \cdot q_2) \,-\,\frac{16}{3}\, (p \cdot
  q_1)^2 \, q_2^2 \right. \nonumber \\
&& 
\qquad\qquad  +\, \frac{8}{3}\, (p \cdot q_1) \, (q_1 \cdot q_2) \, q_2^2
\,+\, \frac{8}{3}\, (p \cdot q_1) \, q_1^2 \, q_2^2
\,\nonumber \\
&& 
\left. \qquad\qquad +\, \frac{8}{3}\, m_\mu^2 \, q_1^2 \, q_2^2
\,-\, \frac{8}{3}\, m_\mu^2 \, (q_1 \cdot q_2)^2 \right]. 
\eea
Recall that the muon momentum is on-shell: $p^2 = m_\mu^2$.

The model-independent weight functions from the three-dimensional
integral representation for $a_\mu^{\HLbLpi}$ in Eqs.~(\ref{amupi0_1})
and (\ref{amupi0_2}) read~\cite{JN_09}  
\bea 
w_1(Q_1,Q_2,\tau) & = & \left(- \frac{2\pi}{3} \right)
\sqrt{1-\tau^2} \, \frac{Q_1^3 Q_2^3}{Q_2^2 + m_\pi^2} \,
I_1(Q_1,Q_2,\tau), \label{w1_App} \\ 
w_2(Q_1,Q_2,\tau) & = & \left(- \frac{2\pi}{3} \right)
\sqrt{1-\tau^2} \, \frac{Q_1^3 Q_2^3}{Q_3^2 + m_\pi^2} \,
I_2(Q_1,Q_2,\tau), \label{w2_App} 
\eea
with
\bea
I_1(Q_1,Q_2,\tau) & = & X(Q_1,Q_2,\tau)\,\biggl(
                       8 \, P_1 \, P_2 \, (Q_1 \cdot Q_2)  
                       -2 \, P_1 \, P_3 \, (Q_2^4/m_\mu^2 - 2 \,
                       Q_2^2)  \nonumber \\  
      &&\qquad\qquad\qquad 
        -2\, P_1 \, ( 2 - Q_2^2/m_\mu^2 
        +2 \, (Q_1 \cdot Q_2) \, / m_\mu^2) 
        +4\, P_2\,P_3 \, Q_1^2  \nonumber \\
      &&\qquad\qquad\qquad  
        -4\, P_2 - 2 \, P_3 \, (4 + Q_1^2/m_\mu^2 - 2\,Q_2^2/m_\mu^2 )
         + 2/m_\mu^2 
       \biggr) \nonumber \\ 
      && -2\, P_1 \, P_2 \, (1 + (1-R_{m1}) \, (Q_1 \cdot Q_2) \, / m_\mu^2)
      \nonumber \\ 
      && + P_1 \, P_3 \, (2 - (1-R_{m1}) \, Q_2^2 / m_\mu^2) 
         + P_1 \, (1-R_{m1})/m_\mu^2 \nonumber \\ 
      && + P_2 \, P_3 \, (2 + (1-R_{m1})^2 \, (Q_1 \cdot Q_2) \, /
         m_\mu^2 ) 
      + 3 \, P_3 \, (1-R_{m1}) / m_\mu^2, \label{I1} 
\eea 
and 
\bea 
I_2(Q_1,Q_2,\tau) & = & X(Q_1,Q_2,\tau)\,\biggl(
                       4\,P_1\,P_2 \, (Q_1\cdot Q_2) 
                       +2\, P_1\,P_3 \, Q_2^2 - 2\, P_1
                       +2\, P_2\,P_3\, Q_1^2 \nonumber \\
      &&\qquad\qquad\qquad 
        -2\, P_2 - 4 P_3 - 4/m_\mu^2 \biggr)  
      \nonumber \\
      && -2\, P_1\,P_2 
         -3 \, P_1 \, (1-R_{m2})/(2 m_\mu^2)
         -3 \, P_2 \, (1-R_{m1})/(2 m_\mu^2) \nonumber \\ 
      && - P_3 \, (2 - R_{m1} - R_{m2}) / (2 m_\mu^2) \nonumber \\
      && + P_1 \, P_3 \, ( 2 + 3\,(1-R_{m2}) \, Q_2^2 / (2 m_\mu^2)
         + (1-R_{m2})^2 \, (Q_1 \cdot Q_2) \, / (2 m_\mu^2))
      \nonumber \\
      && + P_2 \, P_3 \, (2 + 3\,(1-R_{m1})\,Q_1^2/(2 m_\mu^2) 
         + (1-R_{m1})^2 \, (Q_1 \cdot Q_2) \, / (2 m_\mu^2)), 
      \nonumber \\   
      && \label{I2} 
\eea
where\footnote{Except in $Q_1 \cdot Q_2$, we always use the notation
  $Q_i \equiv |(Q_i)_{\mu}|, i=1,2$, for the length of the
  Euclidean four-momenta.}
\bea
Q_3^2 & = & (Q_1 + Q_2)^2 = Q_1^2 + 2 Q_1 \cdot Q_2 + Q_2^2, \\ 
Q_1 \cdot Q_2 & = & Q_1 Q_2 \tau, \\
\tau & = & \cos\theta, 
\eea 
and we introduced the notation $P_1^2 = 1/Q_1^2, P_2^2 = 1/Q_2^2,
P_3^2 = 1/Q_3^2$ for the photon propagators. Furthermore
\bea  
X(Q_1,Q_2,\tau) & = & \frac{1}{Q_1 Q_2\,x} \arctan
  \left(\frac{zx}{1-z\tau}\right), \\ 
x & = & \sqrt{1-\tau^2}, \\ 
z & = & \frac{Q_1 Q_2}{4 m_\mu^2} \left(1-R_{m1}\right)
\left(1-R_{m2}\right), \\  
R_{mi} & = & \sqrt{1+4m_\mu^2/Q_i^2},  \quad i=1,2. 
\eea 
Note that $w_2(Q_1,Q_2,\tau)$ in Eq.~(\ref{w2_App}) and $I_2(Q_1,Q_2,\tau)$
in Eq.~(\ref{I2}) are symmetric under $Q_1 \leftrightarrow Q_2$.

For small momenta, these weight functions have the following behavior
\bea 
\lim_{Q_1 \to 0} w_{1}(Q_1, Q_2, \tau) & = & \frac{16 \pi}{3 m_\mu^2}
\frac{m_\mu Q_2 x^3 + (Q_2^2 - 2 m_\mu^2) x^2 A(Q_2,\tau)}{Q_2^2 +
  m_\pi^2} Q_1^2 + \order \left( Q_1^3 \right), \\  
\lim_{Q_2 \to 0} w_{1}(Q_1, Q_2, \tau) & = & - \frac{32 \pi}{3 m_\pi^2}
x^2 A(Q_1,\tau) Q_2^2 + \order \left( Q_2^3 \right), \\  
\lim_{Q \to 0} w_{1}(Q, Q, \tau) & = & - \frac{16 \pi}{3 m_\pi^2} (1 - \tau)
\arccot \left[ \frac{-1 + \tau}{x} \right] Q^2 + \order \left(
  Q^3 \right), 
\eea    

\bea 
\lim_{Q_1 \to 0} w_{2}(Q_1, Q_2, \tau) & = & - \frac{4 \pi}{3 m_\mu^2}
\frac{Q_2 x (2 m_\mu - Q_2 \tau (1 - R_{m2})) + 2 (Q_2^2 + 2
  m_\mu^2 x^2) A(Q_2,\tau)}{Q_2^2 + m_\pi^2} Q_1^2 \nonumber 
\\ 
& & + \order \left( Q_1^3 \right), \\ 
\lim_{Q_2 \to 0} w_{2}(Q_1, Q_2, \tau) & = & - \frac{4 \pi}{3 m_\mu^2}
\frac{Q_1 x (2 m_\mu - Q_1 \tau (1 - R_{m1})) + 2 (Q_1^2 + 2
  m_\mu^2 x^2) A(Q_1,\tau)}{Q_1^2 + m_\pi^2} Q_2^2 \nonumber 
\\ 
& & + \order \left( Q_2^3 \right), \\ 
\lim_{Q \to 0} w_{2}(Q, Q, \tau) & = & - \frac{8 \pi}{3 m_\pi^2} (1 - \tau)
\arccot \left[ \frac{-1 + \tau}{x} \right] Q^2 + \order \left(
  Q^3 \right), 
\eea
where we introduced the abbreviation 
\be
A(Q_i,\tau) = \arctan\left[ \frac{Q_i \, x \, (1 - R_{mi})}{Q_i \, \tau
    \, (1 - R_{mi}) + 2 m_\mu} \right], \quad i = 1,2. 
\ee

One observes that the weight functions $w_{1,2}(Q_1,Q_2,\tau)$ not only
vanish for small momenta, but also the slopes along the two axis and
along the diagonal $Q_1 = Q_2 = Q$ are zero
\be \label{slopes} 
\left. \frac{\partial w_{1,2}(Q_1,Q_2,\tau)}{\partial Q_1} \right|_{Q_1 =
  0} = \ 
\left. \frac{\partial w_{1,2}(Q_1,Q_2,\tau)}{\partial Q_2} \right|_{Q_2 =
  0} = \ 
\left. \frac{\partial w_{1,2}(Q,Q,\tau)}{\partial Q} \right|_{Q =
  0} = \ 0. 
\ee

\noindent 
On the other hand, for large momenta we get 
\bea
\lim_{Q_1 \to \infty} w_{1}(Q_1, Q_2,\tau) & = & \frac{8\pi}{3 m_\mu^2}
\frac{Q_2^3 \left(Q_2^2 - R_{m2} (Q_2^2 - 2 m_\mu^2) \right)
  x^3}{(Q_2^2 + m_\pi^2)} \frac{1}{Q_1} + \order \left( \frac{1}{Q_1^2} 
\right), \label{w1_large_Q1} \\ 
\lim_{Q_2 \to \infty} w_{1}(Q_1, Q_2,\tau) & = & - \frac{8\pi}{3 m_\mu^2}
Q_1^2 \left(2 m_\mu^2 + Q_1^2 (1 - R_{m1}) \right)
  \tau \, x^3 \frac{1}{Q_2^2} + \order \left( \frac{1}{Q_2^3}
\right), \label{w1_large_Q2} \\ 
\lim_{Q \to \infty} w_{1}(Q, Q,\tau) & = & \frac{8\pi m_\mu^2}{3}
(3-\tau) (1-\tau) x \frac{1}{Q^2}+ \order \left( \frac{1}{Q^4} \right), 
\eea 

\bea
\lim_{Q_1 \to \infty} w_{2}(Q_1, Q_2,\tau) & = & - \frac{8\pi}{9 m_\mu^2}
Q_2^3 \left(m_\mu^2 (3 - R_{m2}) + Q_2^2 (1 - R_{m2}) \right)
  x^3 \frac{1}{Q_1^3} \nonumber \\
& & + \order \left( \frac{1}{Q_1^4} \right), \\ 
\lim_{Q_2 \to \infty} w_{2}(Q_1, Q_2,\tau) & = & - \frac{8\pi}{9 m_\mu^2}
Q_1^3 \left(m_\mu^2 (3 - R_{m1}) + Q_1^2 (1 - R_{m1}) \right)
  x^3 \frac{1}{Q_2^3} \nonumber \\
& & + \order \left( \frac{1}{Q_2^4} \right), \\ 
\lim_{Q \to \infty} w_{2}(Q, Q,\tau) & = & \frac{4\pi m_\mu^4}{9}
\frac{(2-\tau) (1-\tau)^{3/2}}{\sqrt{1+\tau}} \frac{1}{Q^4} + \order
\left( \frac{1}{Q^6} 
\right). \label{w2_large_Q}    
\eea
The slower fall off of $w_1(Q_1,Q_2,\tau)$ for large $Q_1$ in
Eq.~(\ref{w1_large_Q1}), compared to the behavior for large $Q_2$ in
Eq.~(\ref{w1_large_Q2}), leads to the ridge seen in
Fig.~\ref{Fig:w_i_pion} and, for a constant Wess-Zumino-Witten form
factor, to the $\ln^2\Lambda$ divergence for some momentum cutoff
$\Lambda$. Of course, the symmetric function $w_2(Q_1,Q_2,\tau)$
cannot show such a behavior.

Finally, for $\tau \to \pm 1$, the weight functions behave as follows 
\bea
\lim_{\tau \to 1} w_{1}(Q_1,Q_2,\tau) & = & \frac{16 \sqrt{2} \pi}{3
  m_\mu^2} \frac{(1 - R_{m1}) Q_1^3 Q_2^3}{(Q_2^2 + m_\pi^2) (Q_1 +
  Q_2)^2 \left( 4 m_\mu^2 - Q_1 Q_2 (1 - R_{m1}) (1 - R_{m2}) \right)}
\nonumber \\  
& & \quad \times \left[ - 4 m_\mu^2 R_{m2} +
  \left( Q_2 (1 - R_{m2}) \left( Q_1 (1 - R_{m1})  - 2 Q_2 \right)
  \right) \right] (1 - \tau)^{3/2} \nonumber \\  
& & + \order \left( (1 - \tau)^{5/2} \right), 
\\ 
\lim_{\tau \to 1} w_{1}(Q,Q,\tau) & = & - \frac{8\sqrt{2}\pi}{3}
\frac{Q^2 (1 - R_m) }{Q^2 + m_\pi^2} (1 - \tau)^{3/2} 
+ \order \left( (1 - \tau)^{5/2} \right), \\  
\lim_{\tau \to -1} w_{1}(Q_1,Q_2,\tau) & = & \frac{16 \sqrt{2} \pi}{3
  m_\mu^2} \frac{(1 - R_{m1}) Q_1^3 Q_2^3}{(Q_2^2 + m_\pi^2) (Q_1 -
  Q_2)^2 \left( 4 m_\mu^2 + Q_1 Q_2 (1 - R_{m1}) (1 - R_{m2})
  \right)} \nonumber \\  
& & \quad \times \left[ - 4 m_\mu^2 R_{m2} - 
  \left( Q_2 (1 - R_{m2}) \left( Q_1 (1 - R_{m1})  + 2 Q_2 \right)
  \right) \right]
(1 + \tau)^{3/2} \nonumber \\
& & + \order \left( (1 + \tau)^{5/2} \right), 
\\ 
\lim_{\tau \to -1} w_{1}(Q,Q,\tau) & = & \frac{32\sqrt{2}\pi}{3 m_\mu^2} 
\frac{Q^2 \left( 2 m_\mu^4 - 2 m_\mu^2 Q^2 - Q^4 (1 - R_m) \right)}{(Q^2 + 
  m_\pi^2) \left( 4 m_\mu^2 + Q^2 (1 - R_m) \right)} \sqrt{1 + \tau}
\nonumber \\ 
& & + \order \left( (1 + \tau)^{3/2}
\right),  \label{w1_tau_to_-1_diagonal} 
\eea

\bea
\lim_{\tau \to 1} w_{2}(Q_1,Q_2,\tau) & = & - \frac{16 \sqrt{2} \pi}{9
  m_\mu^2} \frac{ Q_1^2 Q_2^2}{((Q_1 + Q_2)^2 + m_\pi^2) (Q_1 +
  Q_2)^3} \nonumber \\
& & \quad \times \left[ Q_1^3 (1 - R_{m1}) + Q_2^3 (1 - R_{m2})
\right. \nonumber \\
& & \left. \quad \quad + m_\mu^2 \left( Q_1 (3 - R_{m1}) + Q_2 (3 -
    R_{m2})  \right) \right] (1-\tau)^{3/2} \nonumber \\  
& & + \order \left( (1 - \tau)^{5/2} \right), 
\\
\lim_{\tau \to 1} w_{2}(Q,Q,\tau) & = & - \frac{4\sqrt{2}\pi}{9 m_\mu^2}
\frac{Q^2 \left( m_\mu^2 (3 - R_m) + Q^2 (1 - R_m) \right)}{4 Q^2 + 
  m_\pi^2} (1 - \tau)^{3/2} \nonumber \\
& & + \order \left( (1 - \tau)^{5/2} \right), \\
\lim_{\tau \to - 1} w_{2}(Q_1,Q_2,\tau) & = & \frac{16 \sqrt{2} \pi}{9
  m_\mu^2} \frac{ Q_1^2 Q_2^2}{((Q_1 - Q_2)^2 + m_\pi^2) (Q_1 -
  Q_2)^3} \nonumber \\
& & \quad \times \left[ Q_1^3 (1 - R_{m1}) - Q_2^3 (1 - R_{m2})
    \right. \nonumber \\
& & \left. \quad \quad + m_\mu^2 \left( Q_1 (3 - R_{m1}) - Q_2 (3 -
    R_{m2})  \right) \right] (1 + \tau)^{3/2} \nonumber \\  
& & + \order \left( (1 + \tau)^{5/2} \right), 
\\
\lim_{\tau \to -1} w_{2}(Q,Q,\tau) & = & \frac{8\sqrt{2}\pi}{3 m_\mu^2
  m_\pi^2} \frac{Q^2 \left( 4 m_\mu^4 + m_\mu^2 Q^2 (5 - 3 R_m) + Q^4
    (1 - R_m) \right)}{4 m_\mu^2 + Q^2} \sqrt{1 + \tau} \nonumber \\ 
& & + \order \left( (1 + \tau)^{3/2} \right),  \label{w2_tau_to_-1_diagonal} 
\eea 
where 
\be 
R_m = \sqrt{1 + 4 m_\mu^2/Q^2}. 
\ee

The weight functions $w_{1,2}(Q_1,Q_2,\tau)$ vanish for $\tau \to \pm
1$. However, for $Q_1 = Q_2 = Q$ and $\tau \to -1$, when the original
four-vectors $(Q_1)_\mu$ and $(Q_2)_\mu$ become more and more
antiparallel, the approach to zero is much steeper, even with infinite
slope, compare Eqs.~(\ref{w1_tau_to_-1_diagonal}) and
(\ref{w2_tau_to_-1_diagonal}) to the other equations, see also
Fig.~\ref{Fig:weightfunctions_1Dplots}.

\section{Form factor models}
\label{App:TFF_models} 

\setcounter{equation}{0}

For illustration, we present in this Appendix briefly the definitions
of the two form factor models used in the main text. Much more details
and all the derivations can be found in Refs.~\cite{KN_EPJC_01, KN_02,
  N_09}. We use these form factor models only for illustration, since
we are interested in the impact of current and future experimental
form factor uncertainties on the pseudoscalar-pole contribution to
HLbL in the muon $g-2$. Those uncertainties have been parametrized by
the functions $\delta_{1, {\rm P}}(Q)$ and $\delta_{2, {\rm
    P}}(Q_1,Q_2)$ introduced in Eqs.~(\ref{delta1}) and
(\ref{delta2}). Therefore we list in the following only the central
values of the model parameters and not their uncertainties, which are
related to their extraction from experimental data or by imposing
theoretical constraints and assumptions.

We first discuss the pion.  The form factor is normalized to the decay
width $\Gamma(\pi^0 \to \gamma\gamma) =
7.63~\mbox{eV}$~\cite{PDG_2015} which is quite well reproduced by the
chiral anomaly (constant Wess-Zumino-Witten form factor)
\be \label{normalization_anomaly}
\FFa(0,0) = \FFa^{\rm WZW}(q_1^2, q_2^2) \equiv -\frac{N_c}{12 \pi^2
  F_\pi} \, ,  
\ee
if one sets $N_c = 3$ and uses the pion decay constant $F_\pi =
92.4~\mbox{MeV}$ obtained from the weak decay of the
charged pion.

\subsection*{LMD+V model} 

The LMD+V model for the pion-photon form factor $\FFa(q_1^2,q_2^2)$ is
rooted in the Minimal Hadronic Approximation (MHA)~\cite{MHA_LMD} to
Green's functions in large-$N_c$ QCD. One starts with an ansatz for
the three point function $\VVP$ and thus the form factor
$\FFa(q_1^2,q_2^2)$ in the chiral limit with one multiplet of the
lightest pseudoscalars (Goldstone bosons) and two multiplets of vector
resonances $\rho$ and $\rho^\prime$: Lowest Meson Dominance (LMD) +
V. The functions $\VVP$ and $\FFa(q_1^2,q_2^2)$ fulfill all leading
and some subleading QCD short-distance constraints from the operator
product expansion (OPE)~\cite{OPE}.\footnote{Recently, the ansatz for
  $\VVP$ was generalized to two-multiplets of pseudoscalars $\pi,
  \pi^\prime$ and two-multiplets of vector mesons $\rho, \rho^\prime$
  in Ref.~\cite{Two_Hadron_Saturation}.}

In particular, in the chiral limit, one obtains from the OPE a
condition for the form factor when both momenta are equal and large in
the Euclidean~\cite{OPE_TFF_1,OPE_TFF_2}
\be \label{OPE_condition} 
\lim\limits_{Q^2 \to \infty} \FFa(-Q^2,-Q^2) = 
- \frac{2 F_\pi}{3} \left\{ \frac{1}{Q^2} - \frac{8}{9}
  \frac{\delta^2}{Q^4} + \order\left( \frac{1}{Q^6} \right) \right\},  
\ee 
where $\order(\alpha_s)$ corrections are neglected and the quantity
$\delta^2 = (0.2 \pm 0.02)~\mbox{GeV}^2$ parametrizes the higher-twist
matrix element in the OPE in the chiral limit. It was determined in
Ref.~\cite{OPE_TFF_2} using QCD sum rules.

Furthermore, one demands that the form factor reproduces the
Brodsky-Lepage (BL)~\cite{Brodsky-Lepage} behavior for the single-virtual
pion-photon transition form factor 
\be \label{BL_condition} 
\lim\limits_{Q^2 \to \infty} \FFa(-Q^2,0) = - \frac{2 F_\pi}{Q^2} +
\order\left( \frac{1}{Q^4} \right). 
\ee

\noindent
The LMD+V form then reads~\cite{KN_EPJC_01,KN_02,N_09}
\bea 
\FFa^{\rm LMD+V}(q_1^2, q_2^2) 
& = & \frac{F_\pi}{3}\, \frac{q_1^2\,q_2^2\,(q_1^2 + q_2^2) +
   h_1\,(q_1^2+q_2^2)^2 + \bar{h}_2 \,q_1^2\,q_2^2 +
   \bar{h}_5 \,(q_1^2+q_2^2) + \bar{h}_7}{(q_1^2-M_{V_1}^2) \,
   (q_1^2-M_{V_2}^2) \, (q_2^2-M_{V_1}^2) \, (q_2^2-M_{V_2}^2)} \, ,
 \nonumber \\  
& & \label{FF_LMD+V} 
\eea 
where we use for the vector meson masses that appear in
Eq.~(\ref{FF_LMD+V}) the values from
Refs.~\cite{KN_EPJC_01,KN_02,N_09} 
\bea
M_{V_1} & = & M_\rho = 775.49~\mbox{MeV}, \\
M_{V_2} & = & M_{\rho^\prime} = 1.465~\mbox{GeV}. 
\eea

On the other hand, the constants $h_i, \bar{h}_i$ are the free
parameters of the LMD+V model and have been determined in
Refs.~\cite{KN_EPJC_01,KN_02,MV_04,N_09} by several experimental and
theoretical constraints: 
\bea 
h_1 & = & 0 \quad \mbox{(to reproduce the BL
  behavior~(\ref{BL_condition}))},  \label{BL} \\ 
\bar{h}_2 & = & -4 (M_{V_1}^2 + M_{V_2}^2) + (16/9) \, \delta^2
= -10.63~\mbox{GeV}^2 
\quad \mbox{(following Ref.~\cite{MV_04})},  \label{h2} \\
\bar{h}_5 & = & (6.93 \pm 0.26)~\mbox{GeV}^4   
\quad \mbox{(from a fit to CLEO data~\cite{CLEO} in
  Ref.~\cite{KN_EPJC_01})},  \\ 
\bar{h}_7 & = & - \frac{N_c M_{V_1}^4 M_{V_2}^4}{4 \pi^2 F_\pi^2} =
-14.83~\mbox{GeV}^6 \quad \mbox{(from chiral anomaly)}. \label{h7}  
\eea

The OPE condition when all momenta in $\VVP$ are large and all the
currents approach one point, uniquely fixes the first term in the
numerator in Eq.~(\ref{FF_LMD+V}). This also shows that the form
factor cannot factorize in QCD: $\FFa(q_1^2,q_2^2) \neq f(q_1^2)
\times f(q_2^2)$. Note that the OPE~(\ref{OPE_condition}) and
BL~(\ref{BL_condition}) conditions cannot be simultaneously satisfied
with only one vector meson multiplet (LMD form factor), see
Ref.~\cite{KN_EPJC_01}. 

We note that unless $\delta^2$ would be much different than the
estimate given below Eq.~(\ref{OPE_condition}), the size and in
particular the negative sign of $\bar{h}_2$ is determined almost
completely by the first term in Eq.~(\ref{h2}) involving the masses
$M_{V_1}$ and $M_{V_2}$. This leads to tensions to reproduce the decay
rate $\pi^0 \to e^+ e^-$, see Refs.~\cite{KN_02,
  HLbL_PS_vs_lepton_pair_decay}.

For completeness, the result for the single-virtual pion-photon
transition form factor is given by 
\be \label{TFF_LMD+V}  
\FFa^{\rm LMD+V}(-Q^2, 0) 
= \frac{F_\pi}{3}\, \frac{1}{M_{V_1}^2
  M_{V_2}^2} \frac{ h_1 Q^4 - \bar{h}_5 Q^2 + \bar{h}_7}{(Q^2 +
  M_{V_1}^2) (Q^2 + M_{V_2}^2)} \, .  
\ee

\subsection*{VMD model} 

The well-known VMD model is given by 
\be \label{VMD} 
\FFa^{{\rm VMD}}(q_1^2,q_2^2) = 
-\frac{N_c}{12 \pi^2 F_\pi} \, 
\frac{M_V^4}{(q_1^2 - M_V^2) (q_2^2 - M_V^2)} \, .  
\ee 
Here the two free model parameters are $F_\pi$ (normalization of the
form factor) and the vector-meson mass $M_V (= M_\rho)$. Note that the
VMD model factorizes ${\cal F}^{\rm
  VMD}_{\pi^0\gamma^\ast\gamma^\ast}(q_1^2, q_2^2) = f(q_1^2) \times
f(q_2^2)$. This might be a too simplifying assumption and also
contradicts the OPE in QCD. Furthermore, the VMD model has a wrong
short-distance behavior:
\be \label{VMD_large_Q}
\FFa^{\rm VMD}(-Q^2,-Q^2) \sim \frac{1}{Q^4} \, , \qquad \mbox{for
  large~}Q^2,  
\ee
i.e.\ it falls off too fast compared to the OPE
prediction in Eq.~(\ref{OPE_condition}).

The single-virtual pion-photon transition form factor is given by 
\be 
\FFa^{\rm VMD}(-Q^2, 0) 
= - \frac{N_c}{12 \pi^2 F_\pi}
\frac{M_V^2}{Q^2 + M_V^2} \, . 
\ee

\subsection*{Comparison of the two form factor models}

\begin{figure}[h!]

\centerline{\hspace*{0.75cm}\includegraphics[width=0.525\textwidth]{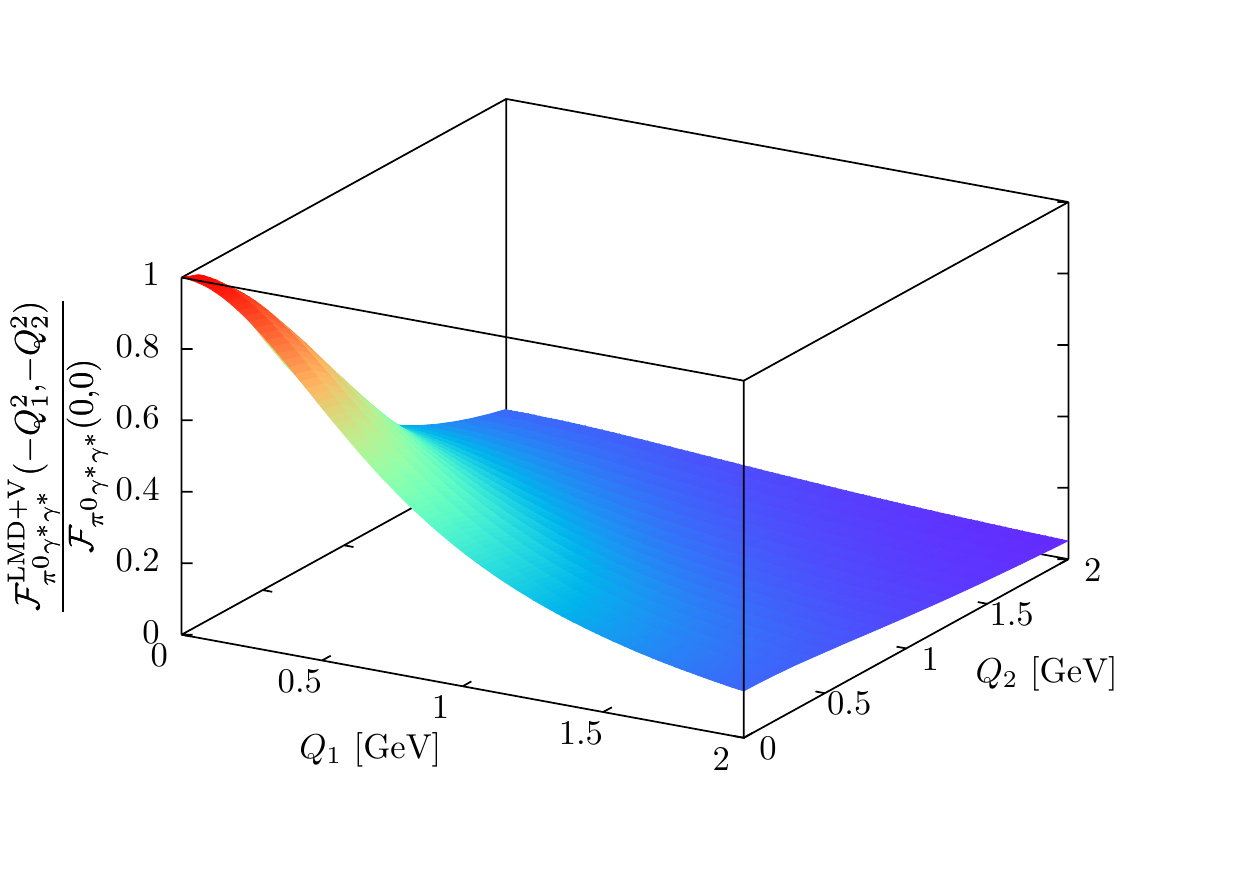}\hspace*{-0.25cm}\includegraphics[width=0.525\textwidth]{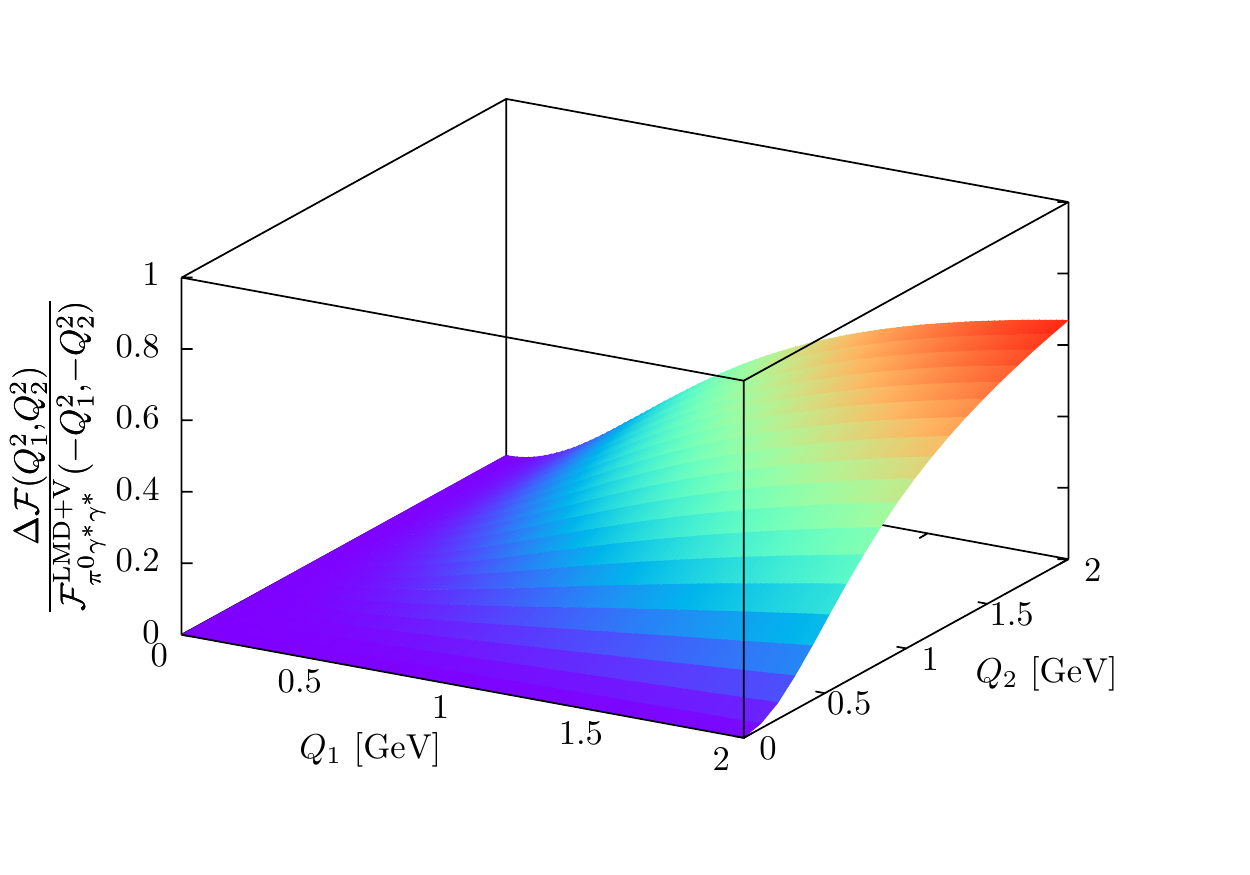}}

\caption{Left panel: normalized LMD+V form factor as a function of
  $Q_1$ and $Q_2$. Note the linear scale in $Q_{1,2}$. Right panel:
  relative difference between the LMD+V and VMD form factors.}

\label{Fig:LMD+V_vs_VMD}

\end{figure}

In order to compare the two form factor models, we introduce the
abbreviation 
\be 
\Delta {\cal F}(Q_1^2,Q_2^2) = \FFa^{{\rm LMD+V}}(-Q_1^2,-Q_2^2) - \FFa^{{\rm
    VMD}}(-Q_1^2,-Q_2^2). 
\ee
In Fig.~\ref{Fig:LMD+V_vs_VMD} we plot the LMD+V form factor
normalized to $\FFa(0,0)$ as a function of $Q_1$ and $Q_2$ for the
given model parameters in the region $Q_{1,2} \leq 2~\mbox{GeV}$ which
is the most relevant for the pion-pole contribution to the muon
$g-2$. As expected, one sees a damping for large momenta. We also plot
the difference between the LMD+V and the VMD form factors, expressed
through $\Delta {\cal F}(Q_1^2,Q_2^2)$, relative to the LMD+V
model. In Table~\ref{Tab:LMD+V_vs_VMD} we give, for a selection of
values of $Q_1$ and $Q_2$, the values for the normalized form factors
for the two models and their comparison.

\begin{table}[h!]

  \caption{Values of the normalized form factors in the LMD+V and VMD
    models and their absolute and relative difference for some
    selected momenta.}  

\label{Tab:LMD+V_vs_VMD}

\vspace*{-0.3cm}
\begin{center}
\renewcommand{\arraystretch}{1.25}
\begin{tabular}{|r@{}l|r@{}l|c|c|r@{.}l|r@{.}l|}
\hline 
\multicolumn{2}{|c|}{$Q_1$ \!\!\! [GeV]} & 
\multicolumn{2}{|c|}{$Q_2$ \!\!\! [GeV]} &
\rule[-0.5cm]{0cm}{1.2cm}$\frac{\FFa^{{\rm 
      LMD+V}}(-Q_1^2,-Q_2^2)}{\FFa(0,0)}$ &  
$\frac{\FFa^{{\rm VMD}}(-Q_1^2,-Q_2^2)}{\FFa(0,0)}$ & 
\multicolumn{2}{|c|}{$\frac{\Delta {\cal F}(Q_1^2,Q_2^2)}{\FFa(0,0)}$} & 
\multicolumn{2}{|c|}{$\frac{\Delta {\cal F}(Q_1^2,Q_2^2)}{\FFa^{{\rm
        LMD+V}}(-Q_1^2,-Q_2^2)}$} \\  
\hline 
~~~0 & .25 & ~~~0 & & 0.906 & 0.906 & ~~0 & 00008 & ~~~~~0 & 00009 \\ 
0 & .5  & 0 &     & 0.707 & 0.706 & 0 & 0002  & 0 & 0003  \\ 
0 & .75 & 0 &     & 0.517 & 0.517 & 0 & 0003  & 0 & 0006  \\ 
1 &     & 0 &     & 0.376 & 0.376 & 0 & 0004  & 0 & 001   \\
1 & .5  & 0 &     & 0.211 & 0.211 & 0 & 0003  & 0 & 002   \\
2 &     & 0 &     & 0.131 & 0.131 & 0 & 0003  & 0 & 002   \\
0 & .25 & 0 & .25 & 0.822 & 0.821 & 0 & 002   & 0 & 002   \\
0 & .5  & 0 & .5  & 0.513 & 0.499 & 0 & 014   & 0 & 027   \\
0 & .75 & 0 & .75 & 0.298 & 0.267 & 0 & 031   & 0 & 10    \\
1 &     & 1 &     & 0.183 & 0.141 & 0 & 042   & 0 & 23    \\
1 & .5  & 1 & .5  & 0.088 & 0.044 & 0 & 043   & 0 & 49    \\
2 &     & 2 &     & 0.052 & 0.017 & 0 & 035   & 0 & 67    \\ 
0 & .5  & 0 & .25 & 0.645 & 0.640 & 0 & 005   & 0 & 007   \\
0 & .75 & 0 & .25 & 0.475 & 0.468 & 0 & 007   & 0 & 015   \\
1 &     & 0 & .25 & 0.349 & 0.340 & 0 & 008   & 0 & 024   \\
\hline 
\end{tabular} 
\end{center}

\end{table}

Note that the form factors in both models reproduce equally well the
CLEO data~\cite{CLEO} for the single-virtual pion photon transition
form factor, once one puts $h_1 = 0$ in the LMD+V model and fits the
constant $\bar{h}_5$~\cite{KN_EPJC_01}. See Ref.~\cite{KLOE2_impact}
for recent fits to more experimental data. Therefore
\be
\FFa^{{\rm LMD+V}}(-Q^2,0) \approx \FFa^{{\rm VMD}}(-Q^2,0). 
\ee

Since the double-virtual LMD+V and VMD form factors
$\FFa(-Q_1^2,-Q_2^2)$ differ for $Q_1 = Q_2 = 1~[1.5]~\mbox{GeV}$ by
23\% [49\%], it might be possible to distinguish the two models
experimentally at BESIII~\cite{BESIII_private}, if the binning is
chosen properly, see also Fig.~\ref{Fig:FF_errors_bins} concerning the
expected (statistical) measurement precision of the double-virtual
form factor. For lower values of $Q_{1,2}$, which are more relevant
for the pion-pole contribution to the muon $g-2$, it might not be
possible to really distinguish the two models with the currently
envisioned data set at BESIII. For instance, for $Q_1 = Q_2 =
0.5~\mbox{GeV}$, the two models differ only by about 3\%.

\subsection*{Form factors for $\eta$ and $\eta^\prime$}

The short-distance analysis in Ref.~\cite{KN_EPJC_01} for the QCD
three point function $\VVP$ and the corresponding transition form
factor ${\cal F}_{{\rm P}\gamma^*\gamma^*}(q_1^2,q_2^2)$ for the LMD+V
ansatz in large-$N_c$ QCD was performed in the chiral limit and
assuming octet symmetry.  These are certainly not good approximations
for the more massive $\eta$ and $\eta^\prime$ mesons, where also the
nonet symmetry, the effect of the $U(1)_A$ anomaly and the
$\eta-\eta^\prime$-mixing have to be taken into account.

Following Refs.~\cite{KN_02,N_09}, we will therefore use a simple VMD
model for $\eta$ and $\eta^\prime$, as for the pion in
Eq.~(\ref{VMD}), but with the decay constant $F_{\rm P}$ fixed from
the decay width $\Gamma({\rm P} \to \gamma\gamma)$ and the value of
$M_V$ obtained from a fit to the CLEO data~\cite{CLEO} for the
single-virtual transition form factor ${\cal F}_{{\rm
    P}\gamma^*\gamma^*}(-Q^2,0)$:
\bea
\eta-\mbox{meson:} & & F_{\eta} = 93.0~\mbox{MeV}, \quad M_V = 774~\mbox{MeV},
\\
\eta^\prime-\mbox{meson:} & & F_{\eta^\prime} = 74.0~\mbox{MeV}, \quad
M_V = 859~\mbox{MeV}.  
\eea

\end{document}